\documentclass[]{aa}
\usepackage{graphics,graphicx}
\usepackage{longtable}
\usepackage{pdflscape}
\usepackage{txfonts}
\usepackage{natbib}
\usepackage{amssymb}
\begin{document} 

   \title{Exploring the formation pathways of formamide}

   \subtitle{near young O-type stars}

   \author{V. Allen\inst{1,2,3}
          \and
          F. F. S. van der Tak\inst{2,3}\fnmsep
          \and A. L{\'{o}}pez-Sepulcre\inst{4,5}
          \and {\'{A}}. S{\'{a}}nchez-Monge\inst{6}
          \and V. M. Rivilla\inst{7}
          \and R. Cesaroni\inst{7}
          }

   \institute{NASA Goddard Space Flight Center, 8800 Greenbelt Road, Greenbelt, MD 20771, USA\\
              \email{veronica.a.allen@nasa.gov}
              \and Kapteyn Astronomical Institute, University of Groningen,
              Landleven 12, 9747 AD Groningen, the Netherlands
             \and SRON, Landleven 12, 9747 AD Groningen, the Netherlands
             \and CNRS, IPAG, Univ. Grenoble Alpes, F-38000 Grenoble, France
             \and IRAM, 300 rue de la Piscine, 38406 Saint-Martin d' H{\`{e}}res, France
             \and I. Physikalisches Institut, Universit{\"{a}}t zu K{\"{o}}ln, Z{\"{u}}lpicher Straße 77, 50937 K{\"{o}}ln, Germany
             \and INAF, Osservatorio Astrofisico di Arcetri, Largo Enrico Fermi 5, I-50125 Florence, Italy
             }

   \date{Received 27 April 2019; accepted 18 December 2019}

 
  \abstract
   {As a building block for amino acids, formamide (NH$_2$CHO) is an important molecule in astrobiology and astrochemistry, but its formation path in the interstellar medium is not understood well.}
   {We aim to find empirical evidence to support the chemical relationships of formamide to HNCO and H$_2$CO.}
   {We examine high angular resolution ($\sim 0.2''$) Atacama Large Millimeter/submillimeter Array (ALMA) maps of six sources in three high-mass star-forming regions and compare the spatial extent, integrated emission peak position, and velocity structure of HNCO and H$_2$CO line emission with that of NH$_2$CHO by using moment maps.  Through spectral modeling, we compare the abundances of these three species.}
   {In these sources, the emission peak separation and velocity dispersion of formamide emission is most often similar to HNCO emission, while the velocity structure is generally just as similar to H$_2$CO and HNCO (within errors).  From the spectral modeling, we see that the abundances between all three of our focus species are correlated, and the relationship between NH$_2$CHO and HNCO reproduces the previously demonstrated abundance relationship.}
   {In this first interferometric study, which compares two potential parent species to NH$_2$CHO, we find that all moment maps for HNCO are more similar to NH$_2$CHO than H$_2$CO in one of our six sources (G24 A1). For the other five sources, the relationship between NH$_2$CHO, HNCO, and H$_2$CO is unclear as the different moment maps for each source are not consistently more similar to one species as opposed to the other.}


\titlerunning{The formation pathways of interstellar formamide}
\authorrunning{V. Allen et al.}

   \maketitle

\section{Introduction}
Formamide (NH$_2$CHO) is an important molecule to study for astrochemistry and astrobiology because its structure and content make it a likely precursor for glycine (NH$_2$CH$_2$COOH), the simplest amino acid and an important building block in the synthesis of prebiotic compounds. \cite{Saladino2012} argue that NH$_2$CHO may have played a key role in creating and sustaining life on the young Earth since it can lead to diversity in biologically relevant chemistry involving amino acids, nucleic acids, and sugars.
   
   \begin{figure}[!ht]
   	\centering
   	\includegraphics[width=0.78\hsize]{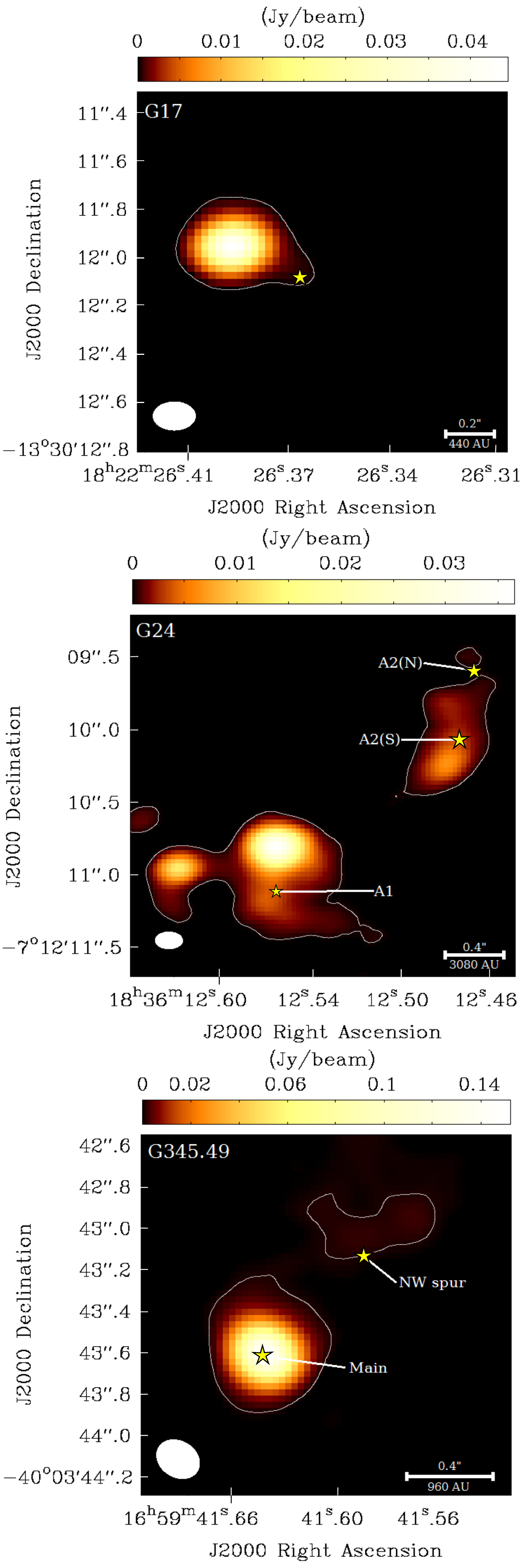}
   	\caption{Images of the 218 GHz continuum emission from Cycle 2 ALMA observations of our three regions focusing on the regions where NH$_2$CHO emission is observed. The color scale indicates the continuum flux as detailed in the color bar above each map. The white contour shows the 5$\sigma$ contour levels for each panel: 1.5, 2.5, and 2.5 mJy/beam.  The stars mark each of the spectral extraction points (coinciding with the zeroth moment peaks of NH$_2$CHO emission) and subsources are labeled.}
   	\label{contpts}
   \end{figure}
   
\par In recent years, two routes to forming NH$_2$CHO, which use common interstellar species, have been studied in depth. NH$_2$CHO forms either on dust grain ice mantles from the hydrogenation of isocyanic acid (HNCO) in the following reaction: HNCO + H + H $\to$ NH$_2$CHO \citep{Charnley1997}. Subsequently, it sublimates into the gas, where we see it in hot cores and hot corinos. The alternative is from reactions between H$_2$CO and NH$_2$ in warm gas (H$_2$CO + NH$_2$ $\to$ NH$_2$CHO + H) \citep{Kahane2013}. It is important to note that NH$_2$ is  especially abundant in photon-dominated regions. Other formation pathways have been tested in the lab \citep{Jones,Fedoseev2016,Skouteris2017}, but we do not investigate them here as the species involved fall outside the frequency range of our dataset. Laboratory studies on these reactions show that both HNCO and H$_2$CO can have a chemical relationship with NH$_2$CHO. An early study by \citet{Raunier2004} found that vacuum ultraviolet irradiation of pure HNCO ice resulted in NH$_2$CHO as a product. Recent laboratory work by \cite{Kanuchova2017} shows that sufficient amounts of NH$_2$CHO can form in cosmic-ray-irradiated ices but the HNCO/NH$_2$CHO ratio does not match observations.  The laboratory study by \cite{Noble2015} finds that hydrogenation of HNCO by deuterium bombardment does not lead to NH$_2$CHO in detectable quantities, while \cite{Barone2015} find that the H$_2$CO+NH$_2$ reaction can reproduce the abundance of NH$_2$CHO in IRAS16293-2422, a Sun-like protostar.  Recent work by \cite{Quenard2018}, which models the formation of HNCO and NH$_2$CHO and other peptide-bearing molecules with the N-C=O group, shows a correlation between the abundances of H$_2$CO and NH$_2$CHO as well as between HNCO and NH$_2$CHO without using hydrogenation. The chemical pathway studied in \citet{Fedoseev2016} indicates that NH$_2$ may be a key precursor to both HNCO and NH$_2$CHO, indicating that they are chemically related, but not in a reactant-product relationship. Recently, further theoretical formation pathways have been investigated through quantum chemical calculations by \citet{Darla2019} (NH$_3$~+~CO or NH$_3$~+~CO$^{+}$) and chemical kinetics by \citet{Vichietti2019} (H$_2$O~+~HCN), but we do not explore those species here.

\par Observational evidence has been found for both chemical relationships.  A tight empirical correlation has been observed using single dish observations between the abundances of HNCO and NH$_2$CHO which spans several orders of magnitude in molecular abundance \citep{Bisschop2007,Ana2015,Mendoza2014}.  This correlation between the abundances of these species is nearly linear, suggesting that the two molecules are chemically related. Atacama Large Millimeter/submillimeter Array (ALMA) observations by \cite{Coutens2016} of IRAS 16293-2422 show that the deuterium fractions in HNCO and NH$_2$CHO are very similar, implying a chemical link. On the other hand, \cite{Codella2017} observed a shock near L1157-B1 using interferometric observations. Through these observations and follow-up chemical modeling, they concluded that NH$_2$CHO is made efficiently in the gas phase from H$_2$CO, at least in this source.  The possibility exists that different types of sources (shocked regions, outflow cavities, accretion disks, protostellar envelopes, etc.) may have different dominant formation routes, but this possibility stands to be examined. A comprehensive review of all research into the formation of interstellar formamide can be found in \citet{AnaFormaReview}.

\par To make progress in the interpretation of the chemical link between these species, interferometric observations around a variety of sources are needed.  We previously used ALMA to study emission extent, peaks, and velocity structure between HNCO and NH$_2$CHO in G35.20-0.74N \citep{Allen}. In the Keplerian disk candidate G35.20-0.74N B, we found that the morphology and velocity structure of HNCO and NH$_2$CHO are almost identical, and the first moment velocity differs by less than 0.5 km s$^{-1}$. While this suggests that HNCO has a relationship to NH$_2$CHO in this source, we could not determine a relationship with H$_2$CO because those observations did not contain spectral windows with H$_2$CO lines for comparison.

\par  In this paper, we investigate the chemical relationships between HNCO, H$_2$CO and NH$_2$CHO using high-angular resolution ($\sim$0.2$''$ beam) ALMA observations to compare the emission morphology (\S\ref{mom0}), velocity structure (\S\ref{mom1}), and velocity dispersion (\S\ref{mom2}) of HNCO, H$_2$CO, and NH$_2$CHO emission in three high-mass star-forming regions (described in \S\ref{sources}).  To complement these observations, we use LTE spectral modeling to determine the column density, excitation temperature, average line width, and central velocity for each of these species in all the sources (\S\ref{xclasssection}).  We discuss the results in \S\ref{discussion} and summarize the main findings in \S\ref{conclusion}.


\section{Observations and method}

\subsection{Source sample}
\label{sources}
We observed three high-mass star forming regions with a high luminosity ($L_\mathrm{bol}= 1-2 \times 10^{5}$ L$_{\odot}$). Our sources (shown in Figure~\ref{contpts}) are a subset of the sample studied and presented in \cite{Cesaroni2017} selected for their potential as O-type (proto)stars harboring circumstellar disks. G17.64+0.16 (hereafter G17, also known as AFGL~2136 and IRAS~18196-1331), shown in the top panel of Figure~\ref{contpts}, is located at a distance of 2.2~kpc, has a bolometric luminosity of 1~$\times 10^{5}$~L$_{\odot}$ and has been well studied from the infrared to the radio.  G17 harbors a millimeter continuum source that is cospatial with weak H30$\alpha$ emission and a molecular plume to the west of the continuum source \citep{Maud2018}. G24.78+0.08 (hereafter G24), which is shown in the middle panel of Figure~\ref{contpts}, is located at a distance of 6.7~kpc \citep{Reid2019}, determined by trigonametric parallax, and has a bolometric luminosity of 1.7~$\times 10^{5}$~L$_{\odot}$.  There are several sources associated with this star-forming region but we focus on the hot molecular cores A1 and A2.  G24 A1 contains a hypercompact HII region ($\sim$1000 au) which has been determined to be expanding through methanol, water maser, and recombination line observations \citep{Beltran2007,Moscadelli2018}. G345.49+1.47 (hereafter G345, also known as IRAS~16562-3959), shown in the bottom panel of Figure~\ref{contpts}, is located at a distance of 2.4~kpc with a bolometric luminosity of 1.5~$\times 10^{5}$~L$_{\odot}$.  G345 has a continuum source associated with strong H30$\alpha$ emission (G345 Main) and a chemically rich region to the northwest of this continuum source (G345 NW spur) (Johnston et al. in prep). The other three sources from the dataset described in \cite{Cesaroni2017} could not be used in these investigations because the formamide lines were strongly blended with other species.

\subsection{Observations}
   The sources were observed with ALMA in Cycle 2 in July and September 2015 (2013.1.00489.S) in Band 6 with baselines from 40 to 1500~m. The observed frequency range was between 216.9 GHz to 236.5 GHz divided into 13 spectral windows.  The flux calibrators were Titan and Ceres and the phase calibrators were J1733-1304 (for G17 and G24) and J1709-3525 (for G345). The rms noise of the continuum maps ranges between 0.2 and 1.0 mJy beam$^{-1}$. The calibration and imaging were carried out using CASA\footnote{Common Astronomy Software Applications is available from http://casa.nrao.edu/}. A statistical method \citep{Alvaro2018} was used within the Python-based tool STATCONT\footnote{STATCONT is freely accessible here: https://hera.ph1.uni-koeln.de/$\sim$sanchez/statcont} for continuum subtraction as there were very few line free channels. The angular resolution is about 0.2$''$ and the spectral resolution in most spectral windows is 488.3 kHz, but higher (244.1 kHz) from 220.303-220.767 GHz and lower (1953.1 kHz) in the spectral window from $\sim$216.976-218.849 GHz.  The bandwidths for all spectral windows are $<$2~GHz with the largest being 1.8~GHz.  For full details on observations and continuum subtraction see \cite{Cesaroni2017}. The continuum intensity for subsources showing H30$\alpha$ emission were corrected for free-free emission with direct measurements for G17 from \citet{Maud2018}, G24 A1 from \citet{Moscadelli2018}, and calculated for G345 using the spectral index fit from \citet{Guzman2016}.
   
\begin{table}[!tbh]
	\centering
	\caption{Source properties}
	\label{sourceTable}
  \begin{tabular}{cccc}
			
			\hline\hline
			Source & v$_\mathrm{LSR}$ & Distance & L$_\mathrm{bol}$ \\
			& (km s$^{-1}$) & (kpc) & (10$^5$ L$_{\odot}$) \\
			\hline
			G17.64+0.16  & 22.5 & 2.2 & 1.0 \\
			G24.78+0.08 & 111.0 & 6.7 & 1.7 \\
			G345.49+1.47 & -12.6 & 2.4 & 1.5 \\
			\hline
		\end{tabular}
		\footnotesize{~~~~~~~~~~~~~~~~~~~~~~~~~~~~~~~~~~~~~~Distance and luminosity values from the RMS database \citep{RMS}\footnote{The Red MSX Source survey database at http://rms.leeds.ac.uk/cgi-bin/public/RMS$\_$DATABASE.cgi which was constructed with support from the Science and Technology Facilities Council of the UK.} except distance and L$_{bol}$ for G24.78+0.08 which is from \citet{Reid2019}.}
\end{table}

\subsection{Line identification}
\label{lineids}
Spectra were extracted from the positions indicated with a star in Figure~\ref{contpts} corresponding with the peak(s) of NH$_2$CHO emission (positions listed in Table~\ref{spectrapoints}) from the continuum subtracted images of each sub-source (except G345 Main) using CASA. We investigate if the NH$_2$CHO transitions are blended by performing simultaneous fits of the species NH$_2$CHO, HNCO, H$_2$CO, and species that were potentially blended with NH$_2$CHO (C$_2$H$_5$OH, CH$_3$CN ($\nu_8$=1), and CH$_3^{18}$OH) via the XCLASS\footnote{Available from: https://xclass.astro.uni-koeln.de/} software \citep{XCLASS} assuming local thermal equilibrium (LTE). This software models the data by solving the radiative transfer equation for an isothermal object in one dimension, taking into account source size and opacity. The observed spectra and the XCLASS fits are shown in Appendix~\ref{xclassfits}.  Using this software, we determine the excitation temperature ($T_\mathrm{ex}$), column density ($N_\mathrm{col}$), line width (FWHM), and velocity offset (v$_\mathrm{LSR}$) for each modeled species (see details in \S\ref{xclasssection}).  The model parameters FWHM and v$_\mathrm{LSR}$ were constrained using Gaussian fits of the observed transitions and allowed to vary $\pm$0.5 km s$^{-1}$ from the measured central velocity.  The $T_\mathrm{ex}$ free parameter was allowed to vary between 50 and 300~K for HNCO and NH$_2$CHO and between 70 and 400~K for H$_2$CO. The temperature of H$_2$CO was modeled using higher temperatures and a source size smaller than the beam size ($\sim0.15''$) in order to fit the emission originating on the small scale.  The range explored for $N_\mathrm{col}$ for each source is equivalent to abundances between 10$^{-13}$ and 10$^{-5}$.  Because G345 Main shows very strong continuum emission (T$_B$ $\sim$90~K) and absorption features, we used spectra extracted from non-continuum subtracted images and also modeled the continuum level within XCLASS.

\begin{table}[!ht]
	\centering
	\caption{Spectral extraction points for line identification and spectral modeling with XCLASS ($\S$\ref{xclasssection}). These points coincide with the NH$_2$CHO peak used for each source.}
	\label{spectrapoints}
	\tiny{
		\begin{tabular}{ccccc}
			
			\hline\hline
			Source & Right Ascension & Declination  & N$_{core}$ & H30$\alpha$ \\
			& (J2000) & (J2000) & (cm$^{-2}$) & \\
			\hline
			G17 & 18:22:26.370 & -13:30:12.06 & 1.0$\times10^{25}$ & \checkmark \\
			G24 A1 & 18:36:12.544 & -07:12:11.14 & 1.3$\times10^{24}$ & \checkmark \\
			G24 A2(N) & 18:36:12.465 & -07:12:09.61 & 9.9$\times10^{23}$ & \\
			G24 A2(S) & 18:36:12.471 & -07:12:10.09 & 8.2$\times10^{23}$ & \\
			G345 Main & 16:59:41.628 & -40:03:43.63 & 9.8$\times10^{25}$ & \checkmark \\
			G345 NW spur & 16:59:41.586 & -40:03:43.15 & 2.3$\times10^{25}$ & \\
			\hline& 
		\end{tabular}
		\tablefoot{N$_{core}$ was determined as in \citet{Alvaro2014} using the continuum intensity at the spectral extraction point assuming a T$_{dust}$ of 100~K, a dust opacity of 1.0 cm$^2$ g$^{-1}$ \citep{Ossenkopf1994}, and a gas-to-dust ratio of 100. Check mark (\checkmark) symbols indicate the detection of H30$\alpha$ emission toward the sub-source. For these sources, the continuum was corrected for free-free emission.}
	}
\end{table}

\begin{table*}[!bht]
  \centering
  \caption{Transition properties from the Cologne Database for Molecular Spectroscopy (CDMS)$^{\mathrm{a}}$ \citep{CDMS}. The last column shows the sources in which this transition appeared. HNCO (3) has a much higher upper energy level than the other transitions, so we consider it cautiously.}
  	  \label{linestats}
  \begin{tabular}{cccccc}

  \hline\hline
  Species & Transition & Frequency & $E_\mathrm{up}$ & $A_\mathrm{ij}$ & Sources\\
   & & (MHz) & (K) & (s$^{-1}$) \\
  \hline
  HNCO (1) & 10$_{0,10}$-9$_{0,9}$ & 219798.27 & 58.0 & 1.47$\times$10$^{-4}$ & G24 \\
  HNCO (2) & 10$_{1,9}$-9$_{1,8}$ & 220584.75 & 101.5 & 1.45$\times$10$^{-4}$ & G17, G345\\
  HNCO (3) & 10$_{3,7}$-9$_{3,6}$ & 219656.77 & 432.9 & 1.20$\times$10$^{-4}$ & G24, G345\\
  NH$_2$CHO (1) & 10$_{1,9}$-9$_{1,8}$ & 218459.21 & 60.8 & 7.47$\times$10$^{-4}$ & G17, G345\\
  NH$_2$CHO (2) & 11$_{2,10}$-10$_{2,9}$ & 232273.64 & 78.9 & 8.81$\times$10$^{-4}$ & G24\\
  H$_2$CO (1) & 3$_{0,3}$-2$_{0,2}$ & 218222.19 & 20.9 & 2.82$\times$10$^{-4}$ & G17, G24, G345\\
  H$_2$CO (2) & 3$_{2,2}$-2$_{2,1}$ & 218475.63 & 68.1 & 1.57$\times$10$^{-4}$ & G17, G24, G345\\
  H$_2$CO (3) & 3$_{2,1}$-2$_{2,0}$ & 218760.07 & 68.1 & 1.58$\times$10$^{-4}$ & G17, G24, G345\\
  \hline 
  \end{tabular}
	\tablefoot{a) The CDMS catalog can be accessed here: https://cdms.astro.uni-koeln.de/}
  \end{table*}

\begin{figure*}[h]
	\centering
	\includegraphics[width=0.8\hsize]{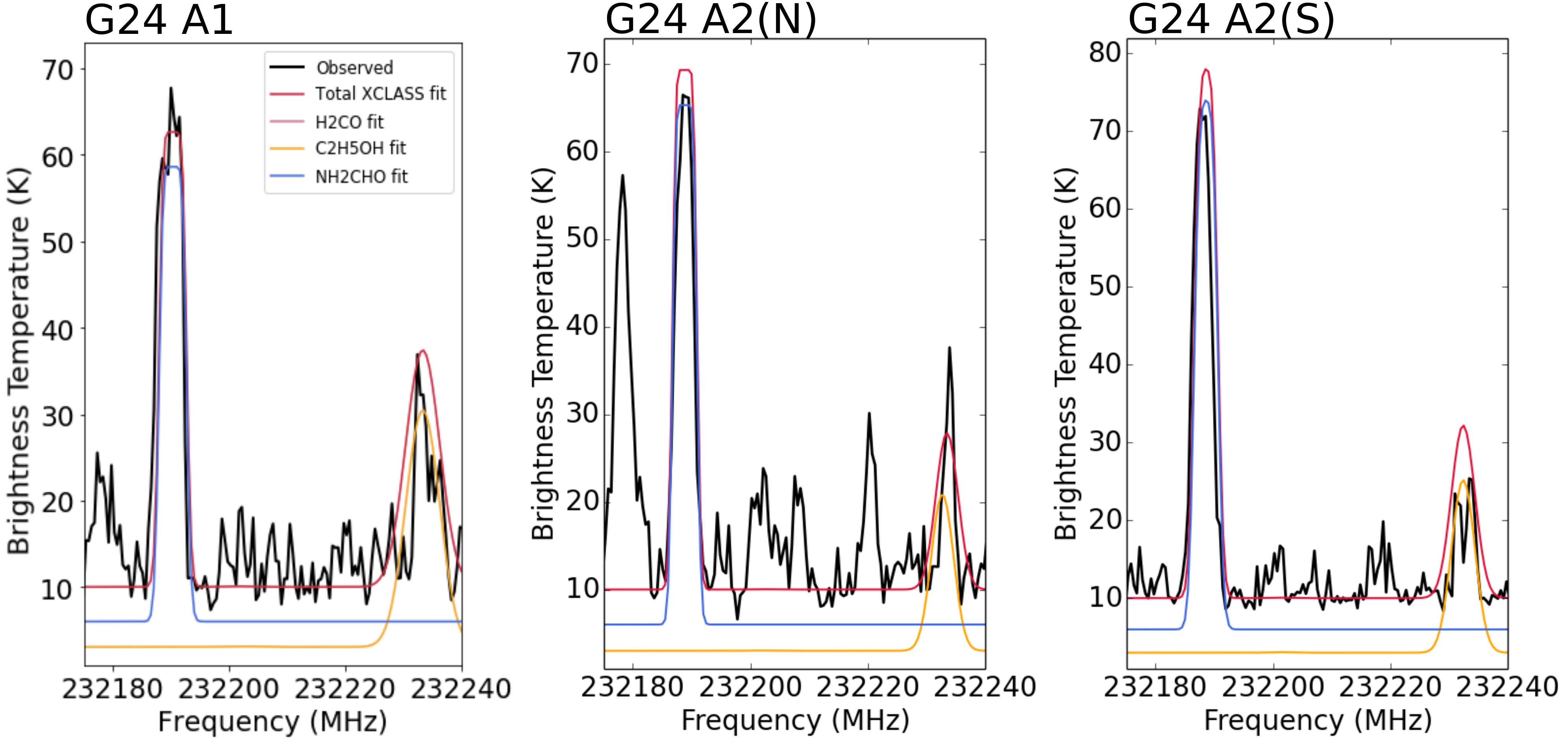}
	\caption{Observed and synthetic spectra of G24 A1 (left), A2(N) (middle), and A2(S) (right) with fits showing NH$_2$CHO (2) (dark blue). The continuum levels are offset for easy viewing.}
	\label{G24}
\end{figure*}
\begin{figure*}[h]
	\centering
	\includegraphics[width=0.8\hsize]{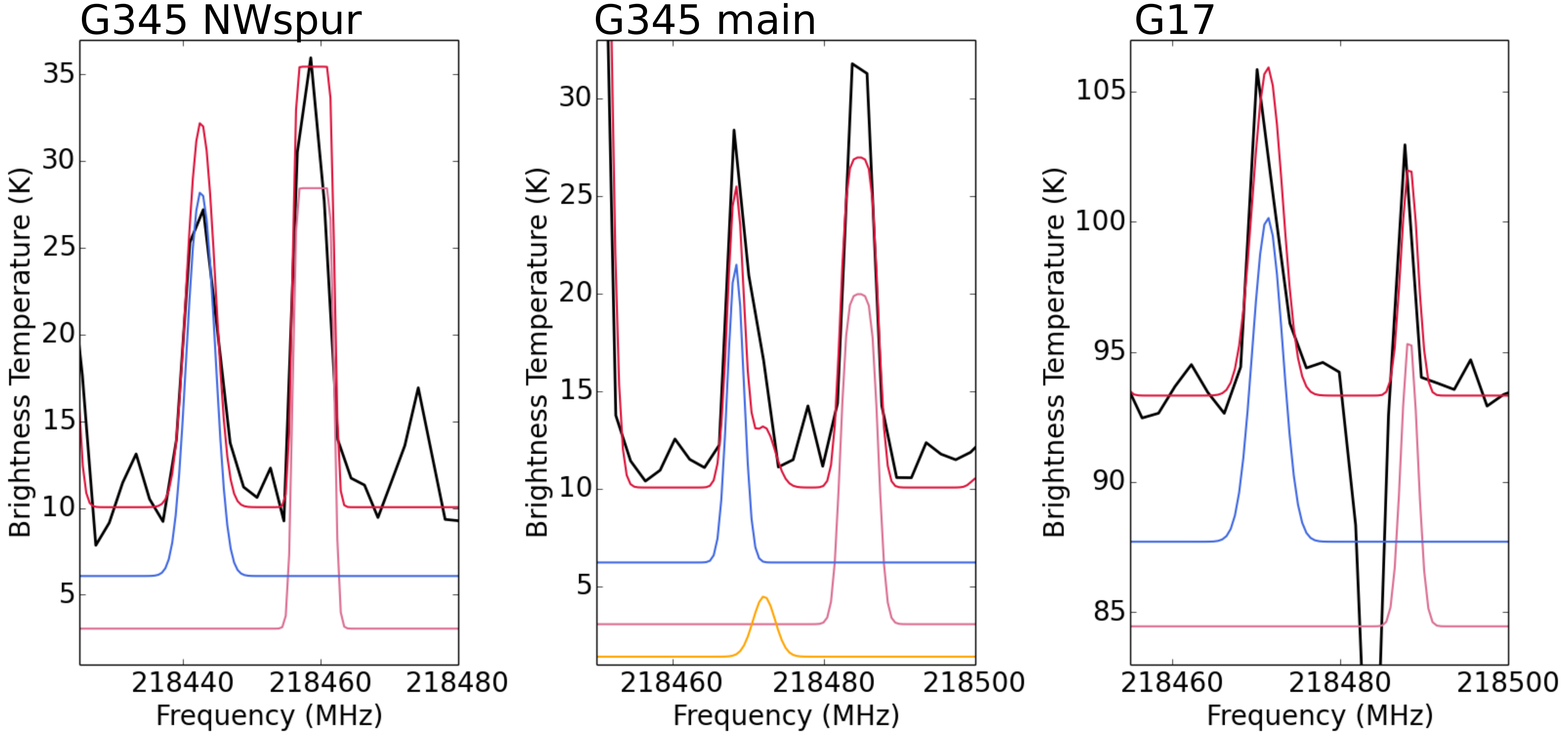}
	\caption{Spectra with fits showing NH$_2$CHO (1) toward G345 NW spur (left), G345 Main (middle) and G17 (right). NH$_2$CHO (1) is weakly blended with C$_2$H$_5$OH in the spectra of G345 NW spur but unblended in G345 Main. The continuum levels are offset for easy viewing.}
	\label{G345G17}
\end{figure*}

\par In general, we compare transitions with similar E$_\mathrm{up}$ values (60-100~K) except where we consider HNCO (3) which has an E$_\mathrm{up}$ of 432.9~K. Additionally, where one species can have strong emission due to larger abundances, another can be undetected because the telescope is not sensitive enough to detect a much weaker signal.  The transitions used in the analysis in this work are listed in Table~\ref{linestats}. There were two different unblended transitions of NH$_2$CHO used: NH$_2$CHO (1) used for G17 and G345 and NH$_2$CHO (2) for G24.  NH$_2$CHO (2) (defined in Table~\ref{linestats}) is the best transition as it is unlikely to be blended (see the best fit spectra in Figure~\ref{G24}) but it only appears within the spectral windows of G24 due to its high v$_\mathrm{LSR}$ (Table~\ref{sourceTable}).  The transitions identified for HNCO and H$_2$CO are generally unblended, but the NH$_2$CHO (1) emission is potentially blended with ethanol (C$_2$H$_5$OH) in G345 (Figure~\ref{G345G17}). In G17, C$_2$H$_5$OH is not detected, so NH$_2$CHO (1) is considered to be unblended. For NH$_2$CHO (1) in G345, we compare the C$_2$H$_5$OH transition that can produce a blend with NH$_2$CHO (at 218461.23~MHz) with a similar transition with the same upper energy level, $E_\mathrm{up}$, (23.9~K) and almost the same Einstein coefficient, $A_\mathrm{ij}$, (6.54$\times10^{-5}$ vs. 6.60$\times10^{-5}$~s$^{-1}$) at 217803.69~MHz.  We use the NH$_2$CHO (1) transition for G345 for three reasons: the emission in G345 from the isolated  C$_2$H$_5$OH transition at 217803.69 MHz is much weaker than the line that is blended with NH$_2$CHO (1), the peak integrated emission of NH$_2$CHO (1) is $\sim$8 times stronger than that of C$_2$H$_5$OH (0.24 vs. 0.03 Jy beam$^{-1}$.km s$^{-1}$), and the two have completely different morphology (see Figure~\ref{notethanol} in Appendix A). 

\begin{figure*}[!thb]
	\centering
	\includegraphics[width=\hsize]{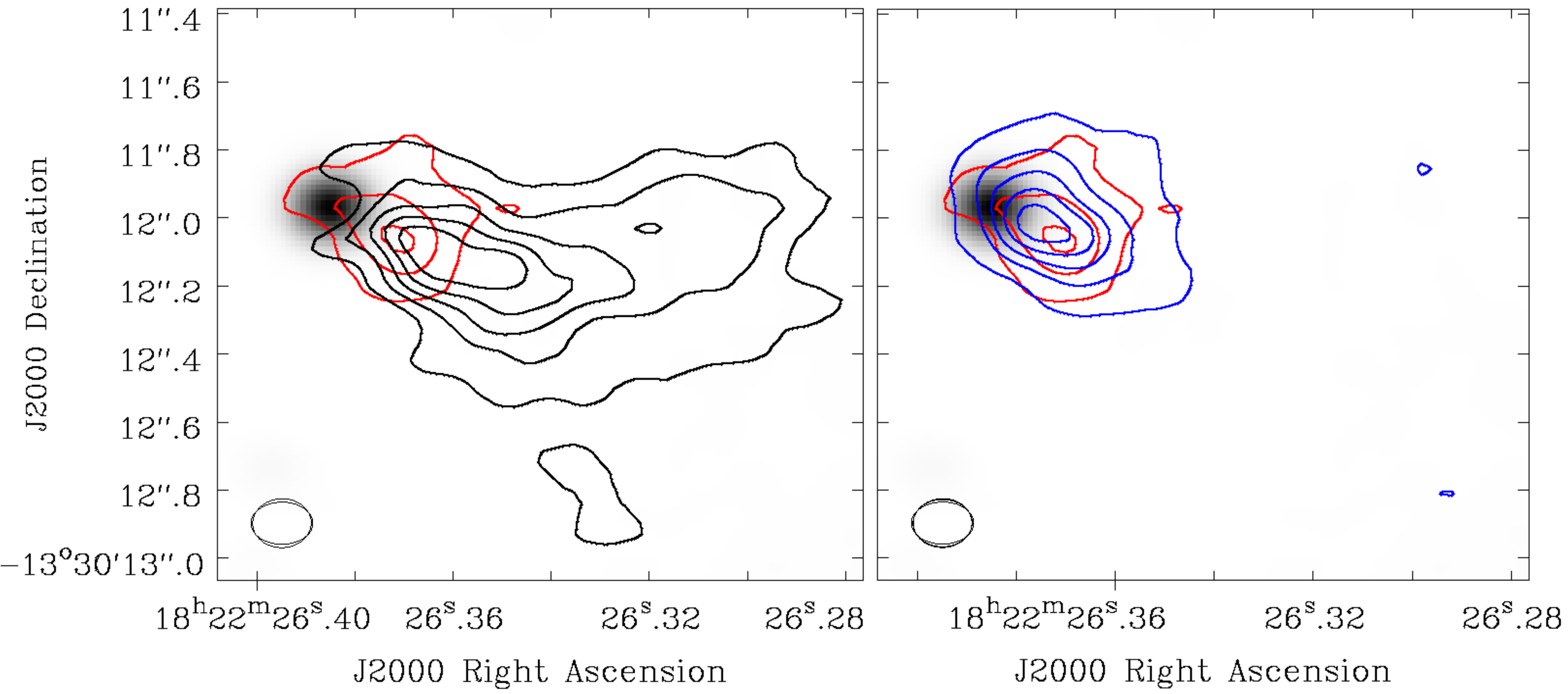}
	\caption{G17 zeroth moment maps (contours) overlaid on the dust continuum (grayscale). \textit{Left:} the black contours show the H$_2$CO (2) transition ($E_\mathrm{up}$=68.1~K) from 5$\sigma$ to a peak of 0.175 Jy/beam km s$^{-1}$ (contour levels 0.021, 0.052, 0.083, 0.113, and 0.144 Jy/beam km s$^{-1}$). The red contours show NH$_2$CHO (1) emission (($E_\mathrm{up}$=60.8~K) from 5$\sigma$ to 0.268 Jy/beam km s$^{-1}$ (contour levels 0.022, 0.071, 0.120, 0.170, and 0.219 Jy/beam km s$^{-1}$). \textit{Right:} the blue contours show the extent of the HNCO (2) emission ($E_\mathrm{up}$=101.5~K) from 5$\sigma$ to 0.146 Jy/beam km s$^{-1}$ (contour levels 0.010, 0.037, 0.064, 0.092, and 0.119 Jy/beam km s$^{-1}$) with the red contours showing NH$_2$CHO (as in the left frame).}
	\label{G17maps1}
\end{figure*}

\section{Comparison of formamide emission to possible chemically-related species}

In this section, we derive gas properties empirically from moment maps.  From the integrated intensity maps (zeroth moment), we locate the peak of line emission to high accuracy with a 2D-Gaussian fit accuracy to 0.01$''$. We assume, then, that if two species peak in the same location, and have the same velocity and line width, then they are in the same gas and are therefore related (either they have been released from the ice around the same time or they have formed in the same gas). From the velocity maps (first moment), we measure the average central velocity for each transition at each pixel and subtract these values from each other. A small difference between these velocities for different species suggests that they are in the same gas as they are moving in the same manner. Peak positions and average velocities can be affected by optical depth, especially when dealing with the main isotope of a species (ie. not isotopologues). Using RADEX \citep{RADEX}, a one dimensional non-LTE radiative transfer code, we have determined that the optical depths for the H$_2$CO lines range from 33-430 while those for HNCO lines are much lower ranging from 0.4-69. The last quantity we derive from moment maps is the velocity dispersion (second moment) map differences.  Velocity dispersion gives the average line width at each pixel which is related to the level of turbulence in the gas. A small difference between velocity dispersion values shows that the gas emitting each transition has a similar turbulence level which suggests that they are in the same gas. 

\subsection{Comparison of spatial distribution}
\label{mom0}
\subsubsection{G17}
Although G17 is not associated with strong emission of typical complex organic molecules (e.g., CH$_3$OCHO, CH$_2$CHCN) (Cesaroni et al. 2017, Maud et al. 2018), it has a clear detection of NH$_2$CHO.  We see in Figure~\ref{G17maps1} that the integrated emission (moment zero) map of NH$_2$CHO is slightly more compact than that of HNCO (0.58 vs. 0.76$''$ or 1275 vs. 1675 au).  Both species are offset from the continuum but the emission peaks of HNCO and NH$_2$CHO are separated by $\sim$ 0.1$''$ (220 au).  For H$_2$CO, the emission is much more extended (up to 1.6$''$ or $\sim$3500 au).  The H$_2$CO (1), (2), and (3) (see Table~\ref{linestats} for line properties) zeroth moment peaks are separated from the NH$_2$CHO peak by 0.07$''$, 0.1$''$, and 0.22$''$ respectively.  The lowest energy ($E_\mathrm{up}$=20.9~K) H$_2$CO (1) peak is slightly closer to the NH$_2$CHO peak than the HNCO peak (0.07$''$~vs.~0.1$''$).

\begin{figure*}[!thb]
	\centering
	\includegraphics[width=\hsize]{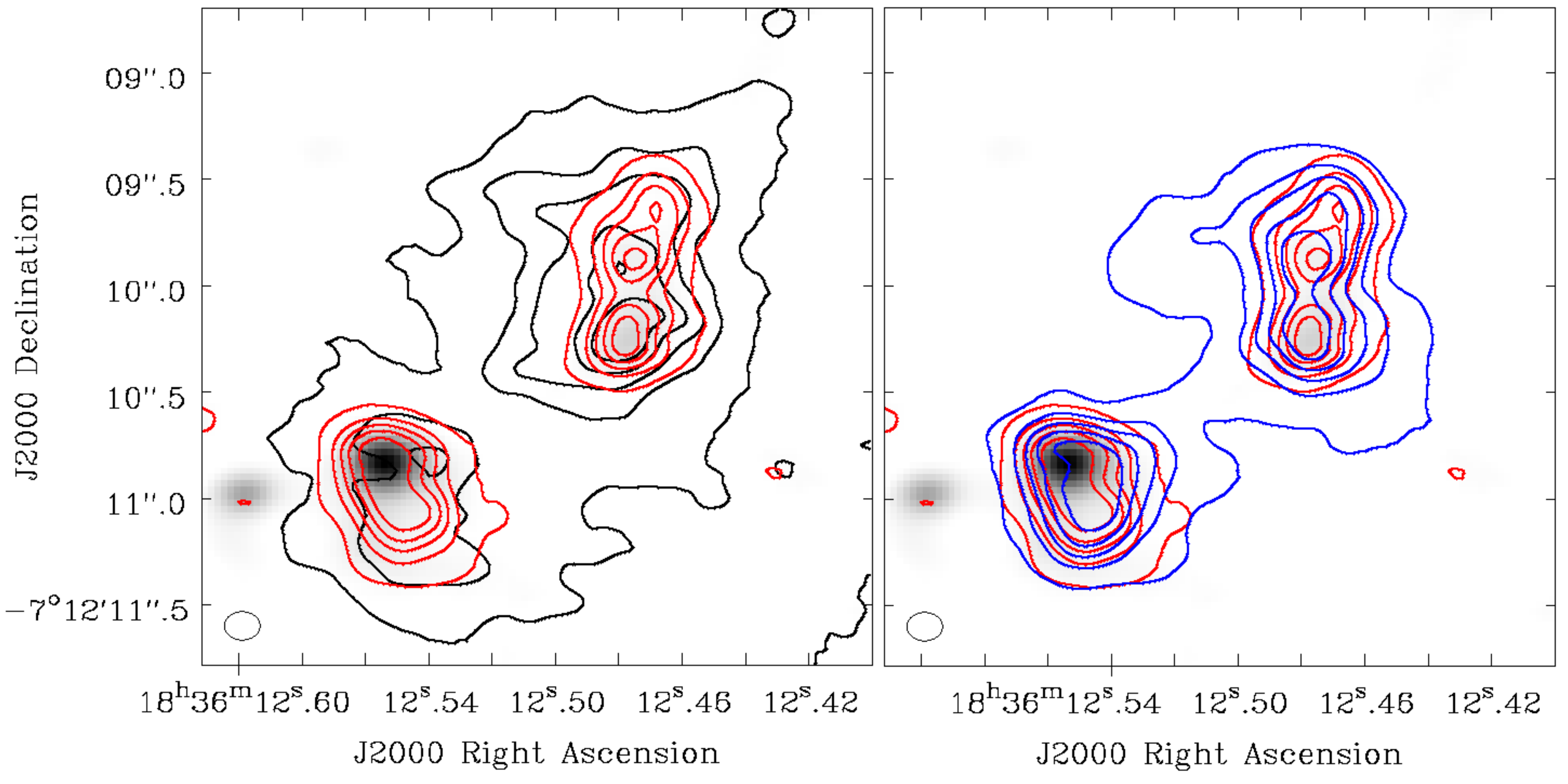}
	\caption{G24 zeroth moment maps (contours) overlaid on the dust continuum (grayscale). \textit{Left:} the black contours show the H$_2$CO (3) transition ($E_\mathrm{up}$=68.1~K) from 5$\sigma$ to a peak of 0.674 Jy/beam km s$^{-1}$ (contour levels 0.03, 0.16, 0.29, 0.42, and 0.55 Jy/beam km s$^{-1}$). The red contours show NH$_2$CHO (2) emission ($E_\mathrm{up}$=78.9~K) from 5$\sigma$ to 0.512 Jy/beam km s$^{-1}$ (contour levels 0.026, 0.123, 0.220, 0.318, and 0.415 Jy/beam km s$^{-1}$). \textit{Right:} the blue contours show the extent of the HNCO (1) emission ($E_\mathrm{up}$=58.0~K) from 5$\sigma$ to 0.738 Jy/beam km s$^{-1}$ (contour levels 0.031, 0.172, 0.314, 0.455, and 0.597 Jy/beam km s$^{-1}$) with the red contours showing NH$_2$CHO (as in the left frame).}
	\label{G24maps1}
\end{figure*}

\begin{figure*}[!thb]
	\centering
	\includegraphics[width=\hsize]{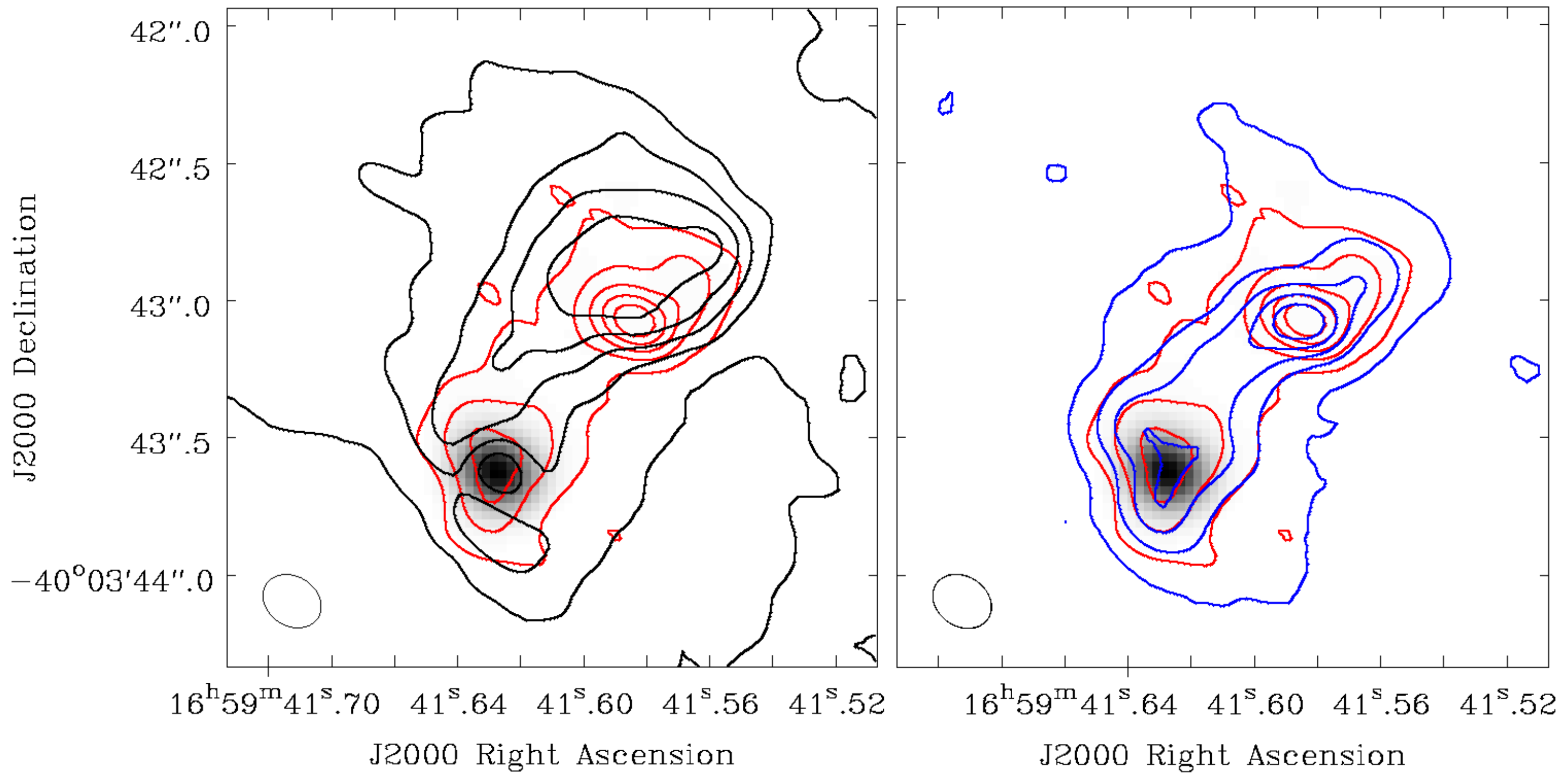}
	\caption{G345 zeroth moment maps (contours) overlaid on the dust continuum (grayscale). \textit{Left:} the black contours show the H$_2$CO (3) transition ($E_\mathrm{up}$=68.1~K) from 5$\sigma$ to a peak of 0.402 Jy/beam km s$^{-1}$ (contour levels 0.027, 0.102, 0.177, 0.232, and 0.327 Jy/beam km s$^{-1}$). The red contours show NH$_2$CHO (1) emission ($E_\mathrm{up}$=60.8~K) from 5$\sigma$ to 0.242 Jy/beam km s$^{-1}$ (contour levels 0.020, 0.064, 0.109, 0.153, and 0.198 Jy/beam km s$^{-1}$). \textit{Right:} the blue contours show the extent of the HNCO (2) emission ($E_\mathrm{up}$=101.5~K) from 5$\sigma$ to 0.428 Jy/beam km s$^{-1}$ (contour levels 0.014, 0.097, 0.180, 0.262, and 0.345 Jy/beam km s$^{-1}$) with the red contours showing NH$_2$CHO (as in the left frame).}
	\label{G34549maps1}
\end{figure*}

\begin{table}[!thb]
	\centering
	\caption{Peak separations between listed transitions and NH$_2$CHO (in arcseconds). The error in the peak position is $\sim~0.01''$.}
	\label{separationsTable}
	{\small  
		\begin{tabular}{ccccccc}
			
			\hline\hline
			Species & G17 & G24 & G24 & G24 & G345 & G345 \\
			&  & A1 & A2(N) & A2(S) & Main & NW spur \\
			\hline
			HNCO (1) & N/A & 0.08 & 0.03 & 0.06  & N/A & N/A \\
			HNCO (2) & 0.1 & N/A & N/A & N/A & 0.04 & 0.03  \\
			HNCO (3) & N/A & 0.04 & 0.04 & 0.03  & 0.21 & 0.03 \\
			H$_2$CO (1) & 0.07 & 0.35 & 0.12 & 0.1 & 0.21 & 0.21 \\
			H$_2$CO (2) & 0.1 & 0.29 & 0.25 & 0.06 & 0.25 & 0.24 \\
			H$_2$CO (3) & 0.22 & 0.27 & 0.07 & 0.03 & 0.25 & 0.23 \\
			\hline 
		\end{tabular}
		\tablefoot{Transitions in column 1 are labeled as in Table~\ref{linestats}. For G24 A2 (N) and (S) the distances are measured from the corresponding northern or southern peaks.}}
\end{table}

\begin{figure}[!thb]
	\centering
	\includegraphics[width=\hsize]{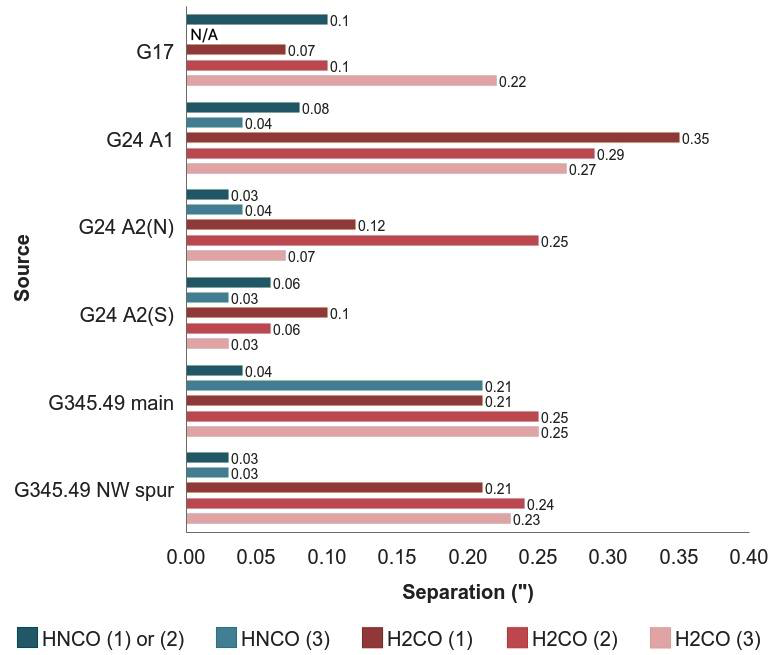}
	\caption{Separations between NH$_2$CHO and each peak of HNCO and H$_2$CO. For the G24 sources HNCO (1) is used instead of HNCO (2) (see Table~\ref{linestats}). The error in the peak position is $\sim~0.01''$.}
	\label{separationsFig}
\end{figure}

\begin{figure*}[!thb]
	\centering
	\includegraphics[width=\hsize]{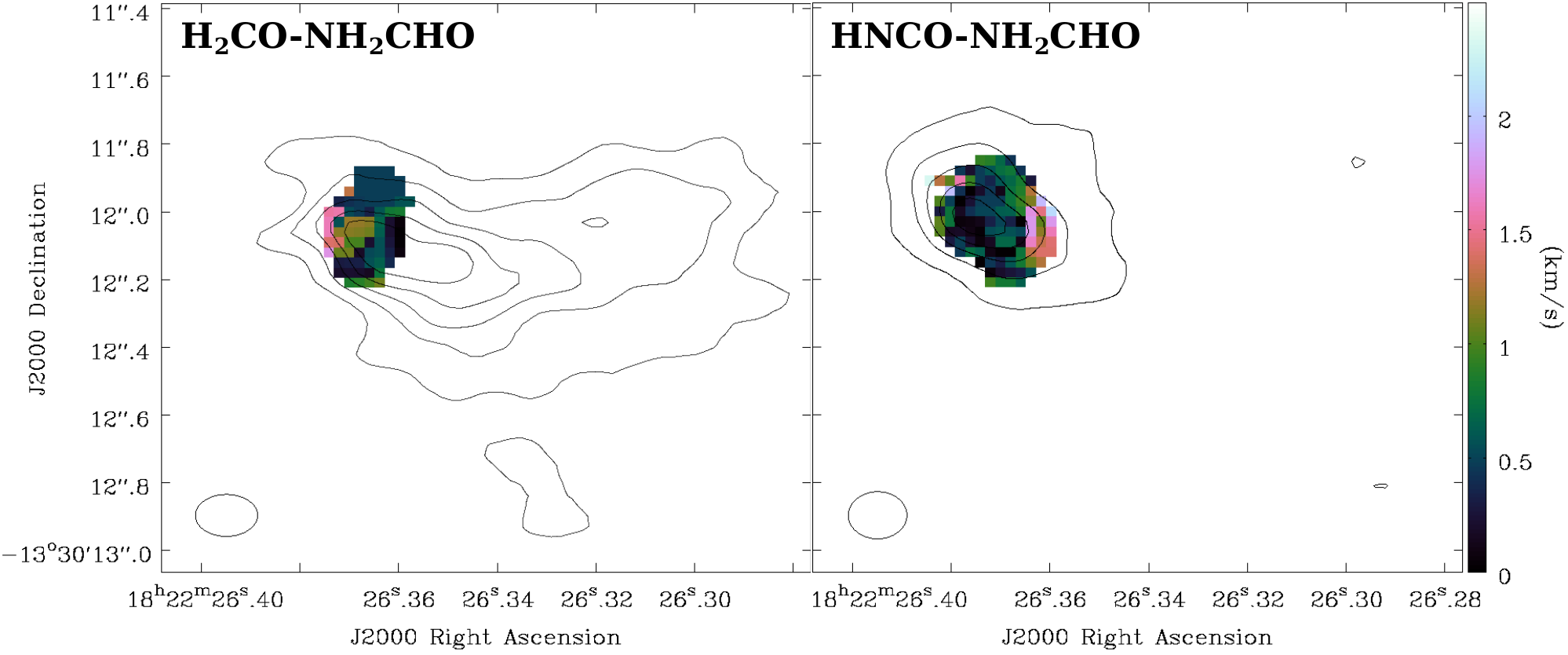}
	\caption{Velocity difference (from first moment maps) at each pixel in G17 between (left) H$_2$CO (2) and NH$_2$CHO (1) and (right) HNCO (2) and  NH$_2$CHO (1). The contours show the integrated intensity maps for H$_2$CO (2) and HNCO (2) as in Figure~\ref{G17maps1}. The velocity scale is the same for both panels.}
	\label{G17maps2}
\end{figure*}

\subsubsection{G24}
G24 has three subsources A1, A2(N), and A2(S). In Figure~\ref{G24maps1} we find that the H$_2$CO emission in G24 is much more extended than the NH$_2$CHO emission.  In G24 A1, the extent of the H$_2$CO emission is 1.71$''$ ($\sim$11500 au) from northeast to southwest whereas the NH$_2$CHO emission extends 0.9$''$ ($\sim$6000 au) in the same direction. In G24 A2, the H$_2$CO emission extends 2.0$''$ ($\sim$13400 au) whereas the NH$_2$CHO emission spans 1.1$''$ ($\sim$7400~au). The extent of HNCO in these sources is 1.1$''$ ($\sim$7400 au) at G24 A1 and 1.5$''$ ($\sim$10050 au) for G24 A2.  The integrated emission for NH$_2$CHO and HNCO (3) breaks off between A1 (to the southeast) and A2 (to the northwest) whereas for the H$_2$CO transitions and HNCO (1) there is some emission between the two continuum sources.  In G24 A1, the separation between all HNCO or H$_2$CO and NH$_2$CHO emission peaks are between 0.04 and 0.35$''$ (270-2350 au) and the closest peak to NH$_2$CHO is that of HNCO (3).  In the case of G24 A1, we must remember that optical depth effects primarily affect H$_2$CO in this source.

\par The NH$_2$CHO, HNCO, and H$_2$CO emission in G24 A2 have two significant NH$_2$CHO integrated intensity peaks of similar strength separated by about 0.35$''$ ($\sim$2350~au) that we refer to as A2(N) and A2(S) (positions of each peak indicated in Figure~\ref{contpts}). The two emission peaks in G24 A2 complicate things slightly, as it is difficult to draw boundaries between the velocity maps of the two peaks.  Nevertheless we can determine the positions of the emission peaks and analyze them separately.  The more northerly H$_2$CO (2) peak was in between the NH$_2$CHO A2(N) and (S) peaks with a distance between H$_2$CO (2) A2(N) and NH$_2$CHO (2) A2(N) of 0.25$''$ ($\sim$1670~au) and between H$_2$CO (2) A2(N) and NH$_2$CHO (2) A2(S) of 0.18$''$ ($\sim$1200~au). The H$_2$CO (3) A2 peaks are nearer to the respective NH$_2$CHO peaks at 0.07$''$ ($\sim$470~au) from A2(N) and 0.03$''$ ($\sim$200~au) from A2(S). In G24 A2(N), the closest peak to the NH$_2$CHO is the lower energy (58~K) HNCO (1) transition.  In A2(S), the HNCO (3) transition and the H$_2$CO (3) peaks are equally separated from NH$_2$CHO peak at 0.03$''$ ($\sim$200~au).

\subsubsection{G345}
The two subsources in G345 (described in Figure~\ref{contpts}) are G345 Main and G345 NW spur.  From the spectra extracted from G345 Main, we note that the chemical composition appears to be affected by a source of strong H30$\alpha$ emission within which is ionizing the region and destroying complex molecular species, but the closest peak to the NH$_2$CHO peak (by far) is HNCO (2). The spectra associated with G345 NW spur show that it is a very chemically diverse region -- possibly an outflow cavity associated with G345 Main.  The HNCO (2) and (3) emission peaks are equally the closest to the NH$_2$CHO peak in G345.49 NW spur (0.03$''$).

\par Figure~\ref{G34549maps1} shows that HNCO (2) and NH$_2$CHO (1) have similar extent and velocity structure at the Main and NW spur positions.  There is little high energy HNCO (3) emission at G345 Main.  The H$_2$CO transitions peak at the NW spur, but there is still emission at Main, without a clear peak.  We take the pixel with the highest intensity on the area designated to Main despite the emission being extended across the two parts of the source.  In Main, the low energy HNCO transition peaks very close to the NH$_2$CHO peak (0.04$''$ away $\sim$100au), but the higher energy HNCO transition and all of the H$_2$CO transitions peak 0.21-0.25$''$ from the NH$_2$CHO peak.  In the NW spur, both HNCO transitions peak very near the NH$_2$CHO peak (0.03$''$ $\sim$75au) whereas all three H$_2$CO transitions are farther at 0.21-0.24$''$ (500-575 au).

\subsubsection{Summary of spatial distribution comparison}
For our six subsources in these regions (see Table~\ref{separationsTable} and Figure~\ref{separationsFig}), it is clear that the integrated emission peaks of HNCO are closer to the peaks of NH$_2$CHO than the H$_2$CO peaks.  The morphology of the HNCO emission is also more similar to NH$_2$CHO, as the H$_2$CO emission tends to be much more extended and even the brightest emission (see higher intensity contours in Figures~\ref{G17maps1}, \ref{G24maps1}, and \ref{G34549maps1}) have a different shape to the NH$_2$CHO emission. The lack of NH$_2$CHO emission in the more extended regions indicates that it can be more efficiently made from H$_2$CO (in the gas phase) near the continuum peaks than farther out in these cases. It is clear from the H$_2$CO emission toward G24 A1 and G345 Main, that these transitions are suffering by optical depth effects.

\subsection{Comparison of the velocity field}
\label{mom1}
The velocity field of each molecule was investigated by creating the first order moment map for each transition listed in Table~\ref{linestats}. These maps were then subtracted from each other to determine the difference between the gas velocities for each species.  Where possible, two transitions from the same species were also compared to determine the "internal error", as the velocity difference of transitions within the same gas implies a lower limit for accuracy. Histograms were made for the absolute values of each velocity difference map showing the number of pixels within each bin (see Appendix~\ref{histograms}). The average value and standard deviation of these histograms were used to determine which species was most similar to NH$_2$CHO. Results are detailed per source below and summarized in Table \ref{histTable} and Figure~\ref{mom1Fig}.

\begin{figure*}[!thb]
	\centering
	\includegraphics[width=\hsize]{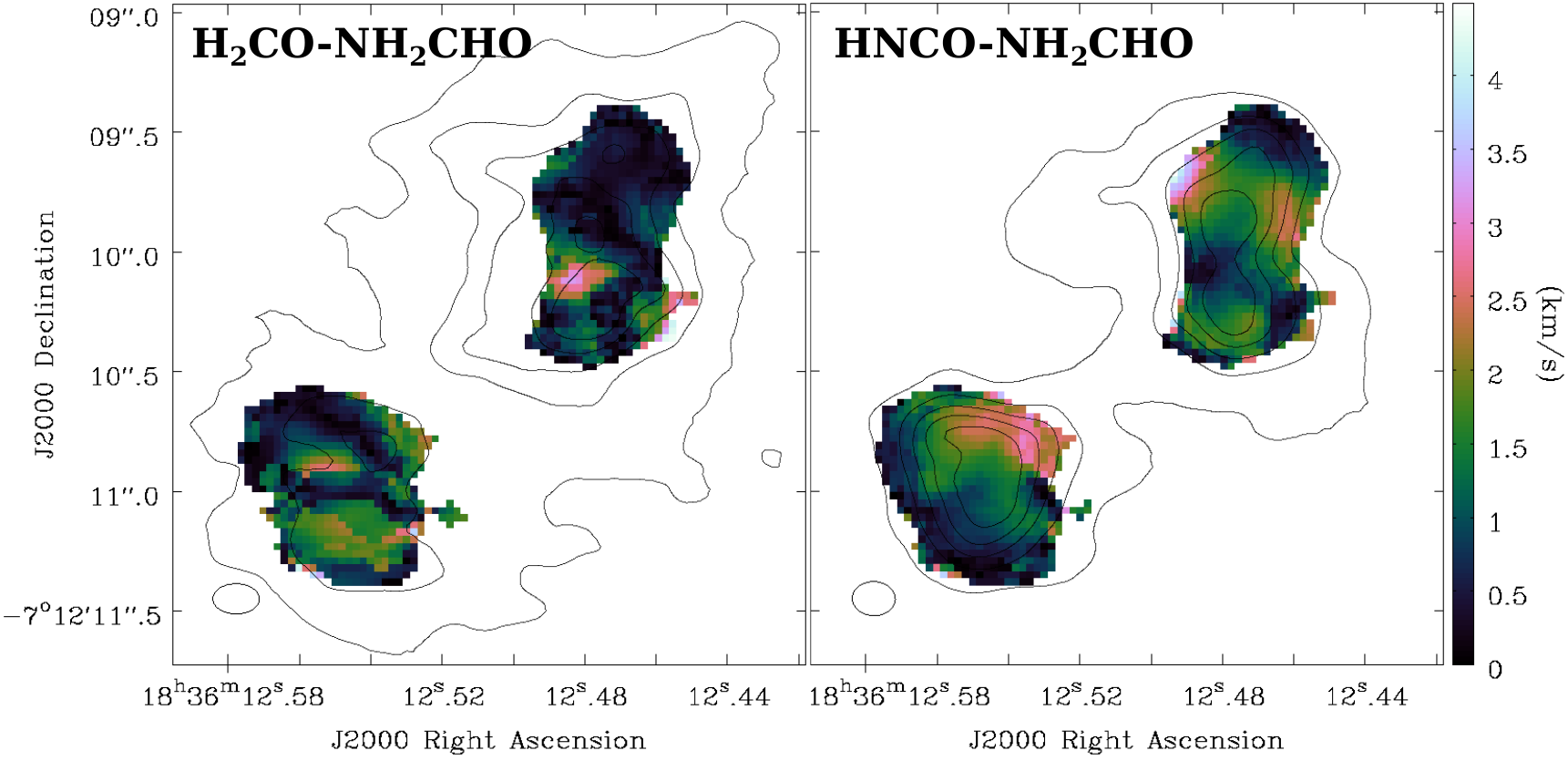}
	\caption{Velocity difference (from first moment maps) at each pixel in G24 between (left) H$_2$CO (3) and NH$_2$CHO (2) and (right) HNCO (1) and  NH$_2$CHO (2). The contours show the integrated intensity maps for H$_2$CO (3) and HNCO (1) as in Figure~\ref{G24maps1}. The velocity scale is the same for both panels.}
	\label{G24maps2}
\end{figure*}

\subsubsection{G17}
Figure \ref{G17maps2} shows that the velocity differences between HNCO and NH$_2$CHO and H$_2$CO and NH$_2$CHO are not significantly different.  The average velocity difference for HNCO (2) is 0.78 km s$^{-1}$, whereas the differences for H$_2$CO (2) and (3) are 0.72 and 0.67 km s$^{-1}$, respectively.  For G17 overall the H$_2$CO transitions are on average more similar to NH$_2$CHO.  HNCO (3) is not detected toward G17.

\subsubsection{G24}
Figure \ref{G24maps2} shows that the range of velocity differences in G24~A2(N) and A2(S) (to the northwest) are greater between HNCO and NH$_2$CHO than H$_2$CO and NH$_2$CHO with the smallest average difference between H$_2$CO (3) and NH$_2$CHO for A2(N) and between H$_2$CO (2) and NH$_2$CHO for A2(S) at 0.53 and 1.13~km s$^{-1}$, respectively. It is less obvious visually for G24 A1 (to the southeast), but we can see from the average values listed in Table~\ref{histTable} that the average difference closest to zero is between HNCO (3) and NH$_2$CHO at 1.01~km s$^{-1}$.

\begin{figure*}[!thb]
	\centering
	\includegraphics[width=\hsize]{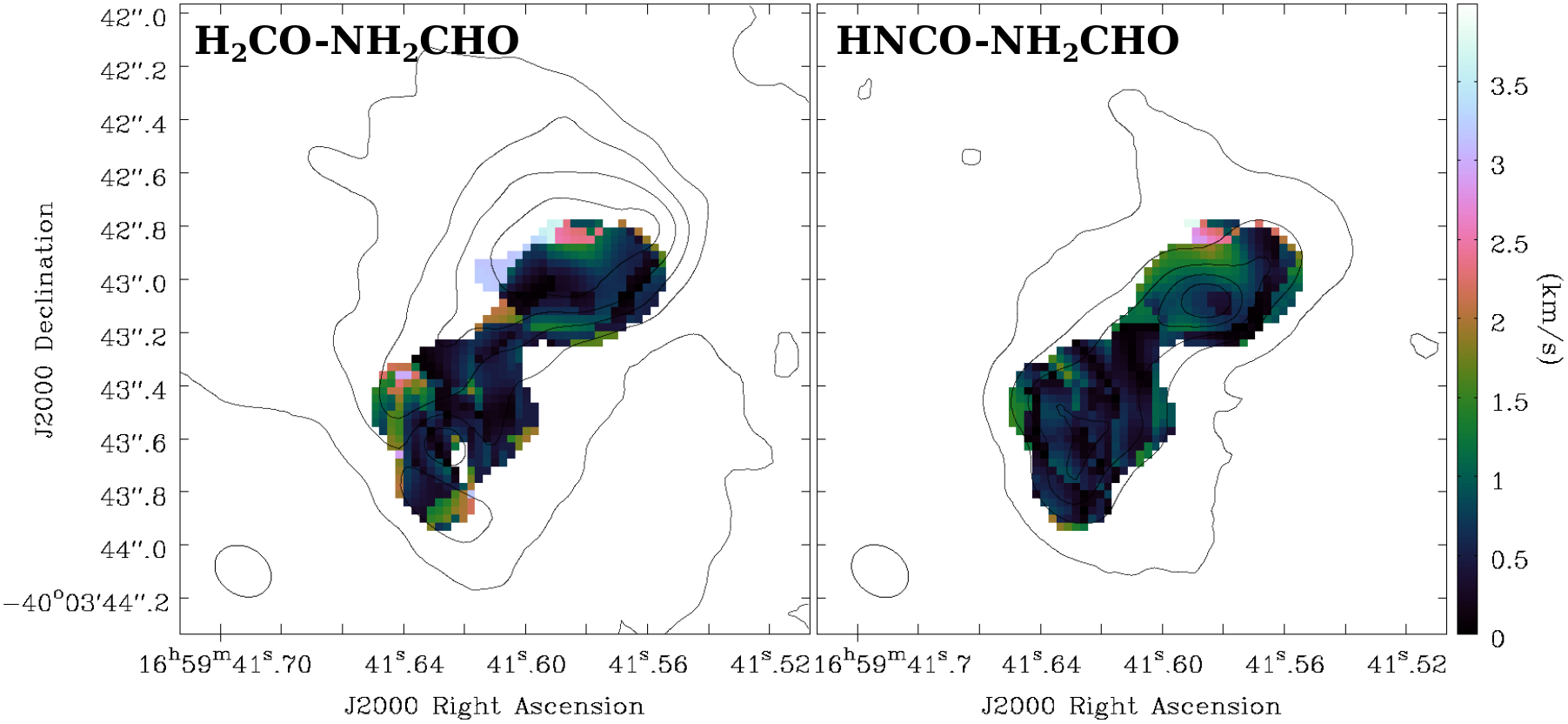}
	\caption{Velocity difference (from first moment maps) at each pixel in G345 between (left) H$_2$CO (2) and NH$_2$CHO (1) and (right) HNCO (2) and  NH$_2$CHO (1). The contours show the integrated intensity maps for H$_2$CO (2) and HNCO (2) as in Figure~\ref{G34549maps1}. The velocity scale is the same for both panels.}
	\label{G34549maps2}
\end{figure*}

\begin{figure}[!thb]
	\centering
	\includegraphics[width=\hsize]{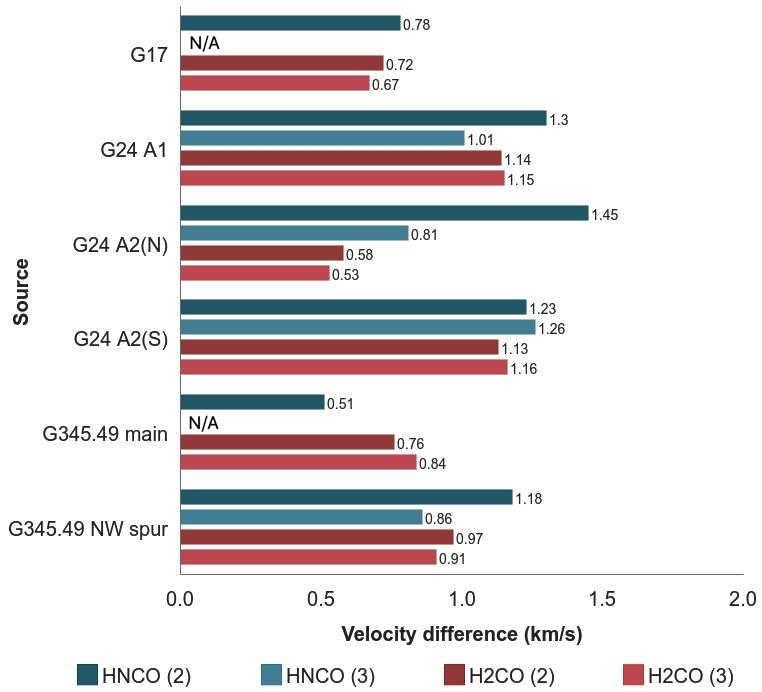}
	\caption{Average velocity difference between NH$_2$CHO and transitions HNCO (2) and (3) and H$_2$CO (2) and (3).  For the G24  sources HNCO (1) is used instead of HNCO (2) (see Table~\ref{linestats}).}
	\label{mom1Fig}
\end{figure}

\subsubsection{G345}
Figure \ref{G34549maps2} shows the velocity differences in G345 Main (to the southeast) and NW spur (to the northwest).  The range of values for the velocity difference is smaller for HNCO and NH$_2$CHO for both subsources. The smallest average velocity difference for G345 NW spur is between HNCO (3) and NH$_2$CHO at 0.86~km s$^{-1}$, closely followed by H$_2$CO (3) and (2) at 0.91 and 0.97 km s$^{-1}$, respectively. For G345 Main, the smallest average difference is between HNCO (2) and NH$_2$CHO at 0.51~km s$^{-1}$ with H$_2$CO (2) and (3) averages of 0.76 and 0.84~km s$^{-1}$. Within errors, the velocity fields of the two precursors are equally similar to that of NH$_2$CHO. HNCO (3) is not detected toward G345 Main.

\begin{table*}[ht]
	\centering
	\setlength{\tabcolsep}{3pt}
	\caption{Average values (with the standard deviation in parentheses) of the histograms (see Appendix~\ref{histograms}) of each velocity (first moment) difference map. All units are km/s.}
	\label{histTable}
	\begin{tabular}{ccccccc}
		\hline\hline
		Transitions & G17 & G24 A1 & G24 A2(N) & G24 A2(S) & G345 NW spur & G345 Main \\
		\hline
		H$_2$CO (2) – H$_2$CO (3) & 0.42 (0.58) & 0.27 (0.54) & 0.22 (0.50) & 0.34 (0.61) & 0.27 (0.39) & 0.43 (0.43) \\
		HNCO (2) – HNCO (3)       & N/A & 0.73 (0.44) & 0.74 (0.48) & 0.44 (0.56) & 0.25 (0.18) & N/A \\
		H$_2$CO (2) – NH$_2$CHO   & 0.72 (0.54) & 1.14 (1.14) & 0.58 (0.46) & 1.13 (1.06) & 0.97 (0.90) & 0.76 (0.62) \\
		H$_2$CO (3) – NH$_2$CHO   & 0.67 (0.38) & 1.15 (1.11) & 0.53 (0.38) & 1.16 (1.10) & 0.91 (0.89) & 0.84 (0.73) \\
		HNCO (2) – NH$_2$CHO      & 0.78 (0.74) & 1.30 (0.92) & 1.45 (0.75) & 1.23 (0.82) & 1.18 (1.07) & 0.51 (0.38) \\
		HNCO (3) – NH$_2$CHO      & N/A & 1.01 (0.49) & 0.81 (0.28) & 1.26 (0.85) & 0.86 (0.35) & N/A \\
		\hline
	\end{tabular}
	\tablefoot{For G24, HNCO (1) is used instead of HNCO (2).  G17 and G345 Main have only one HNCO transition, so the internal error for HNCO transitions cannot be determined.}
\end{table*}

\subsubsection{Summary of the velocity field comparison}
We see in the Table~\ref{histTable} and Figure~\ref{mom1Fig} that there are an equal number of subsources where the average velocity difference is closest to zero for each of our related species.  For a few sources, the range of average differences between different transitions is very small.  For G17 in particular, the averages are 0.67, 0.72, and 0.78 km s$^{-1}$ for H$_2$CO (3), H$_2$CO (2) and HNCO (2), respectively.  For G24 A2(N) the difference is clearer with average velocity differences of 0.53, 0.58, 0.81, and 1.45 km s$^{-1}$ for H$_2$CO (3), H$_2$CO (2), HNCO (3) and HNCO (1), respectively. In the cases of G24 A1 and G24 A2(N), the internal error for HNCO is much larger than that of H$_2$CO (0.73 and 0.74 vs. 0.27 and 0.22 km s$^{-1}$ ). This indicates that, in terms of errors, the difference between NH$_2$CHO and H$_2$CO is stronger than the difference between NH$_2$CHO and HNCO in these sources.

\begin{figure*}[!thb]
	\centering
	\includegraphics[width=\hsize]{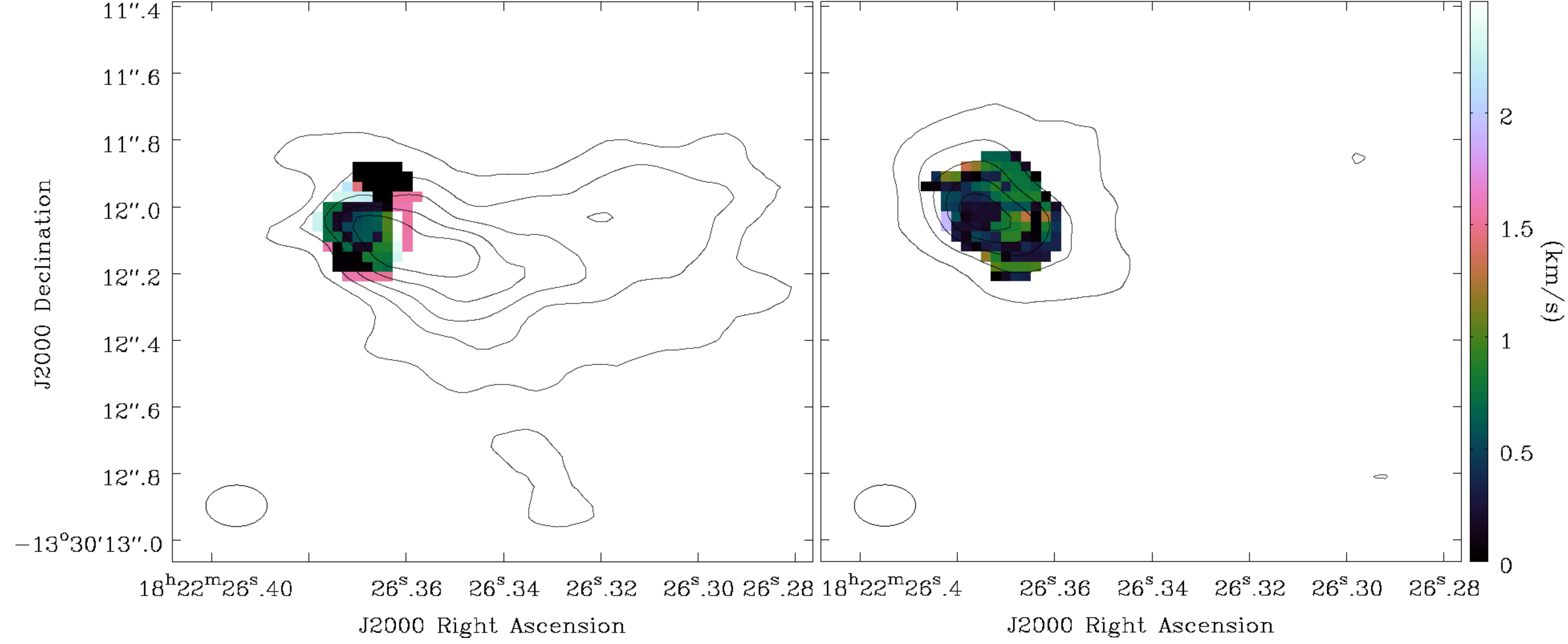}
	\caption{Velocity dispersion difference (from second moment maps) at each pixel in G17 between (left) H$_2$CO (2) and NH$_2$CHO (1) and (right) HNCO (2) and  NH$_2$CHO (1). The contours show the integrated intensity maps for H$_2$CO (2) and HNCO (2) as in Figure~\ref{G17maps1}.  The velocity scale is the same for both panels.}
	\label{G17mom2}
\end{figure*}

\subsection{Comparison of the velocity dispersion}
\label{mom2}
Second order moment maps were made for each of the transitions studied for each star-forming region.  These maps were then subtracted from each other to determine the difference between the velocity dispersion for each species. Though it may be affected by optical depth, similar line widths between species can suggest that they are in the same gas.  As in \S\ref{mom1}, transitions from the same species were compared to determine internal error.  Histograms were made of the absolute values of each dispersion difference map showing the number of pixels within each velocity bin. The average value and standard deviation of these histograms were used to determine which species was most similar to NH$_2$CHO. Results are detailed per source in the text and summarized in Table~\ref{histTable2} and Figure~\ref{mom2Fig}.  The histograms for this analysis are shown in Appendix~\ref{histograms}.

\subsubsection{G17}
Figure~\ref{G17mom2} shows the difference at each pixel between the second order moment maps of H$_2$CO and NH$_2$CHO and HNCO and NH$_2$CHO toward G17. It is clear that the difference between HNCO and NH$_2$CHO is smaller and we determine that the average difference is 0.52~km s$^{-1}$ for HNCO (2), whereas for H$_2$CO (2) and (3) the average differences are 0.68 and 0.86~km s$^{-1}$, respectively.  HNCO (3) is not detected toward G17.

\begin{figure*}[!thb]
	\centering
	\includegraphics[width=\hsize]{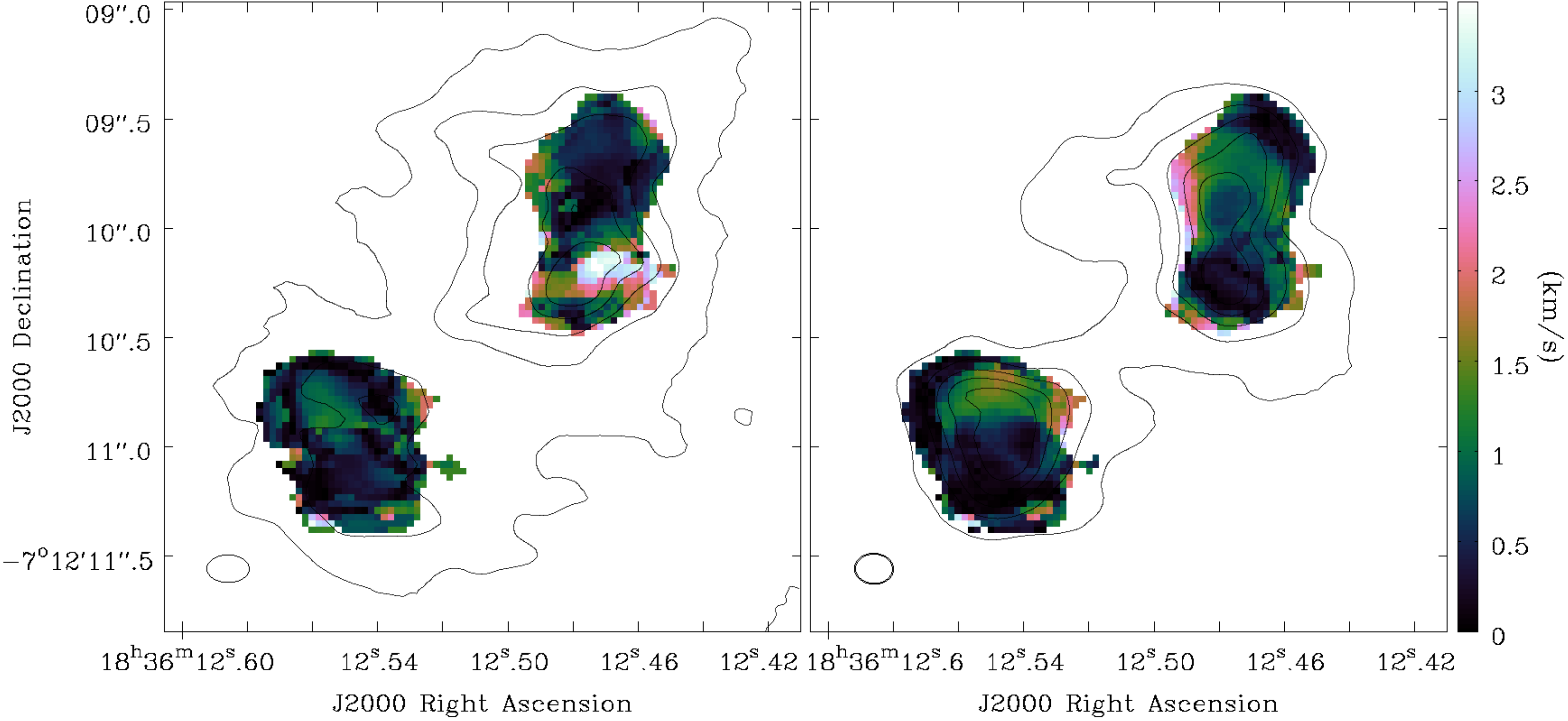}
	\caption{Velocity dispersion difference (from second moment maps) at each pixel in G24 between (left) H$_2$CO (3) and NH$_2$CHO (2) and (right) HNCO (1) and NH$_2$CHO (2). The contours show the integrated intensity maps for H$_2$CO (3) and HNCO (1) as in Figure~\ref{G24maps1}.  The velocity scale is the same for both panels.}
	\label{G24mom2}
\end{figure*}

\begin{figure*}[!thb]
	\centering
	\includegraphics[width=\hsize]{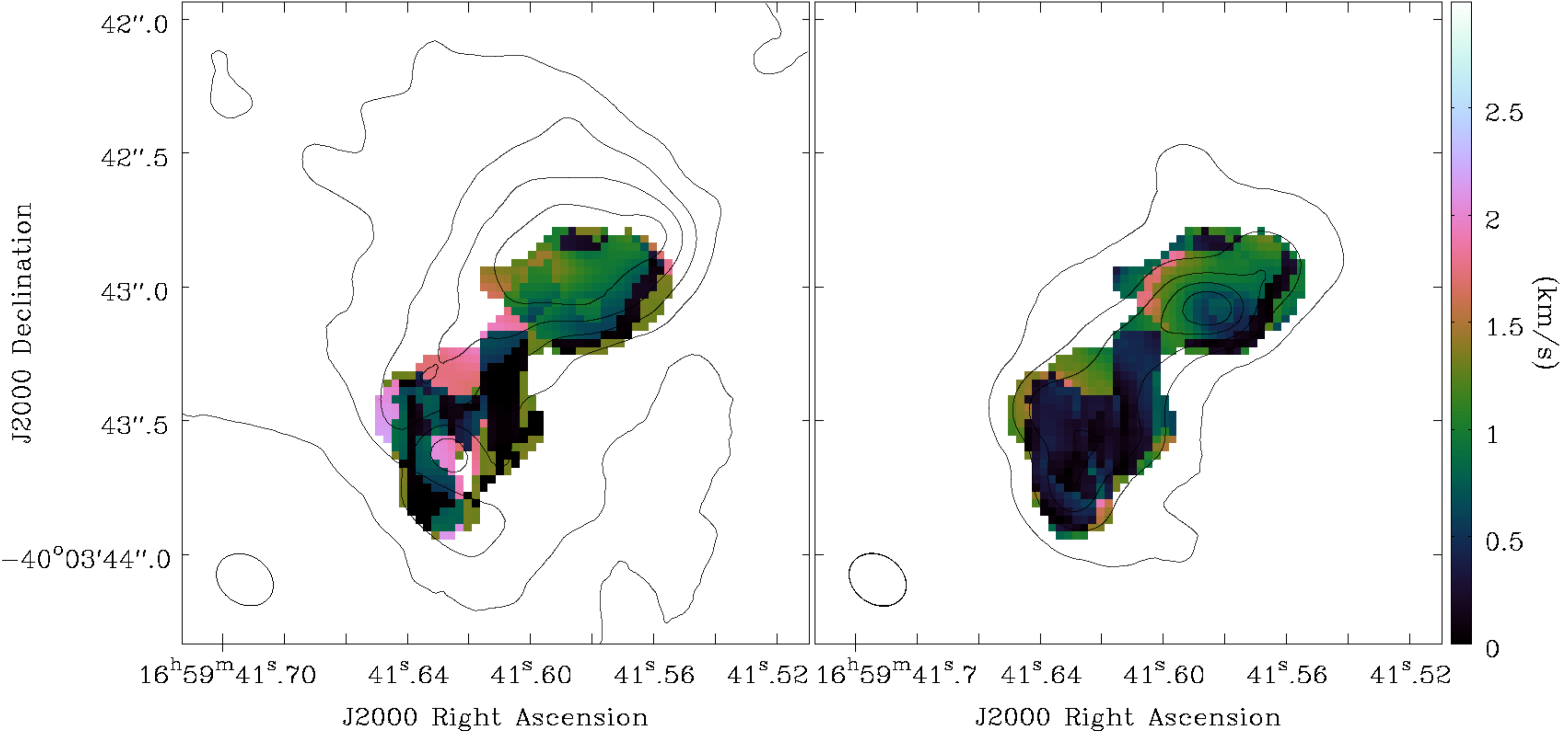}
	\caption{Velocity dispersion difference (from second moment maps) at each pixel in G345 between (left) H$_2$CO (2) and NH$_2$CHO (1) and (right) HNCO (2) and NH$_2$CHO (1). The contours show the integrated intensity maps for H$_2$CO (2) and HNCO (2) as in Figure~\ref{G34549maps1}. The velocity scale is the same for both panels.}
	\label{G34549mom2}
\end{figure*}

\begin{table*}[ht]
	\centering
	\setlength{\tabcolsep}{3pt}
	\caption{Average values (with the standard deviation in parentheses) of the histograms of each dispersion (second moment) difference map. All units are km s$^{-1}$.}
	\label{histTable2}
	\begin{tabular}{ccccccc}
		\hline\hline
		Transitions & G17 & G24 A1 & G24 A2(N) & G24 A2(S) & G345 NW spur & G345 Main \\
		\hline
		H$_2$CO (2) – H$_2$CO (3) & 0.08 (0.60) & 0.12 (0.58) & 0.17 (0.61) & 0.13 (0.67) & 0.17 (0.58) & 0.07 (0.56) \\
		HNCO (2) – HNCO (3)       & N/A & 0.99 (0.53) & 1.11 (0.58) & 0.98 (0.67) & 0.67 (0.23) & N/A \\
		H$_2$CO (2) – NH$_2$CHO   & 0.68 (0.68) & 0.75 (0.55) & 0.79 (0.48) & 1.59 (0.86) & 0.92 (0.44) & 0.80 (0.73) \\
		H$_2$CO (3) – NH$_2$CHO   & 0.86 (0.69) & 0.62 (0.51) & 0.65 (0.47) & 1.63 (0.91) & 0.80 (0.50) & 0.77 (0.70) \\
		HNCO (2) – NH$_2$CHO      & 0.52 (0.39) & 0.67 (0.62) & 0.95 (0.59) & 0.81 (0.68) & 0.81 (0.41) & 0.53 (0.45) \\
		HNCO (3) – NH$_2$CHO      & N/A & 0.41 (0.41) & 0.21 (0.19) & 0.47 (0.47) & 1.32 (0.49) & N/A \\
		\hline
	\end{tabular}
	\tablefoot{For G24, HNCO (1) is used instead of HNCO (2).  G17 and G345 Main have only one HNCO transition, so the internal error for HNCO transitions cannot be determined.}
\end{table*}

\subsubsection{G24}
Figure~\ref{G24mom2} shows the difference at each pixel between the second order moment maps of H$_2$CO and NH$_2$CHO and HNCO and NH$_2$CHO toward G24 A1 (to the southeast) and G24 A2(N) and A2(S) (to the northwest). HNCO (3)-NH$_2$CHO has the smallest average velocity dispersion difference for all three subsources of G24 at 0.41~km s$^{-1}$ for A1, 0.21~km s$^{-1}$ for A2(N), and 0.47~km s$^{-1}$ for A2(S).

\subsubsection{G345}
Figure~\ref{G34549mom2} shows the difference at each pixel between the second order moment maps of H$_2$CO and NH$_2$CHO and HNCO and NH$_2$CHO toward G345 Main (to the southeast) and G345 NW spur (to the northwest).  For G345 Main, it is clear from the figure and Table~\ref{histTable2} that HNCO (2) has the smallest average difference between velocity dispersion values at 0.53~km s$^{-1}$. The average second order moment map differences for H$_2$CO (2) and (3) are 0.80 and 0.77~km s$^{-1}$, respectively, and there is no HNCO (3) emission toward G345 Main. The average difference between H$_2$CO (3) and NH$_2$CHO in G345 NW spur is smallest at 0.80~km s$^{-1}$, but the average difference for HNCO (2)-NH$_2$CHO is 0.81~km s$^{-1}$, so these two maps are equally similar within errors.

\subsubsection{Summary of velocity dispersion comparison}
As a measure of the similarity between the motions of the gas containing each species, the line width test comes out in favor of HNCO for five out of six subsources.  In the sixth (G345 NW spur), the difference between H$_2$CO (3)-NH$_2$CHO and HNCO (2)-NH$_2$CHO is only 0.01~km s$^{-1}$.  In the five subsources that show the velocity dispersion of HNCO as definitively closest to NH$_2$CHO, the average values are also consistent with zero if we consider the difference between H$_2$CO (2) and (3) as the error for these measurements.

\begin{figure}[!thb]
	\centering
	\includegraphics[width=\hsize]{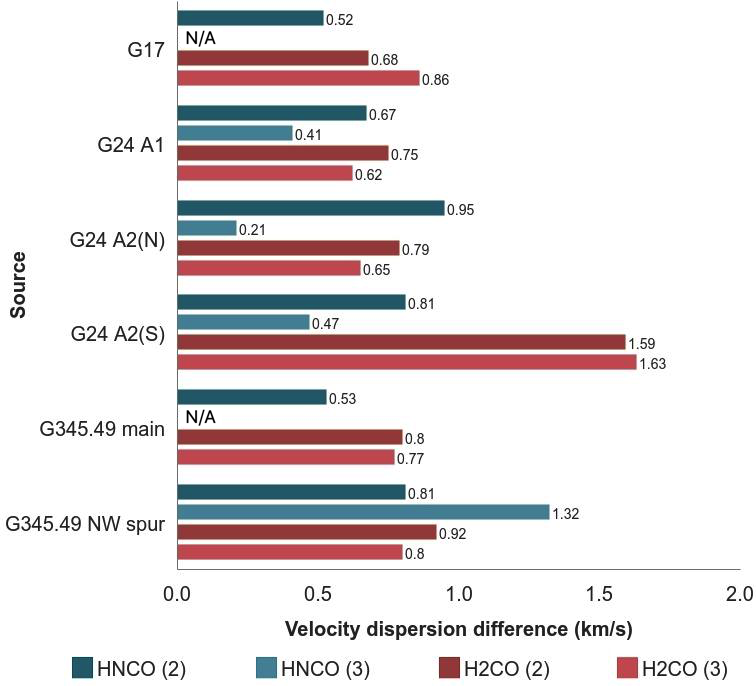}
	\caption{Average velocity dispersion difference between NH$_2$CHO and transitions HNCO (2) and (3) and H$_2$CO (2) and (3). For the G24 sources HNCO (1) is used instead of HNCO (2).}
	\label{mom2Fig}
\end{figure}

\begin{table*}[!htb]
	\centering
	\caption{Summary of results from map analyses. The check symbol (\checkmark) indicates the species with the emission peak closest to the NH$_2$CHO peak, velocity-difference histogram center nearest to zero, or dispersion-difference histogram center nearest to zero. Equals signs (=) indicate that the parameters were equal for both HNCO and H$_2$CO within errors.}
	\label{checklist}
	\begin{tabular}{c|ccc|ccc}
		\hline\hline
		& \multicolumn{3}{c|}{\textbf{HNCO}} & \multicolumn{3}{c}{\textbf{H$_2$CO}} \\
		\textbf{Source} & \textbf{Peak} & \textbf{Velocity} & \textbf{Dispersion} & \textbf{Peak} & \textbf{Velocity} & \textbf{Dispersion} \\ 
		\hline
		G17          &            &  =          & \checkmark & \checkmark & = &            \\
		G24 A1       & \checkmark & \checkmark & \checkmark &            &            &            \\
		G24 A2(N)    & \checkmark &            & \checkmark &            & \checkmark &            \\
		G24 A2(S)    & =          &  =          & \checkmark & =          & = &            \\
		G345 Main    & \checkmark & = & \checkmark &            &  =          & \\
		G345 NW spur & \checkmark & = & =          &            & =           &   =  \\ 
		\hline
	\end{tabular}
\end{table*}

\subsection{Comparison of column densities and excitation temperatures}
\label{xclasssection}

Using the XCLASS LTE spectral modeling software described in \S\ref{lineids}, we determine excitation temperature ($T_\mathrm{ex}$), column densities ($N_\mathrm{col}$), line width (FWHM), and velocity (v$_\mathrm{LSR}$) for spectra extracted from single pixels (indicated in Figure~\ref{contpts}).  Modeled $N_\mathrm{col}$ values were divided by the H$_2$ column densities listed in Table~\ref{spectrapoints} to obtain abundances for comparison, and output v$_\mathrm{LSR}$ were subtracted from the v$_\mathrm{LSR}$ of the sources listed in Table~\ref{sourceTable} to obtain velocity offsets. The full modeling results are presented in Tables~\ref{xclassformaerr}, \ref{xclasshncoerr}, and \ref{xclassh2coerr}. The errors shown were determined using the \textit{errorestim$\_$ins} algorithm using the Markov chain Monte Carlo (MCMC) method built into the XCLASS software. A detailed description of this method is included in the XCLASS manual\footnote{Manual downloadable from: https://xclass.astro.uni-koeln.de/sites/xclass/files/pdfs/XCLASS-Interface$\_$Manual.pdf}. 

\par
\begin{figure}[!htb]
	\centering
	\includegraphics[width=0.9\hsize]{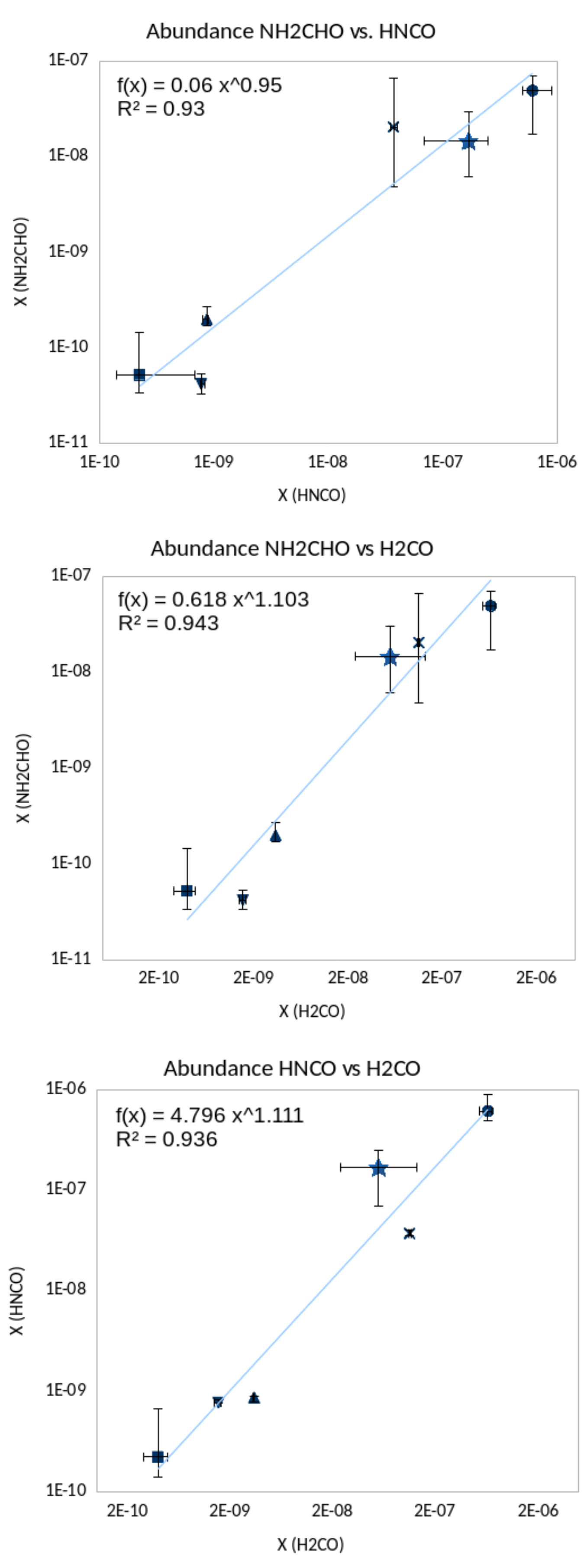}
	\caption{XCLASS determined abundance comparison between NH$_2$CHO and HNCO (top left), NH$_2$CHO and H$_2$CO (top right), and HNCO and H$_2$CO (bottom). The symbols correspond to different regions as follows: G17 is an upward triangle, G24 A1 is an 'x', G24 A2(N) is a star, G24 A2(S) is a circle, G345 Main is a square, and G345 NW spur is a downward triangle.}
	\label{abundances}
\end{figure}

 Figure~\ref{abundances} shows the modeled abundance values (\textit{X}) for NH$_2$CHO, H$_2$CO, and HNCO plotted against each other for all subsources.  The relationships between each of the species pairs all have good fits, all with an $R^2$ (a statistical measurement of linear correlation where 1 is best and values over 0.7 are considered well correlated.) greater than 0.93 with a slightly better correlation between H$_2$CO and NH$_2$CHO is best at 0.943.  The fit for NH$_2$CHO vs. H$_2$CO is [\textit{X}(NH$_2$CHO)=0.618 \textit{X}(H$_2$CO)$^{1.103}$] and the fit for NH$_2$CHO vs. HNCO is [\textit{X}(NH$_2$CHO)=0.06 \textit{X}(HNCO)$^{0.95}$] (R$^2$=0.93).  The abundances of HNCO and H$_2$CO are also correlated with a fit of [\textit{X}(HNCO)=4.796 \textit{X}(H$_2$CO)$^{1.111}$] (R$^2$=0.936). The errors on the abundance are less than one order of magnitude, as seen in Figure~\ref{abundances}.

\par The $T_\mathrm{ex}$ and FWHM values do not show any correlation between any of the pairs of species, but both of these parameters have a very narrow range of results for NH$_2$CHO.  The $T_\mathrm{ex}$ range for NH$_2$CHO is 50-150~K, whereas for HNCO it is 75-200~K and for H$_2$CO it is 70-375~K. The largest errors in $T_\mathrm{ex}$ are for HNCO around G345 Main at 63~K, but the average error in $T_\mathrm{ex}$ is 12.7~K. The FWHM for NH$_2$CHO range from $\sim$2.3-5.7~km s$^{-1}$, while for the other two the range is 2.8-6.5~km s$^{-1}$. The errors associated with the FWHM fits are very small with an average of 0.2~km s$^{-1}$ with the largest error being 0.7~km s$^{-1}$.

\par
\begin{figure}[!htb]
	\centering
	\includegraphics[width=0.9\hsize]{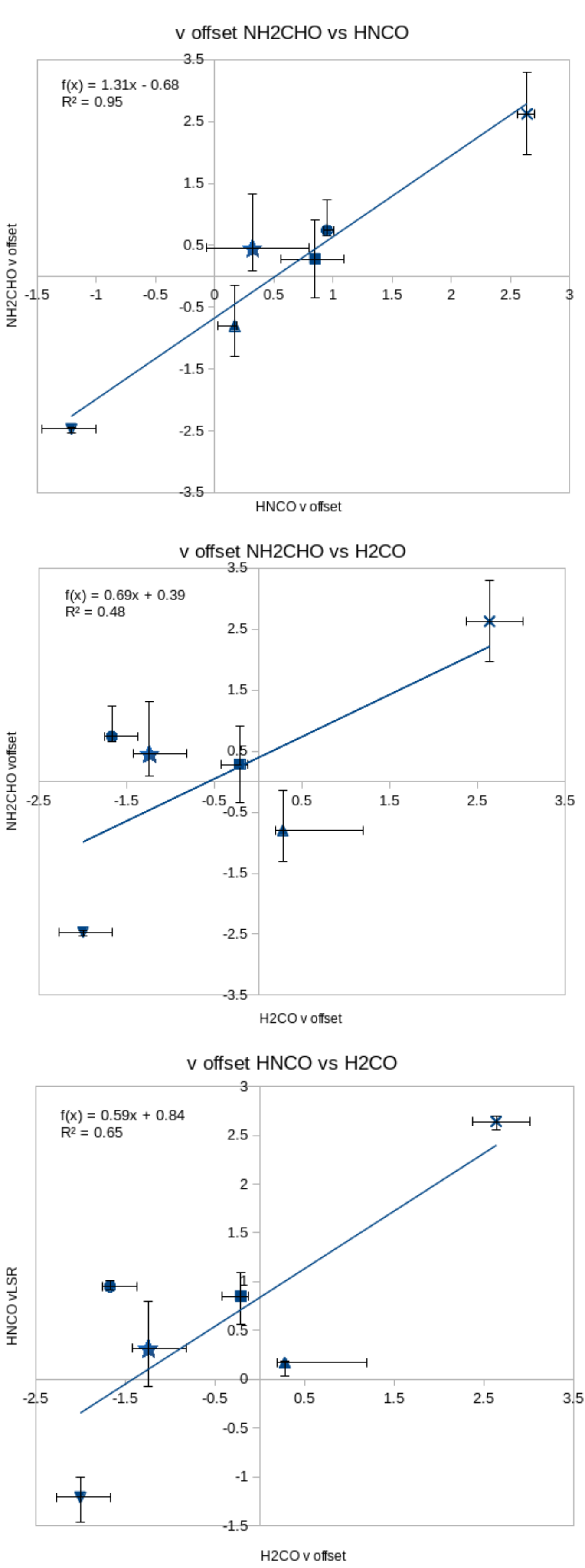}
	\caption{XCLASS determined velocity offset comparison between NH$_2$CHO and HNCO (top left), NH$_2$CHO and H$_2$CO (top right), and HNCO and H$_2$CO (bottom). The symbols are as in Figure~\ref{abundances}.}
	\label{offsets}
\end{figure}

 Figure~\ref{offsets} shows the velocity offset values for NH$_2$CHO, HNCO, and H$_2$CO plotted against each other for all subsources.  The scatter of velocity offset values for NH$_2$CHO is smaller with HNCO than with H$_2$CO (R$^2$ of 0.95 vs. 0.48).  The slope of the NH$_2$CHO vs. HNCO velocity offset plot is nearly 1, but the intercept is not zero (V$_\mathrm{NH2CHO}$=1.31 V$_\mathrm{HNCO}$-0.68) whereas the slope of NH$_2$CHO vs. H$_2$CO is closer to 0.7 but the intercept is nearer to zero (V$_\mathrm{NH2CHO}$=0.69 V$_\mathrm{H2CO}$+0.39). The errors on the model fit of velocity offset are also small where the largest is 0.9~km s$^{-1}$ but the average error for all sources and species is 0.3~km s$^{-1}$.

\section{Discussion}
\label{discussion}

\subsection{Overall map trends}
We see from the summary of map analysis results in Table~\ref{checklist} that the peak positions and dispersion maps favor HNCO slightly over H$_2$CO in similarity with NH$_2$CHO and the velocity dispersion maps for HNCO are almost always most similar to NH$_2$CHO. There are two sources which favor HNCO over H$_2$CO in all three moment map tests: G24 A1 and G345 Main.  While the integrated emission peaks of HNCO are generally much closer to NH$_2$CHO (by 0.1-0.3$''$), differences of less than 0.2$''$ are smaller than the beam. We measure the 2D-Gaussian peaks of the lines with an error of 0.01$''$ so the similarity between HNCO and NH$_2$CHO peaks is significant. The gas velocity structure of NH$_2$CHO is closer to HNCO in half of the sources (G24 A1, G345 Main, and G345 NW spur), and closer to H$_2$CO in the other half (G17, G24 A2(N) and A2(S)), but the difference in gas velocities between H$_2$CO-NH$_2$CHO and HNCO-NH$_2$CHO is generally less than 0.2~km s$^{-1}$ (as depicted in Figure~\ref{mom1Fig} and Table~\ref{histTable}). With an error of the central velocity measurement of 0.4 km s$^{-1}$ for NH$_2$CHO (1) (from Gaussian fits of this transition in each source), this difference is not significant for G17, and G345 Main and NW spur. For the subsources in G24 which use NH$_2$CHO (2), the error on the velocity measurement is 0.1 km s$^{-1}$ (also from Gaussian fits), which makes the similarity between NH$_2$CHO and H$_2$CO in G24 A2(N) and between NH$_2$CHO and HNCO (3) in G24 A1 significant. The velocity dispersion values for HNCO are closer to NH$_2$CHO for five sources but closer to H$_2$CO for one source and they typically span a larger range of velocities for H$_2$CO.  From these overall results, it seems that HNCO has a slightly stronger relationship with NH$_2$CHO.

\par The opacity of the H$_2$CO transitions investigated here cannot be discounted. It is possible that the greater differences in spatial distribution, velocity, and dispersion between H$_2$CO and NH$_2$CHO compared to HNCO arise from optical depth issues. This is being investigated in a follow-up study involving isotopologues.

\subsection{XCLASS analysis}
The result of our XCLASS analysis shows no relationship between the widths of lines of different species or between the gas temperatures ($T_\mathrm{ex}$) of any of the species.  The velocity offset relationship is strongest between HNCO and NH$_2$CHO with a nearly linear fit and a small scatter.

\par There is a correlation between abundances for all three pairs of species but the best fit is between H$_2$CO and NH$_2$CHO.  Most interesting is the relationship between the abundances of HNCO and NH$_2$CHO in this work is almost exactly the same as that reported in \citet{Ana2015}.  In their paper, the best power-law fit was \textit{X}(NH$_2$CHO) = 0.04 \textit{X}(HNCO)$^{0.93}$ and the best fit in this work is \textit{X}(NH$_2$CHO)=0.06($\pm$0.03) \textit{X}(HNCO)$^{0.95 (\pm0.05)}$. The correlation between abundances of all three pairs of species suggests that such a correlation is not a good indicator of a direct chemical relationship.

\section{Conclusions}
\label{conclusion}
We present an observational study of two species potentially chemically related (HNCO and H$_2$CO) to NH$_2$CHO.  Our study improves upon previous studies using single dish observations by including map analysis made possible using highly-sensitive interferometric observations. The different moment maps that we employ indicate whether the gas containing NH$_2$CHO has the same properties as the gas containing its potential precursors. The spectral analysis performed in \citet{Ana2015} and \citet{Bisschop2007} used significantly more transitions and the rotational diagram method in order to determine abundances. While we had fewer unblended transitions available to us, the XCLASS LTE spectral modeling method is more rigorous and does not require the assumption of optically thin transitions as that would be incorrect. A few interferometric studies involving NH$_2$CHO have been done (e.g., \citet{Codella2017} - L1157-B1, \citet{Coutens2016} - IRAS 16293-2422), but only involving one of its precursors, so we also improve upon that method by investigating the three species together in several star forming regions.

\par In our spectral modeling, we confirm the single dish relationship between the abundances of HNCO and NH$_2$CHO demonstrated in \citet{Bisschop2007} and \citet{Ana2015} using interferometric observations. Our map analyses favor HNCO as chemically related to NH$_2$CHO.  The abundance correlation between H$_2$CO and NH$_2$CHO is slightly stronger than the correlation between HNCO and NH$_2$CHO but both are very well correlated. It is possible that both formation processes are important in creating this species, or that different environments favor one process over the other.  Dedicated studies using more transitions and isotopologues in a more diverse selection of sources (high- and low-mass protostars, young stellar objects with disks, outflow regions, etc.) would shed light on this relationship.

\begin{acknowledgements}
	This paper makes use of the following ALMA data: ADS/JAO.ALMA 2013.1.00489.S (P.I. Riccardo Cesaroni). ALMA is a partnership of ESO (representing its member states), NSF (USA) and NINS (Japan), together with NRC (Canada) and NSC and ASIAA (Taiwan), in cooperation with the Republic of Chile. The Joint  ALMA Observatory is operated by ESO, AUI/NRAO and NAOJ.\\
	This paper made use of information from the Red MSX Source survey database at http://rms.leeds.ac.uk/cgi-bin/public/RMS$\_$DATABASE.cgi which was constructed with support from the Science and Technology Facilities Council of the UK.\\
	The PhD project of V. Allen was funded by NWO and SRON. V. Allen's research is supported by an appointment to the NASA Postdoctoral Program at the NASA Goddard Space Flight Center, administered by Universities Space Research Association under contract with NASA.
	V.M.R. is funded by the European Union's Horizon 2020 research and innovation programme under the Marie Sk\l{}odowska-Curie grant agreement No 664931.
\end{acknowledgements}

\begin{table*}[!h]
	\centering
	\caption{XCLASS LTE spectral modeling results for formamide (NH$_2$CHO). Columns show modeled excitation temperature ($T_\mathrm{ex}$), column density ($N_\mathrm{col}$), line width ($\Delta$ v), and line velocity (v$_\mathrm{LSR}$) for each of our key species. The columns indicated by a minus sign (-) indicate the error to the left of the result and those indicated by a plus sign (+) indicate the error to the right.}
	\label{xclassformaerr}
	\begin{tabular}{c|ccc|ccc|ccc|ccc}
		\hline \hline
		\textbf{NH$_2$CHO} &          &            &             &           &             &             &            &            &             &          &            &             \\ \hline
		Source             & T$_{ex}$ & - & + & N$_{col}$ & lower limit & upper limit & $\Delta$v & - & + & v$_\mathrm{LSR}$  & - & + \\
		G17                & 56.23    & 0.29       & 0.44        & 2.09$\times10^{15}$  & 1.78$\times10^{15}$    & 2.88$\times10^{15}$    & 5.7        & 0.3        & 0.2         & 23.3     & 0.5        & 0.65        \\
		G24 A1             & 65       & 14         & 26          & 2.77$\times10^{16}$  & 6.49$\times10^{15}$    & 8.96$\times10^{16}$    & 3.1        & 0.2        & 0.2         & 108.4    & 0.7        & 0.7         \\
		G24 A2(N)          & 79       & 13         & 15          & 1.45$\times10^{16}$  & 6.06$\times10^{15}$    & 2.97$\times10^{16}$    & 3.3        & 0.1        & 0.1         & 110.5    & 0.4        & 0.9         \\
		G24 A2(S)          & 89       & 20         & 2           & 2.42$\times10^{15}$  & 1.41$\times10^{15}$    & 5.75$\times10^{15}$    & 2.5        & 0.1        & 0.3         & 110.3    & 0.1        & 0.5         \\
		G345 Nain       & 152      & 33         & 27          & 5.13$\times10^{15}$  & 3.31$\times10^{15}$    & 1.41$\times10^{16}$    & 2.4        & 0.6        & 0.5         & -17.8    & 0.6        & 0.6         \\
		G345 NW spur    & 93       & 4          & 10          & 9.68$\times10^{14}$  & 7.69$\times10^{14}$    & 1.22$\times10^{15}$    & 3.3        & 0.2        & 0.2         & -12.03   & 0.06       & 0.03  \\
		\hline     
	\end{tabular}
\end{table*}

\begin{table*}[!h]
	\centering
	\caption{XCLASS best fit results as in Table~\ref{xclassformaerr} but for HNCO.}
	\label{xclasshncoerr}
	\begin{tabular}{c|ccc|ccc|ccc|ccc}
		\hline \hline
		\textbf{HNCO}   &          &            &             &           &             &             &            &            &             &          &            &             \\ \hline
		Source          & T$_{ex}$ & - & + & N$_{col}$ & lower limit & upper limit & $\Delta$v & - & + & v$_\mathrm{LSR}$  & - & + \\
		G17             & 86       & 4          & 1           & 9.12$\times10^{15}$  & 8.32$\times10^{15}$    & 9.33$\times10^{15}$    & 6.17       & 0.03       & 0.19        & 22.67    & 0.14       & 0.02        \\
		G24 A1          & 192.0    & 0          & 1           & 5.01$\times10^{16}$  & 4.79$\times10^{16}$    & 5.37$\times10^{16}$    & 6.43       & 0.06       & 0.31        & 108.4    & 0.08       & 0.06        \\
		G24 A2(N)       & 177.0    & 18         & 27          & 1.66$\times10^{17}$  & 6.77$\times10^{16}$    & 2.46$\times10^{17}$    & 3.28       & 0.05       & 0.13        & 110.7    & 0.4        & 0.5         \\
		G24 A2(S)       & 114.9    & 0.7        & 3.7         & 5.02$\times10^{17}$  & 4.08$\times10^{17}$   & 7.43$\times10^{17}$    & 3.63       & 0.01       & 0.1         & 110.1    & 0.03       & 0.06        \\
		G345 Main    & 173.0    & 27         & 63          & 2.19$\times10^{16}$  & 1.38$\times10^{16}$    & 6.61$\times10^{16}$    & 5.0        & 0.7        & 0.3         & -18.4    & 0.3        & 0.2         \\
		G345 NW spur & 198.7    & 0.4        & 0.5         & 1.78$\times10^{16}$  & 8.00$\times10^{14}$    & 2.17$\times10^{15}$    & 4.3        & 0.2        & 0.1         & -13.3    & 0.3        & 0.2        \\
		\hline
	\end{tabular}
\end{table*}

\begin{table*}[!h]
	\centering
	\caption{XCLASS best fit results as in Table~\ref{xclassformaerr} but for H$_2$CO.}
	\label{xclassh2coerr}
	\begin{tabular}{c|ccc|ccc|ccc|ccc}
		\hline \hline
		\textbf{H$_2$CO} &          &            &             &           &             &             &            &            &             &          &            &             \\ \hline
		Source           & T$_{ex}$ & - & + & N$_{col}$ & lower limit & upper limit & $\Delta$v & - & + & v$_\mathrm{LSR}$ & - & + \\
		G17              & 110    & 1        & 3         & 3.58$\times10^{16}$  & 3.51$\times10^{16}$    & 3.59$\times10^{16}$    & 4.26       & 0.06       & 0.23        & 22.78    & 0.08       & 0.92        \\
		G24 A1           & 374       & 5          & 6           & 1.51$\times10^{17}$  & 1.47$\times10^{17}$    & 1.55$\times10^{17}$    & 6.4        & 0.1        & 0.3         & 108.4    & 0.3        & 0.4         \\
		G24 A2(N)        & 76       & 4          & 9           & 5.49$\times10^{16}$  & 2.34$\times10^{16}$    & 1.29$\times10^{17}$    & 5.1        & 0.1        & 0.2         & 112.2    & 0.2        & 0.4         \\
		G24 A2(S)        & 138     & 30        & 46        & 5.33$\times10^{17}$  & 4.42$\times10^{17}$    & 5.89$\times10^{17}$    & 4.6       & 0.3       & 0.2         & 112.38   & 0.03       & 0.75        \\
		G345 Main     &  188    & 4         & 33          & 3.86$\times10^{16}$ & 2.81$\times10^{16}$    & 4.77$\times10^{16}$  & 2.9        & 0.6        & 0.7         & -17.29   & 0.21       & 0.09        \\
		G345 NW spur  & 70       & 2          & 4           & 3.52$\times10^{16}$  & 3.19$\times10^{16}$    & 3.76$\times10^{16}$    & 4.00       & 0.01       & 0.31        & -12.5    & 0.3        & 0.3        \\
		\hline
	\end{tabular}
\end{table*}

\bibliography{bib} 

\begin{appendix} 
	\begin{onecolumn}
\section{Formamide and ethanol in G345 NWspur}
   \begin{figure*}[!thb]
	\centering
	\includegraphics[width=\hsize]{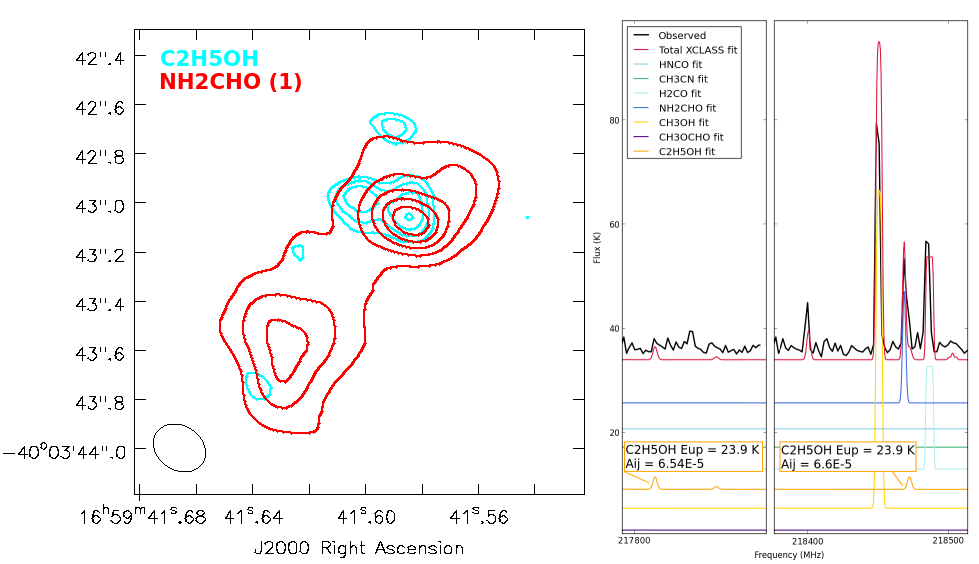}
	\caption{\textit{Right:} Spectrum of G345 NW spur showing a C$_2$H$_5$OH transition (5$_{3,3}$-4$_{2,2}$) at 217803~MHz with the same $E_\mathrm{up}$ and nearly equal $A_\mathrm{ij}$ as the C$_2$H$_5$OH transition  (5$_{3,2}$-4$_{2,3}$) that is blended with the NH$_2$CHO (1) transition (at 218461~MHz). \textit{Left:} Contours of the integrated intensity map of this C$_2$H$_5$OH line is overlaid on the map of the NH$_2$CHO (1) transition to show that the strength and spatial extent is different.}
	\label{notethanol}
\end{figure*}
\clearpage
\section{XCLASS fits}
\label{xclassfits}

   \begin{figure*}[!thb]
   \centering
      \includegraphics[width=\hsize]{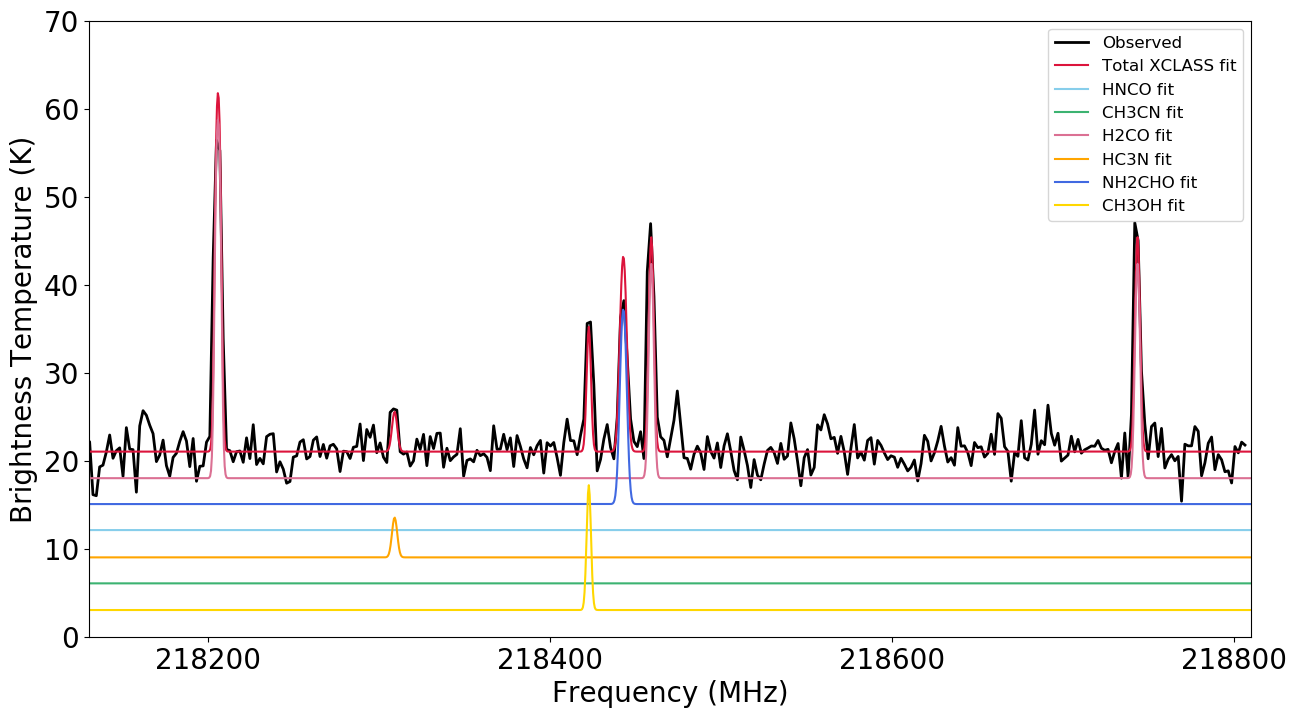}
      \caption{Spectral window from 218.1-218.8 GHz for G17 containing NH$_2$CHO (1) and H$_2$CO (1), (2), and (3).}
         \label{G17spw0}
   \end{figure*}

   \begin{figure*}[!thb]
   \centering
      \includegraphics[width=\hsize]{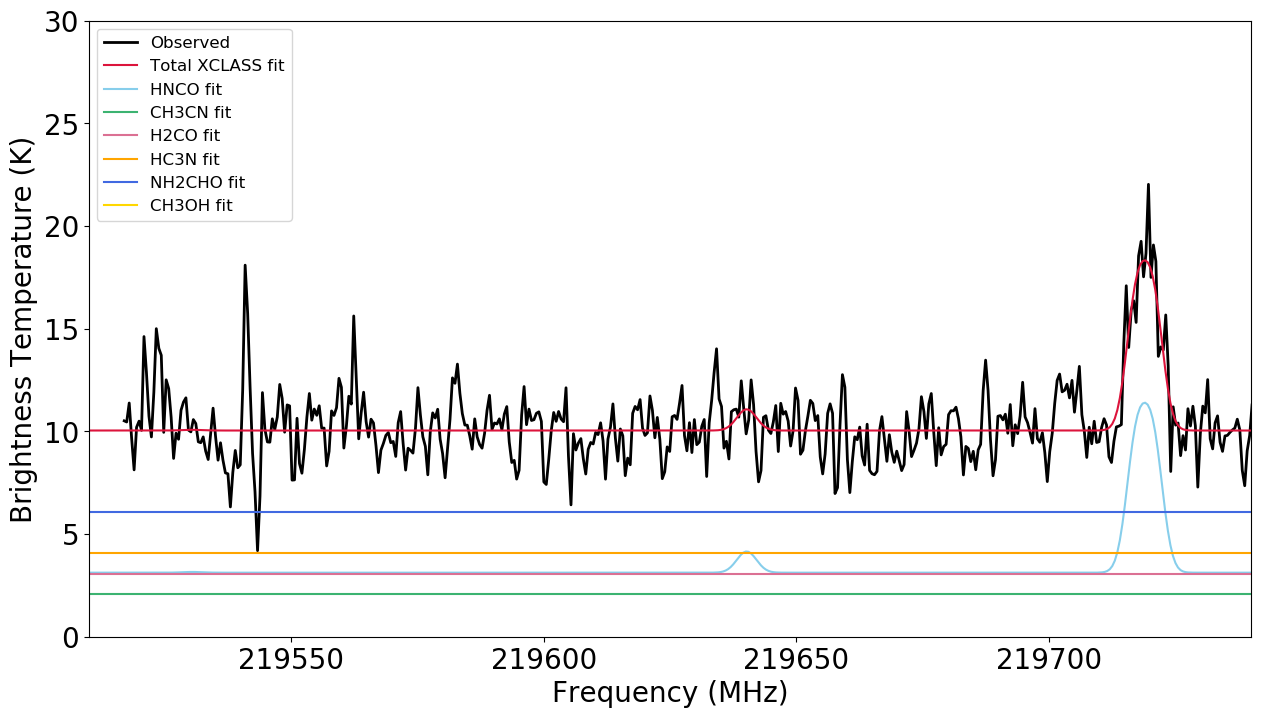}
      \caption{Spectral window from 219.5-219.8 GHz for G17 containing HNCO (3).}
         \label{G17spw1}
   \end{figure*}

\clearpage

   \begin{figure*}[!thb]
   \centering
      \includegraphics[width=\hsize]{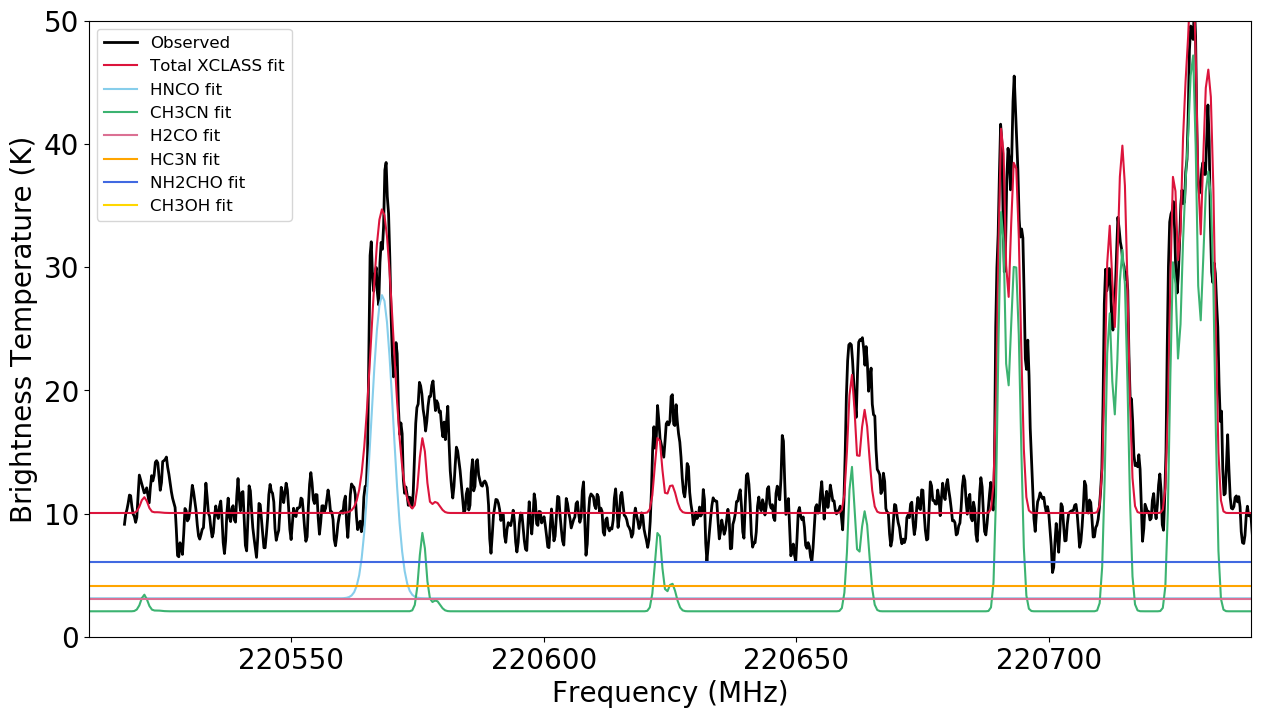}
      \caption{Spectral window from 220.5-220.75 GHz for G17 containing HNCO (2). We modeled two components for the CH$_3$CN emission (green) toward this source.}
         \label{G17spw2}
   \end{figure*}

   \begin{figure*}[!thb]
   \centering
      \includegraphics[width=\hsize]{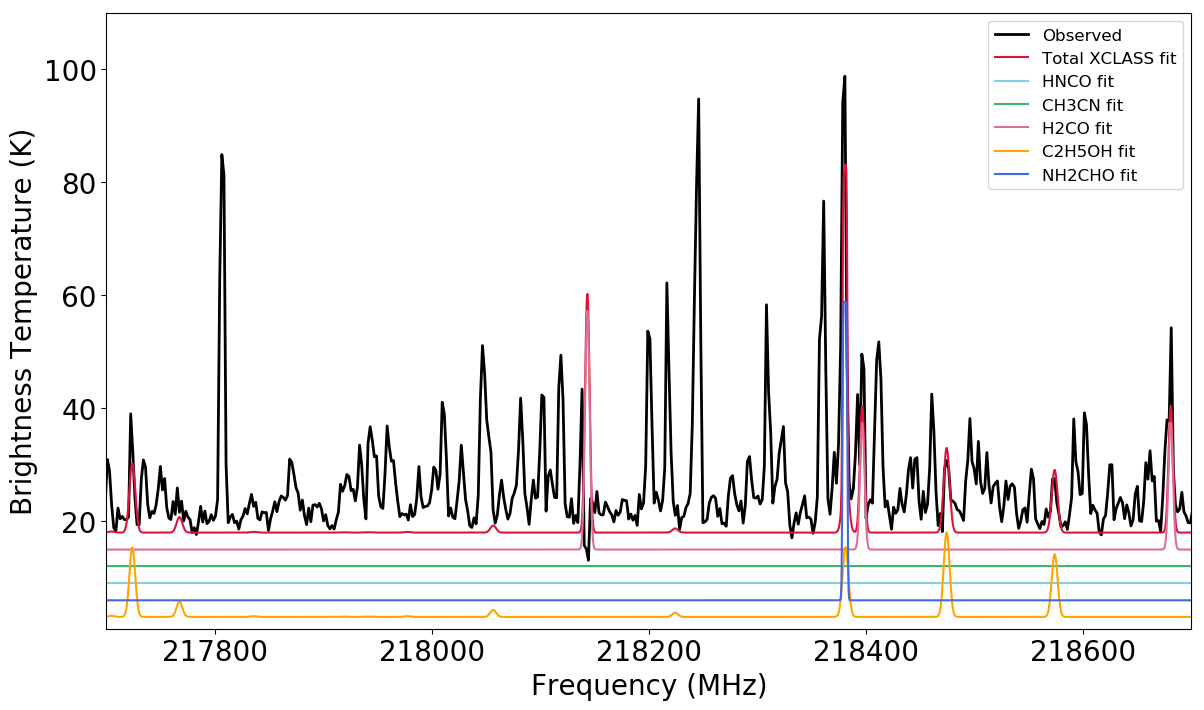}
      \caption{Spectral window from 217.7-218.7 GHz for G24 A1 containing NH$_2$CHO (1) and H$_2$CO (1), (2), and (3). NH$_2$CHO (1) is blended with a transition of C$_2$H$_5$OH.}
         \label{G24A1spw0}
   \end{figure*}

\clearpage

   \begin{figure*}[!thb]
   \centering
      \includegraphics[width=\hsize]{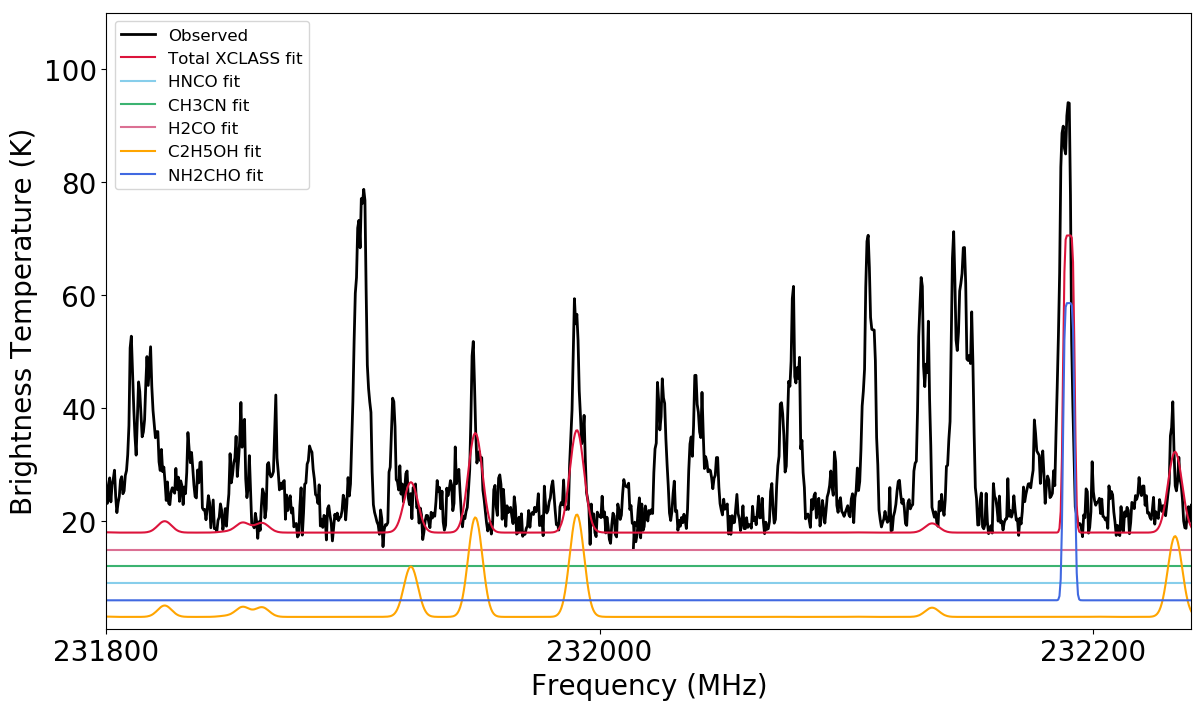}
      \caption{Spectral window from 231.8-232.3 GHz for G24 A1 containing NH$_2$CHO (2).}
         \label{G24A1spw6}
   \end{figure*}

   \begin{figure*}[!thb]
   \centering
      \includegraphics[width=\hsize]{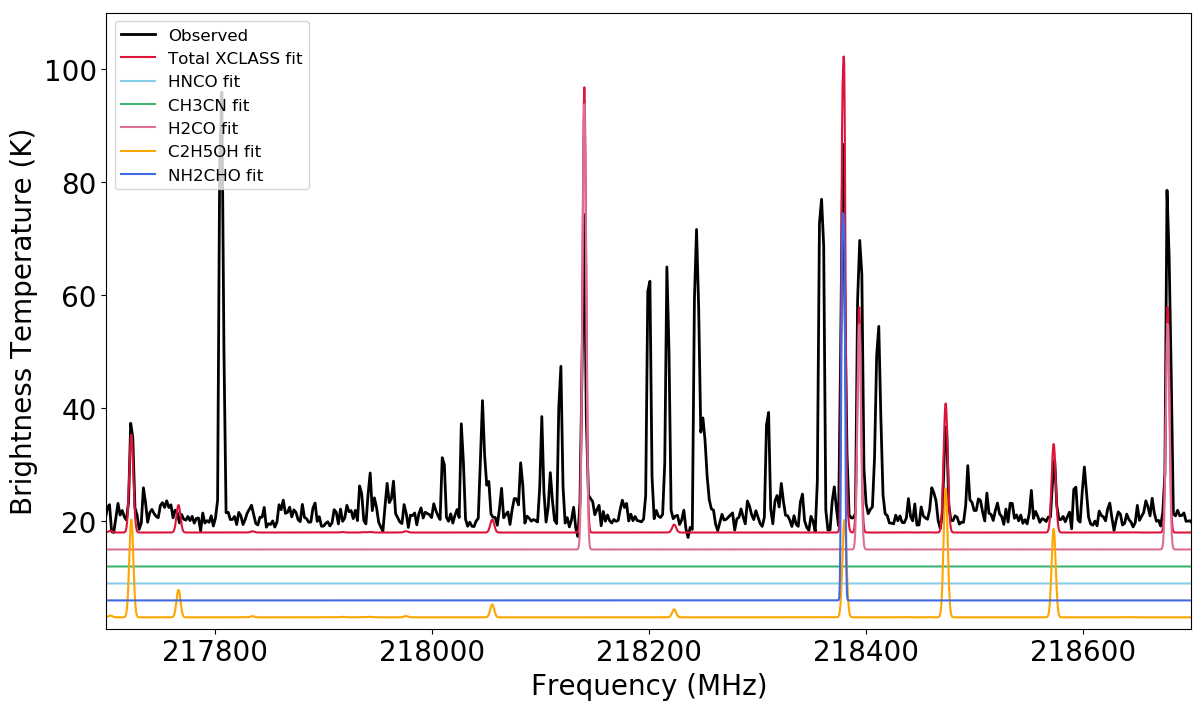}
      \caption{Spectral window from 217.7-218.7 GHz for G24 A2(N) containing NH$_2$CHO (1) and H$_2$CO (1), (2), and (3). NH$_2$CHO (1) is blended with a transition of C$_2$H$_5$OH.}
         \label{G24A2Nspw0}
   \end{figure*}

\clearpage

   \begin{figure*}[!thb]
   \centering
      \includegraphics[width=\hsize]{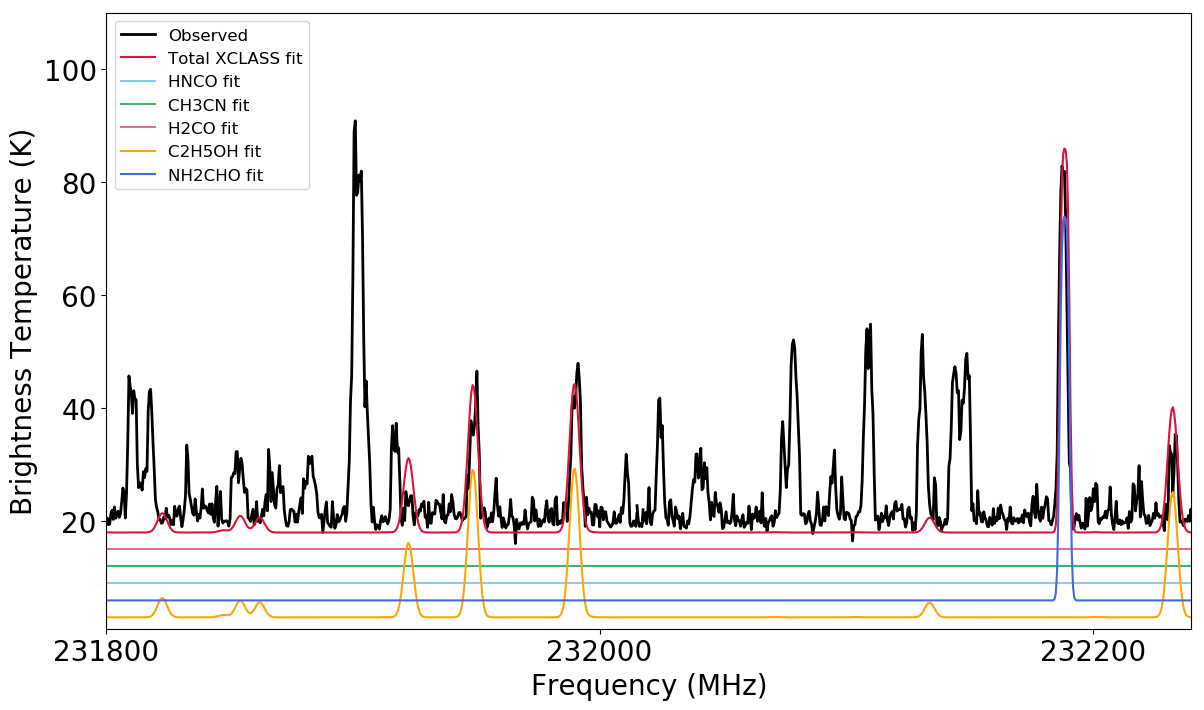}
      \caption{Spectral window from 231.8-232.3 GHz for G24 A2(N) containing NH$_2$CHO (2).}
         \label{G24A2Nspw6}
   \end{figure*}

   \begin{figure*}[!thb]
   \centering
      \includegraphics[width=\hsize]{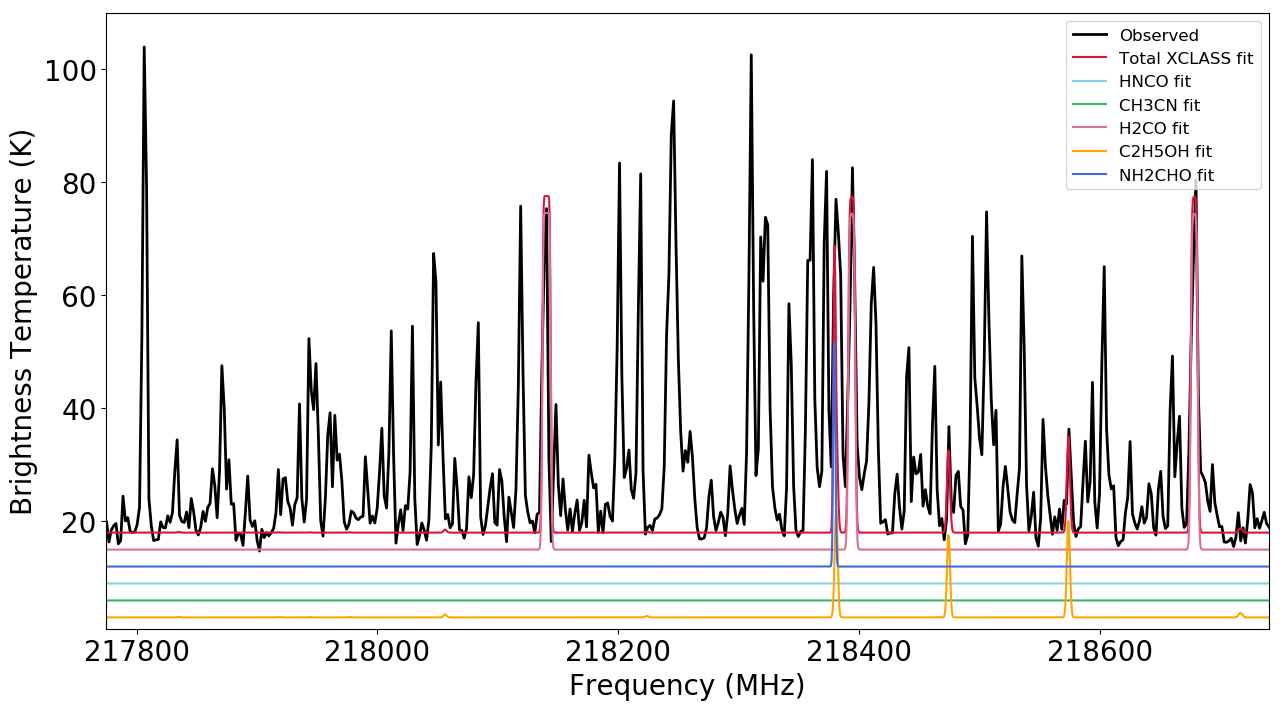}
      \caption{Spectral window from 217.7-218.7 GHz for G24 A2(S) containing NH$_2$CHO (1) and H$_2$CO (1), (2), and (3). NH$_2$CHO (1) is blended with a transition of C$_2$H$_5$OH.}
         \label{G24A2Sspw0}
   \end{figure*}

\clearpage

   \begin{figure*}[!thb]
   \centering
      \includegraphics[width=\hsize]{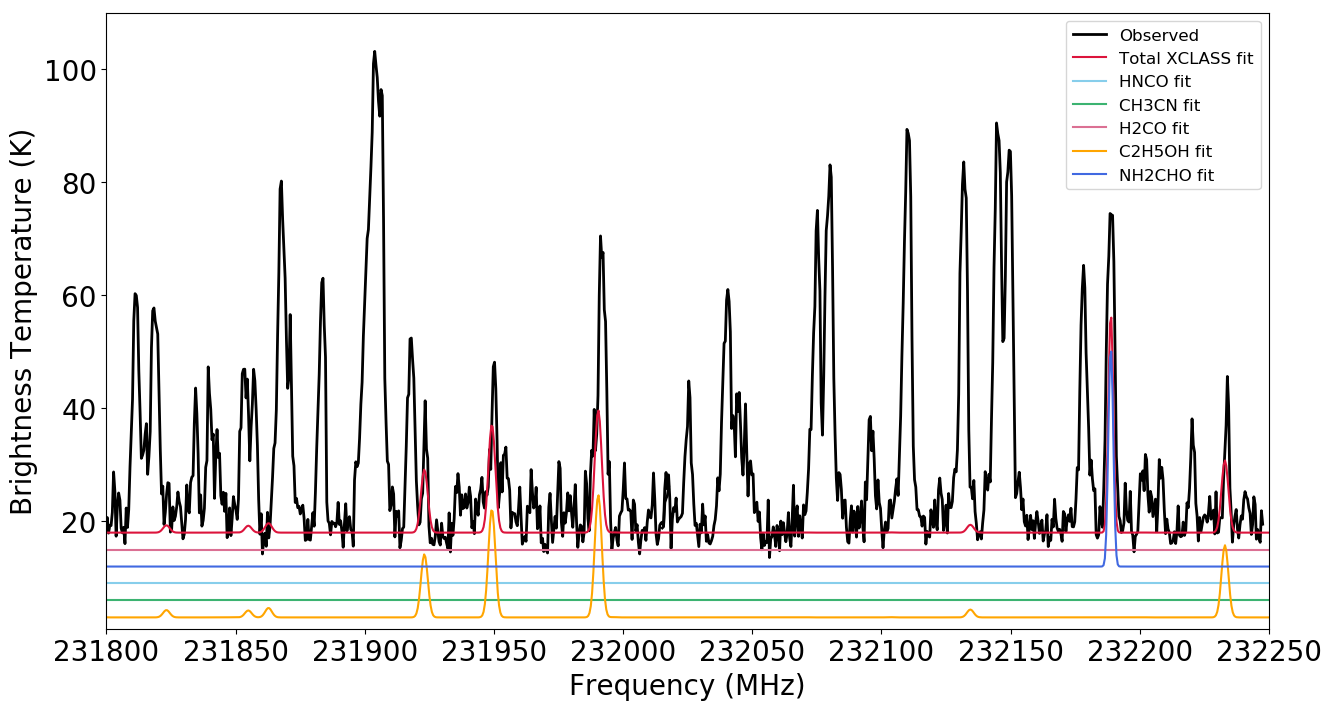}
      \caption{Spectral window from 231.8-232.3 GHz for G24 A2(S) containing NH$_2$CHO (2).}
         \label{G24A2Sspw6}
   \end{figure*}

   \begin{figure*}[!thb]
   \centering
      \includegraphics[width=\hsize]{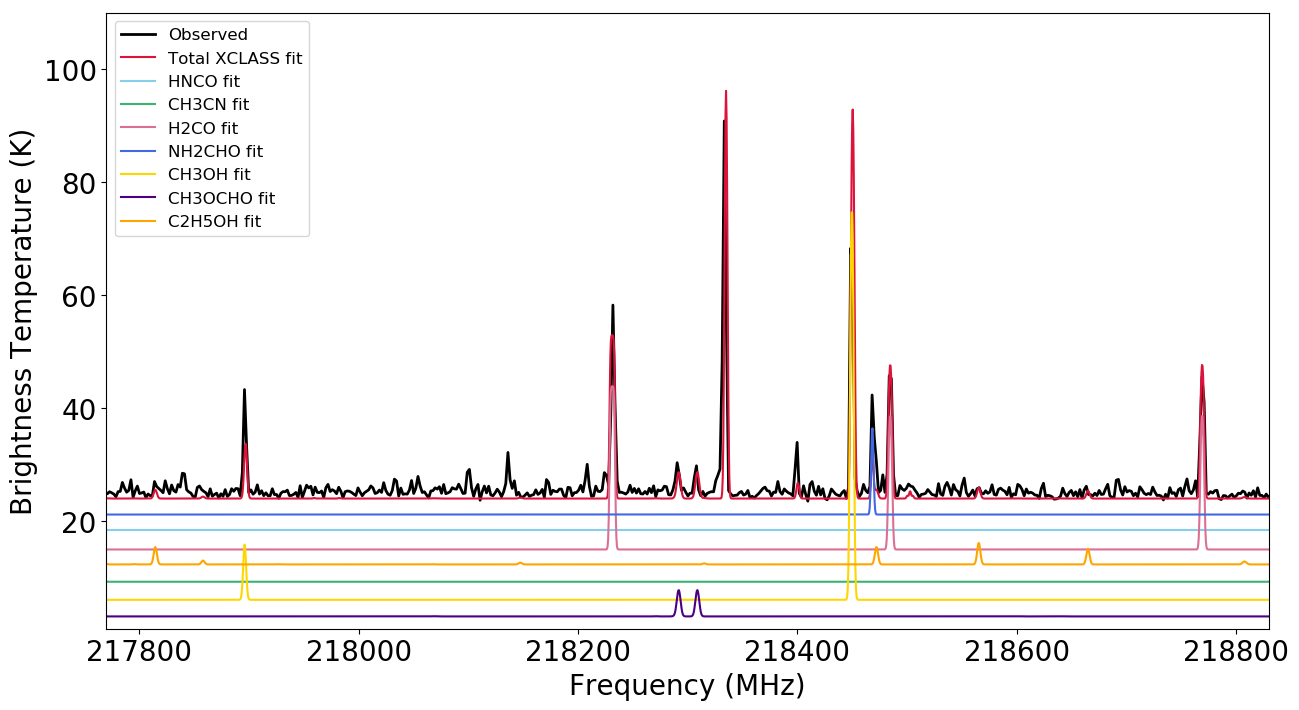}
      \caption{Spectral window from 217.8-218.8 GHz for G345 NW containing NH$_2$CHO (1) and H$_2$CO (1), (2), and (3).}
         \label{G34549nwspw0}
   \end{figure*}

\clearpage

   \begin{figure*}[!thb]
   \centering
      \includegraphics[width=\hsize]{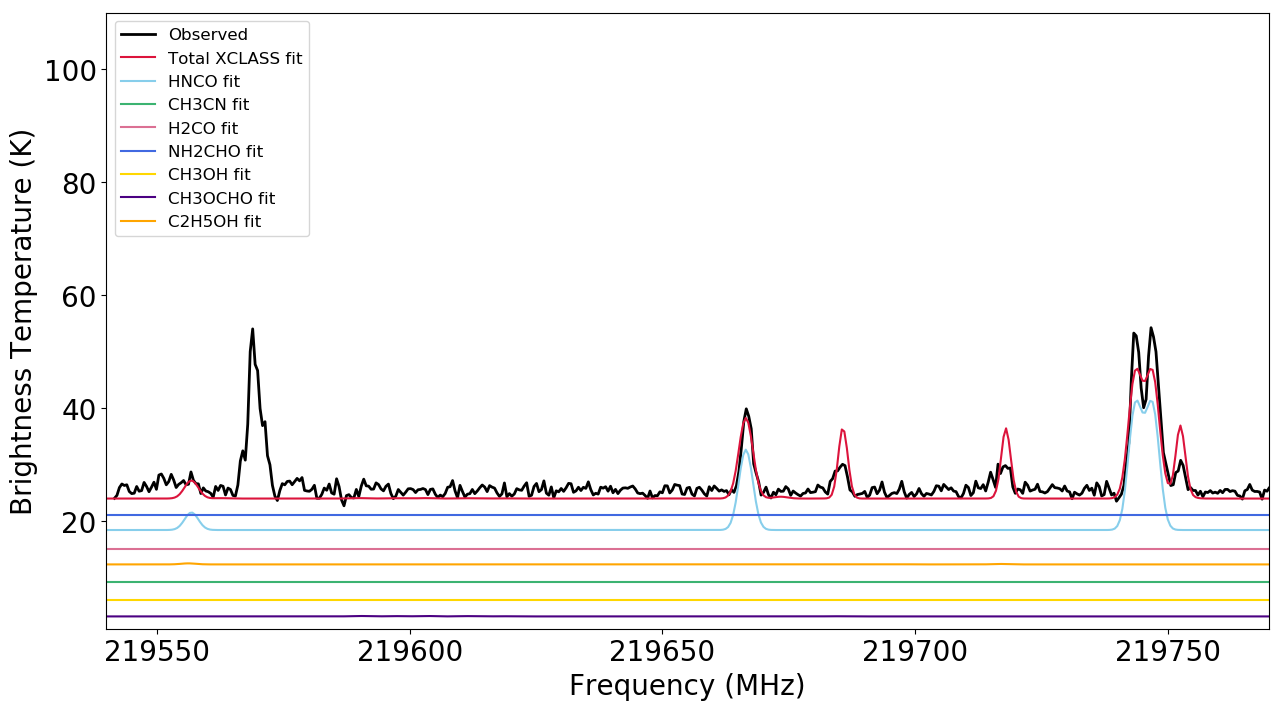}
      \caption{Spectral window from 219.5-219.8 GHz for G345 NW containing HNCO (3). The additional transitions shown in the total XCLASS fit are HC$_3$N (219675 MHz), C$_{2}$H$_{5}$CN (219699 MHz), and CH$_{3}$CHO (219756 MHz).}
         \label{G34549nwspw1}
   \end{figure*}

   \begin{figure*}[!thb]
   \centering
      \includegraphics[width=\hsize]{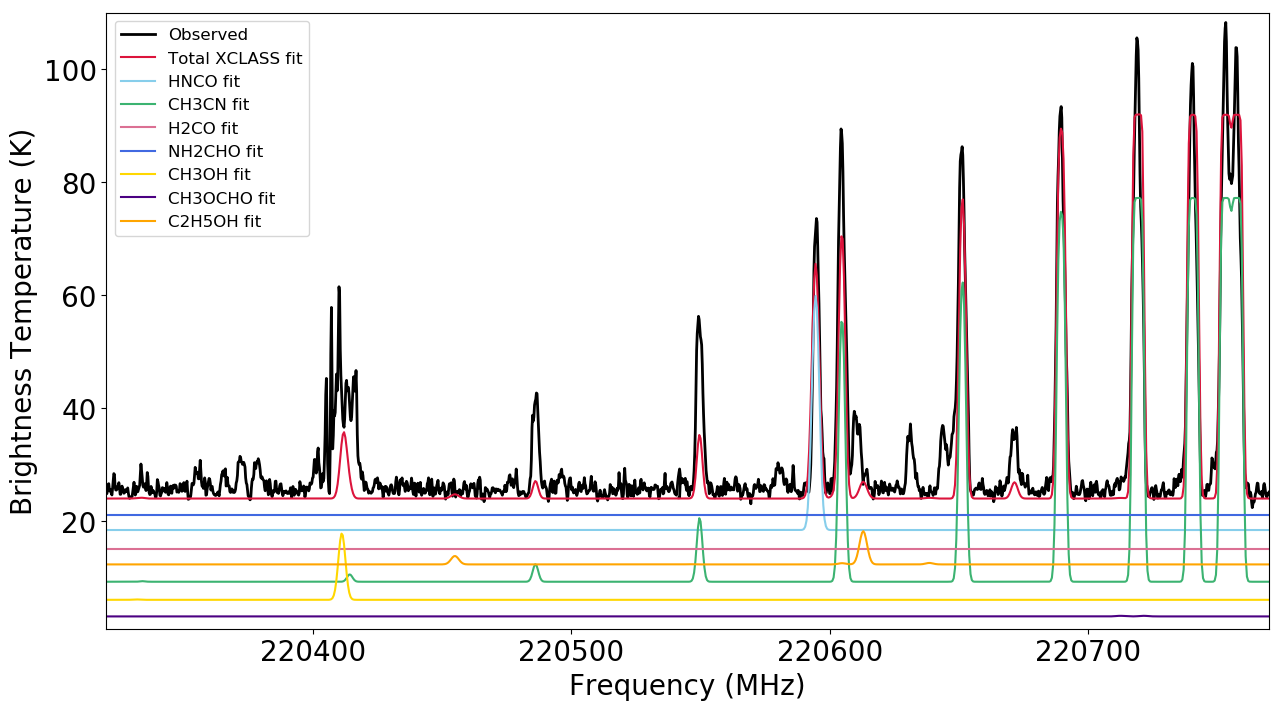}
      \caption{Spectral window from 220.3-220.8 GHz for G345 NW containing HNCO (2).}
         \label{G34549nwspw2}
   \end{figure*}

\clearpage

   \begin{figure*}[!thb]
   \centering
      \includegraphics[width=\hsize]{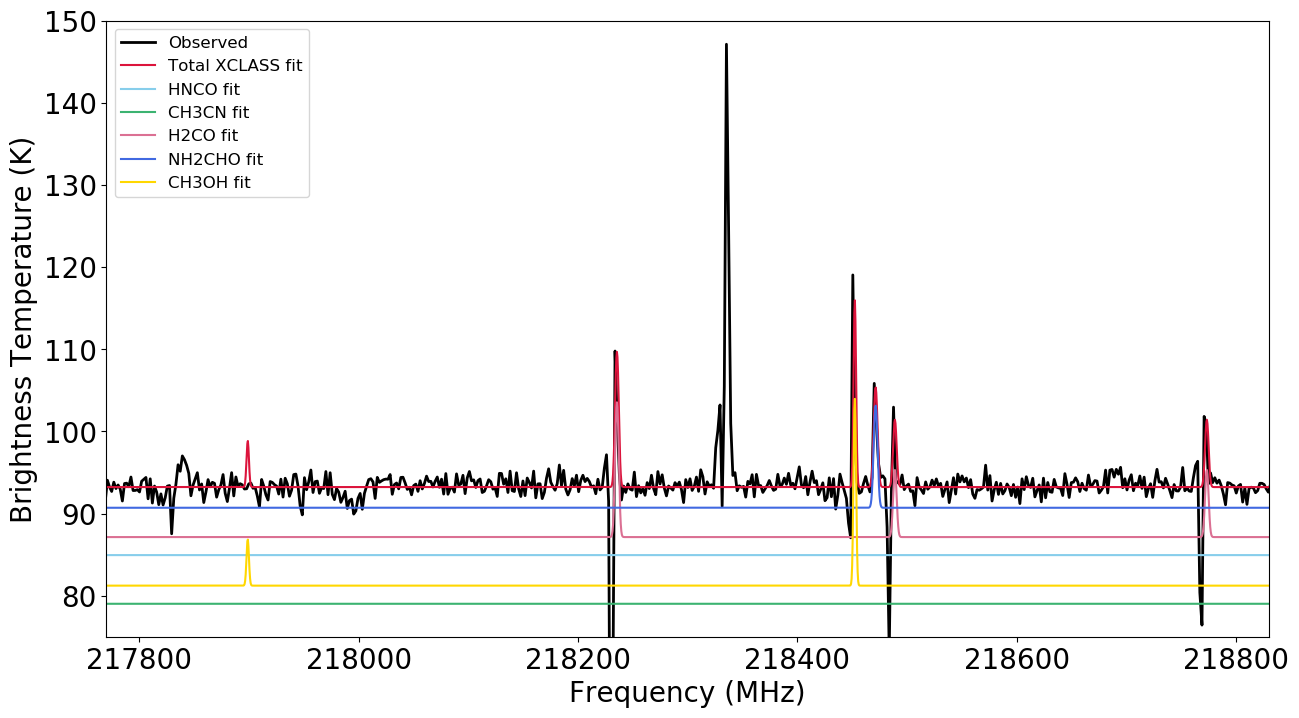}
      \caption{Spectral window from 217.8-218.8 GHz for G345 Main containing NH$_2$CHO (1) and H$_2$CO (1), (2), and (3).}
         \label{G34549mainspw0}
   \end{figure*}

   \begin{figure*}[!thb]
   \centering
      \includegraphics[width=\hsize]{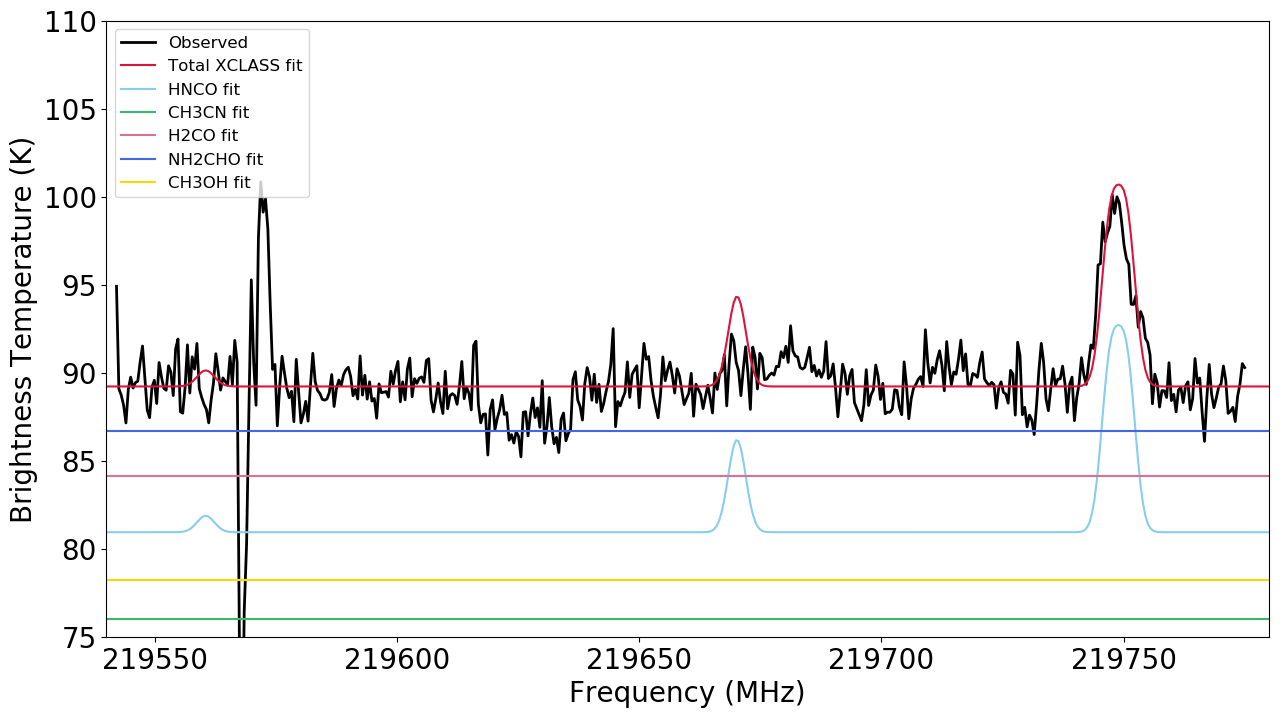}
      \caption{Spectral window from 219.5-219.8 GHz for G345 Main containing HNCO (3). }
         \label{G34549mainspw1}
   \end{figure*}

\clearpage

   \begin{figure*}[!thb]
   \centering
      \includegraphics[width=\hsize]{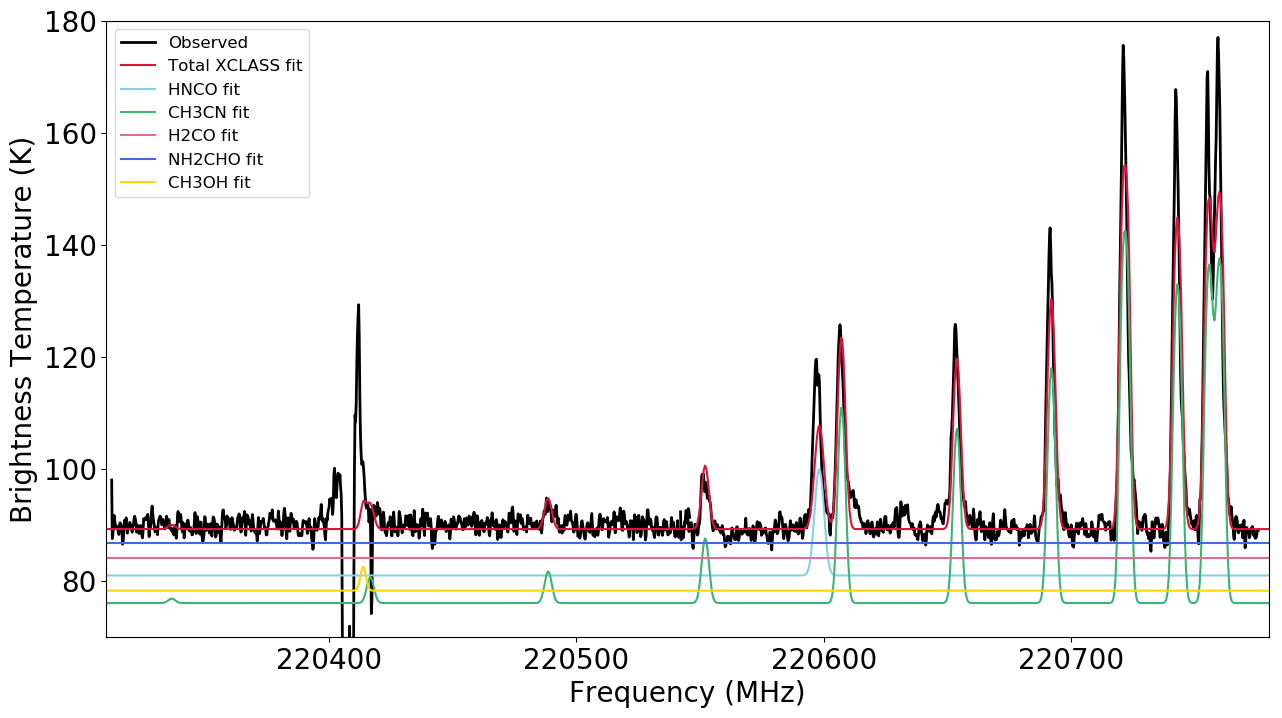}
      \caption{Spectral window from 220.3-220.8 GHz for G345 Main containing HNCO (2).}
         \label{G34549mainspw2}
   \end{figure*}

\clearpage

\section{Histograms}
\label{histograms}

   \begin{figure}[!thb]
   \centering
      \includegraphics[width=0.85\hsize]{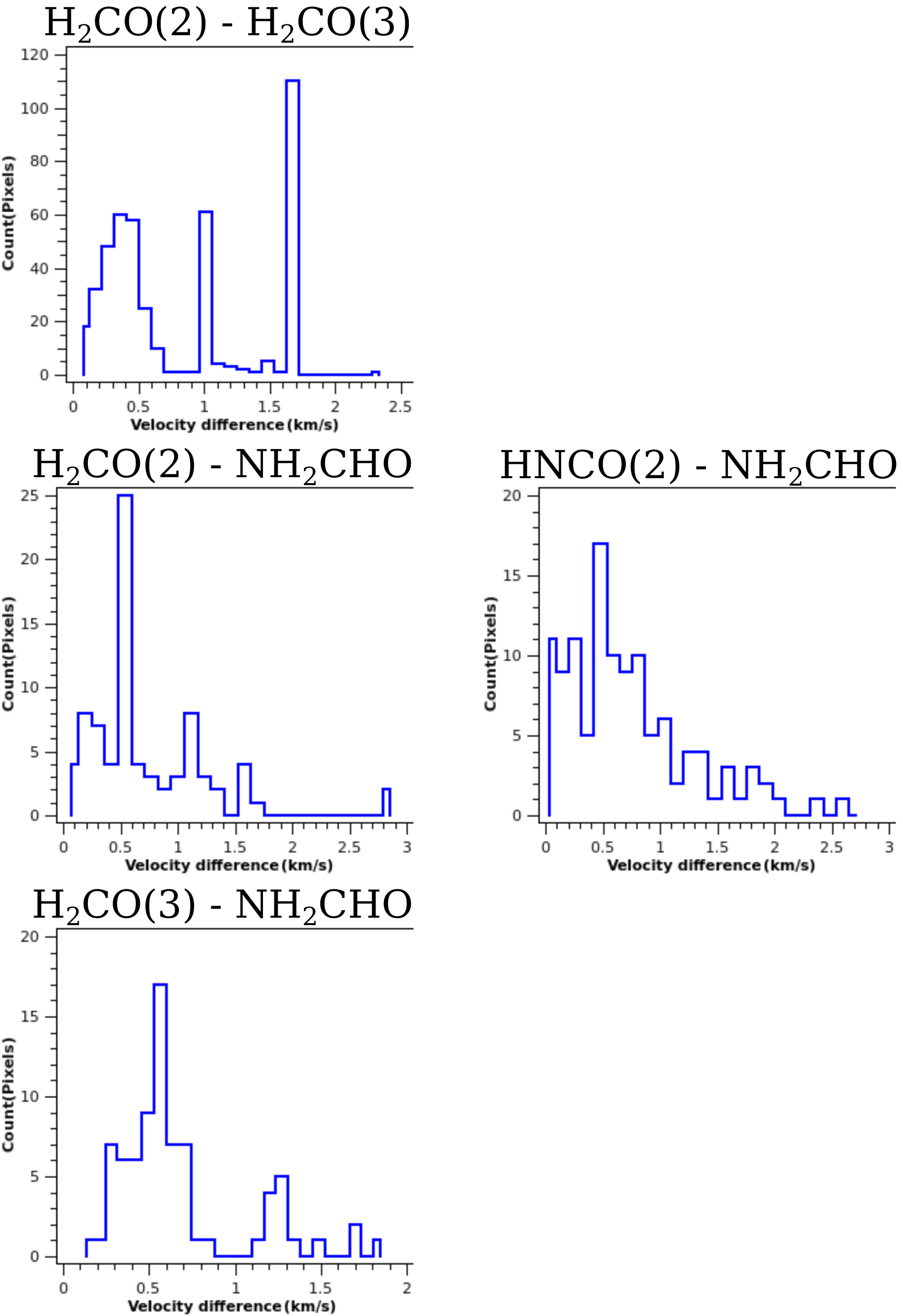}
      \caption{G17 first moment difference histogram.}
         \label{G17hist}
   \end{figure}

\clearpage

   \begin{figure*}[!thb]
   \centering
      \includegraphics[width=0.85\hsize]{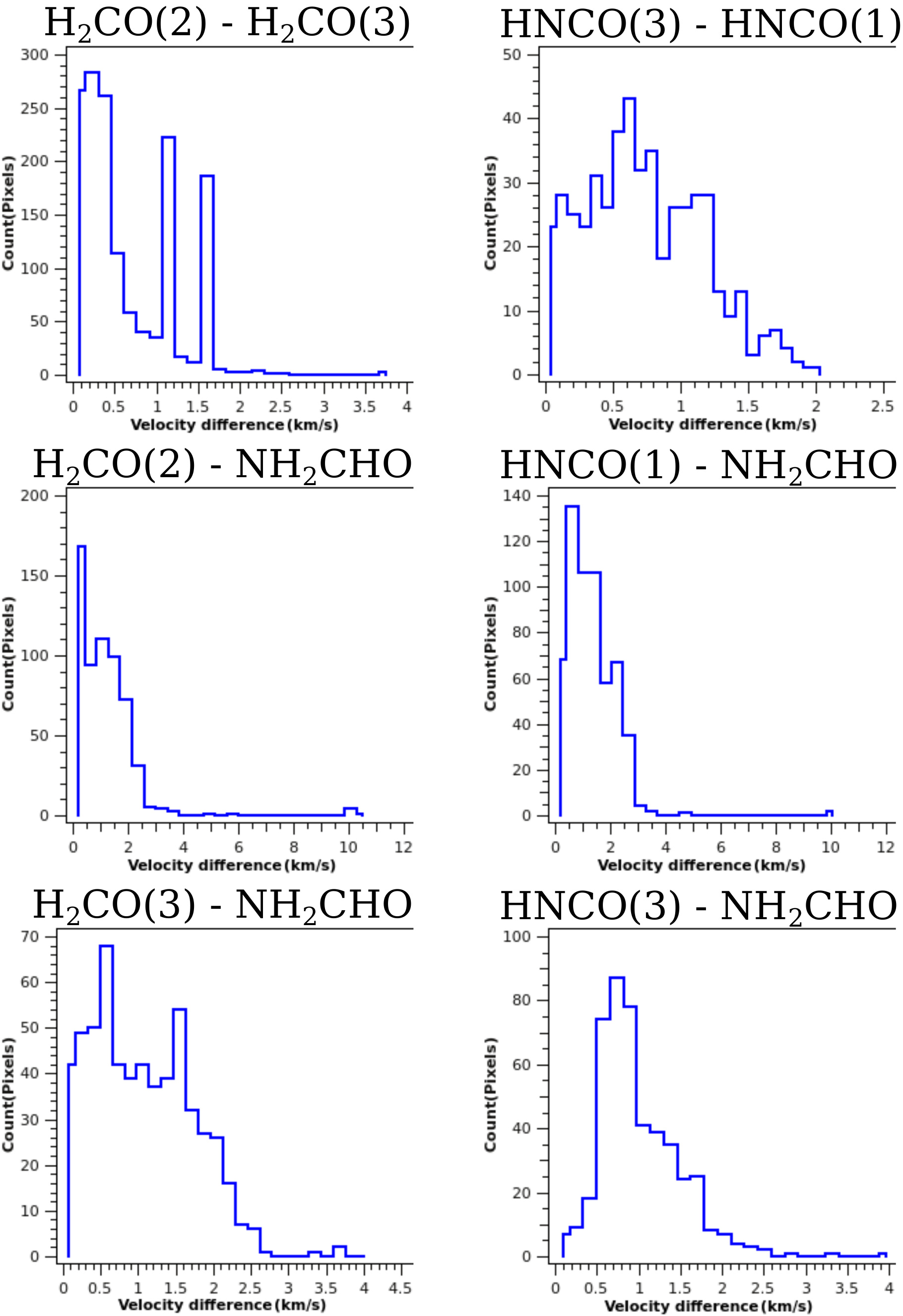}
      \caption{G24A1 first moment difference histogram.}
         \label{G24A1hist}
   \end{figure*}

   \begin{figure*}[!thb]
   \centering
      \includegraphics[width=0.85\hsize]{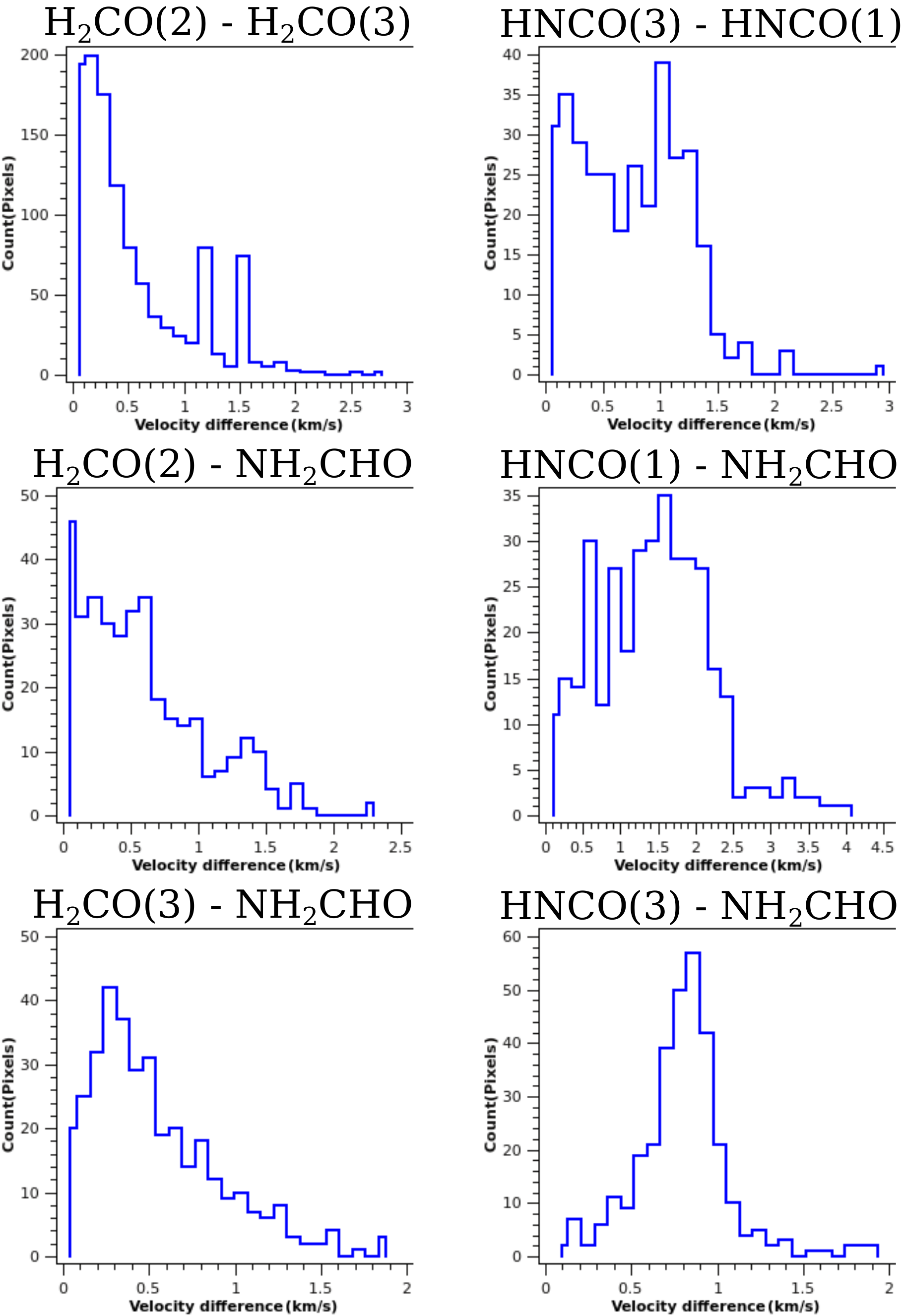}
      \caption{G24A2(N) first moment difference histogram.}
         \label{G24A2Nhist}
   \end{figure*}

   \begin{figure*}[!thb]
   \centering
      \includegraphics[width=0.85\hsize]{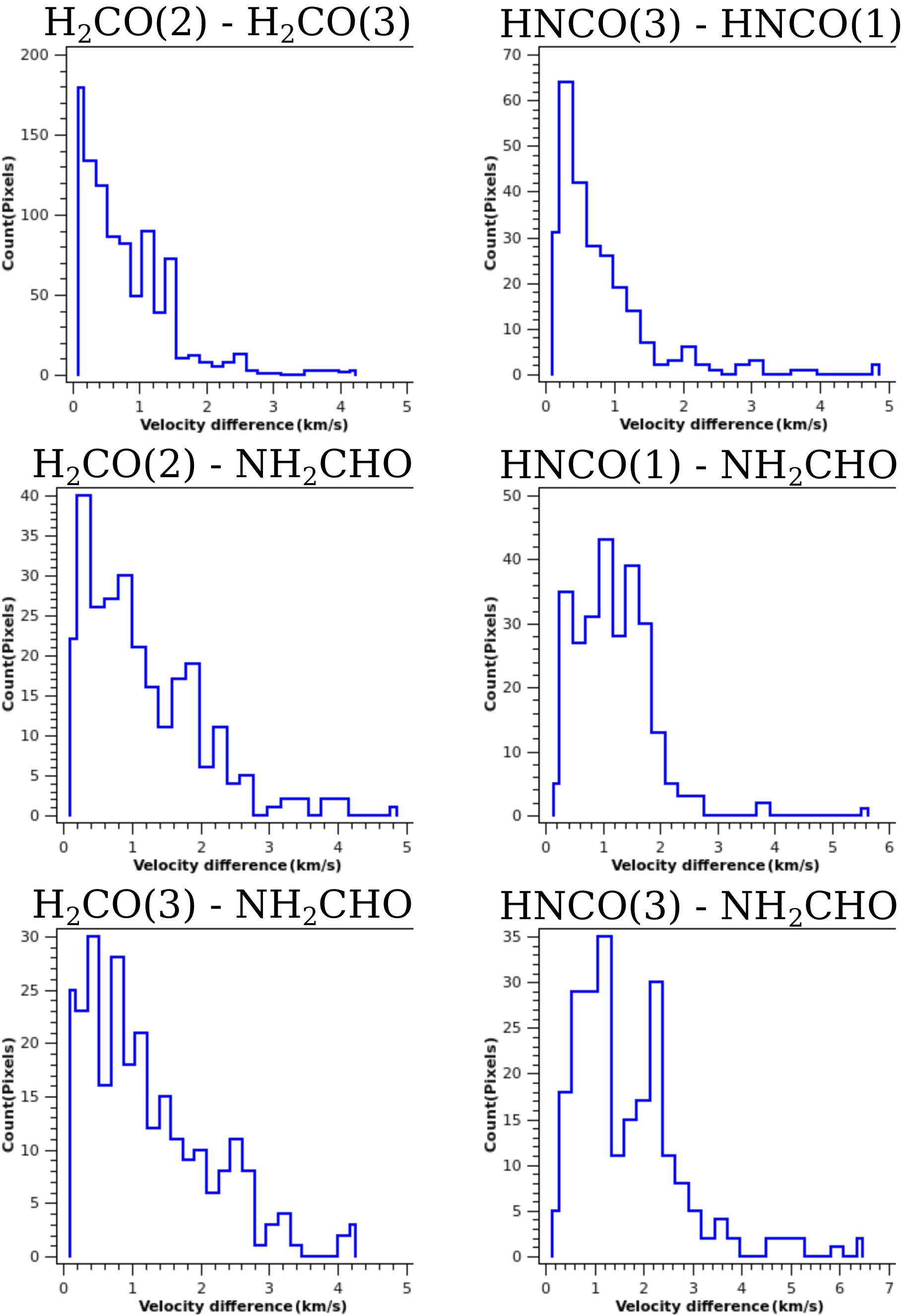}
      \caption{G24A2(S) first moment difference histogram.}
         \label{G24A2Shist}
   \end{figure*}

   \begin{figure*}[!thb]
   \centering
      \includegraphics[width=0.85\hsize]{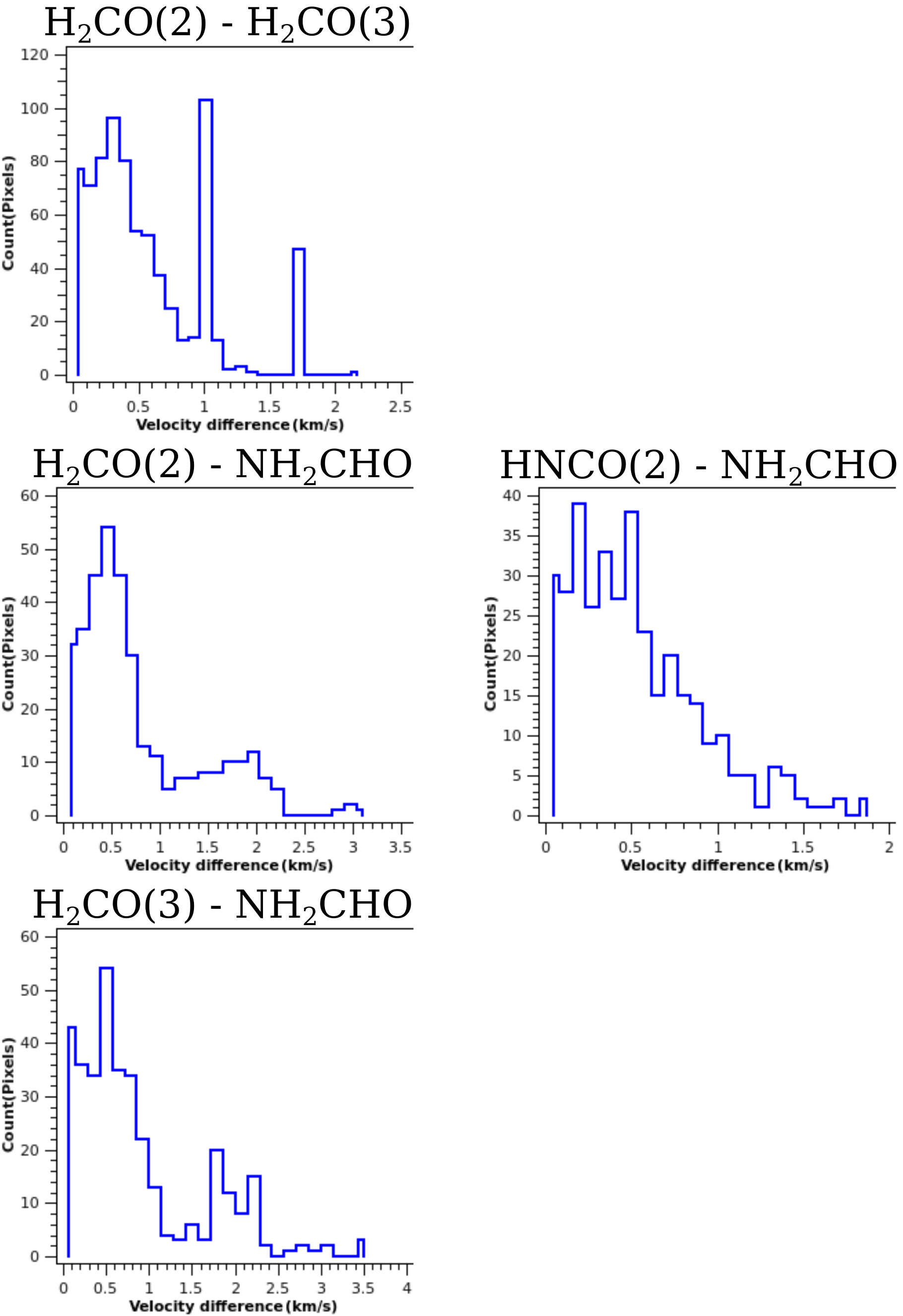}
      \caption{G345 Main first moment difference histogram.}
         \label{G34549mainhist}
   \end{figure*}

   \begin{figure*}[!thb]
   \centering
      \includegraphics[width=0.85\hsize]{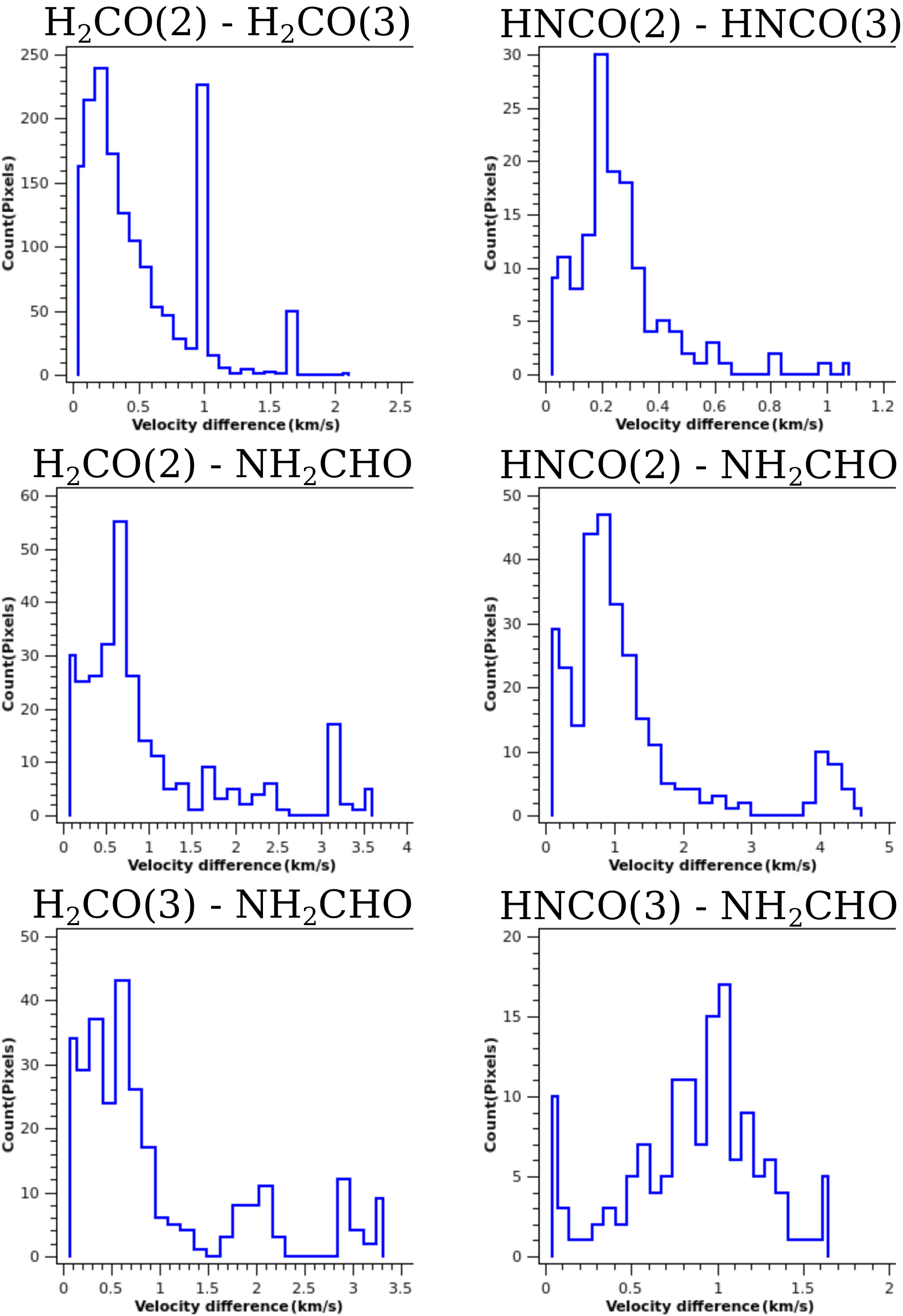}
      \caption{G345 NW spur first moment difference histogram.}
         \label{G34549nwspurhist}
   \end{figure*}

   \begin{figure*}[!thb]
   \centering
      \includegraphics[width=0.85\hsize]{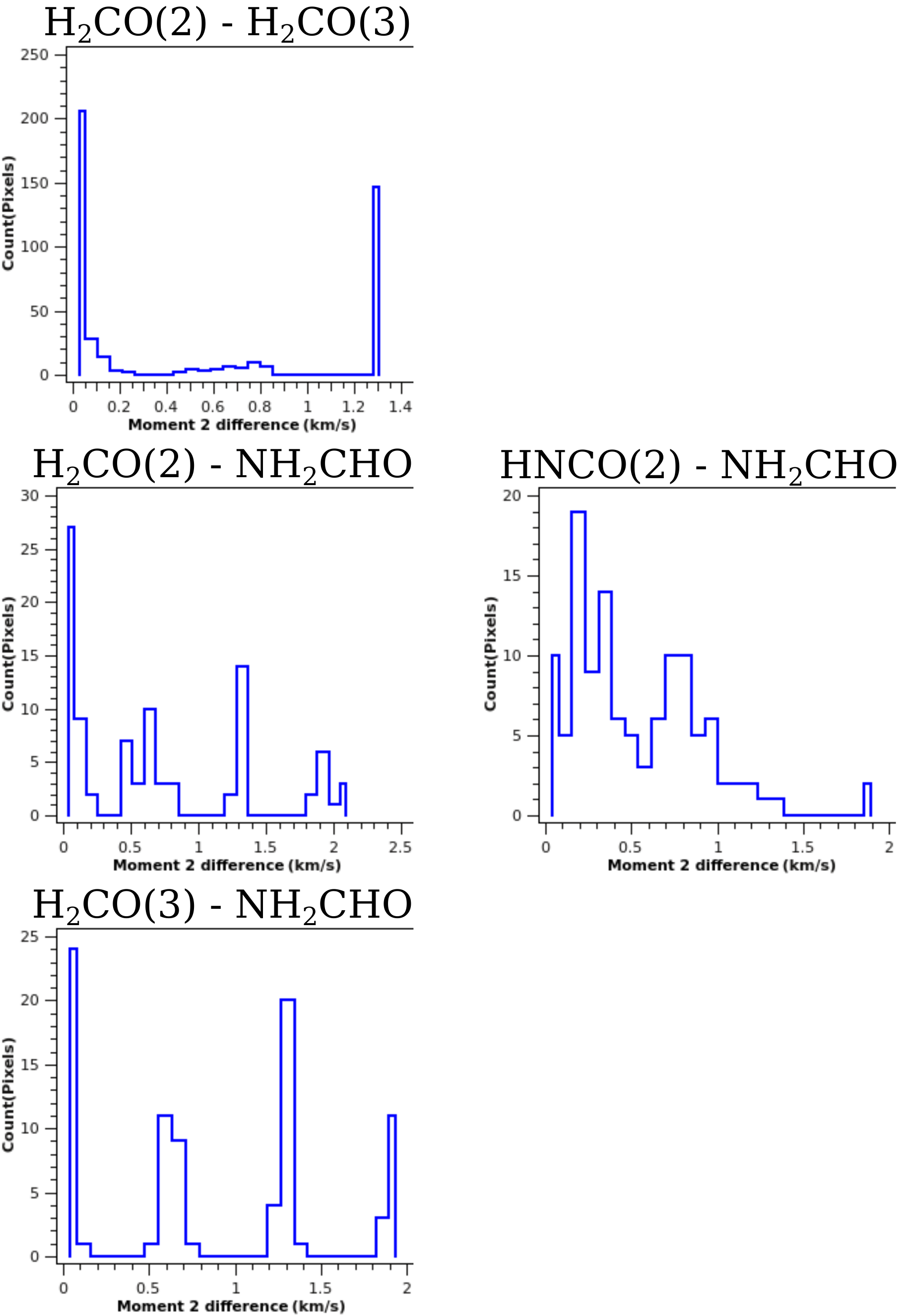}
      \caption{G17 second moment difference histogram.}
         \label{G17hist2}
   \end{figure*}

   \begin{figure*}[!thb]
   \centering
      \includegraphics[width=0.85\hsize]{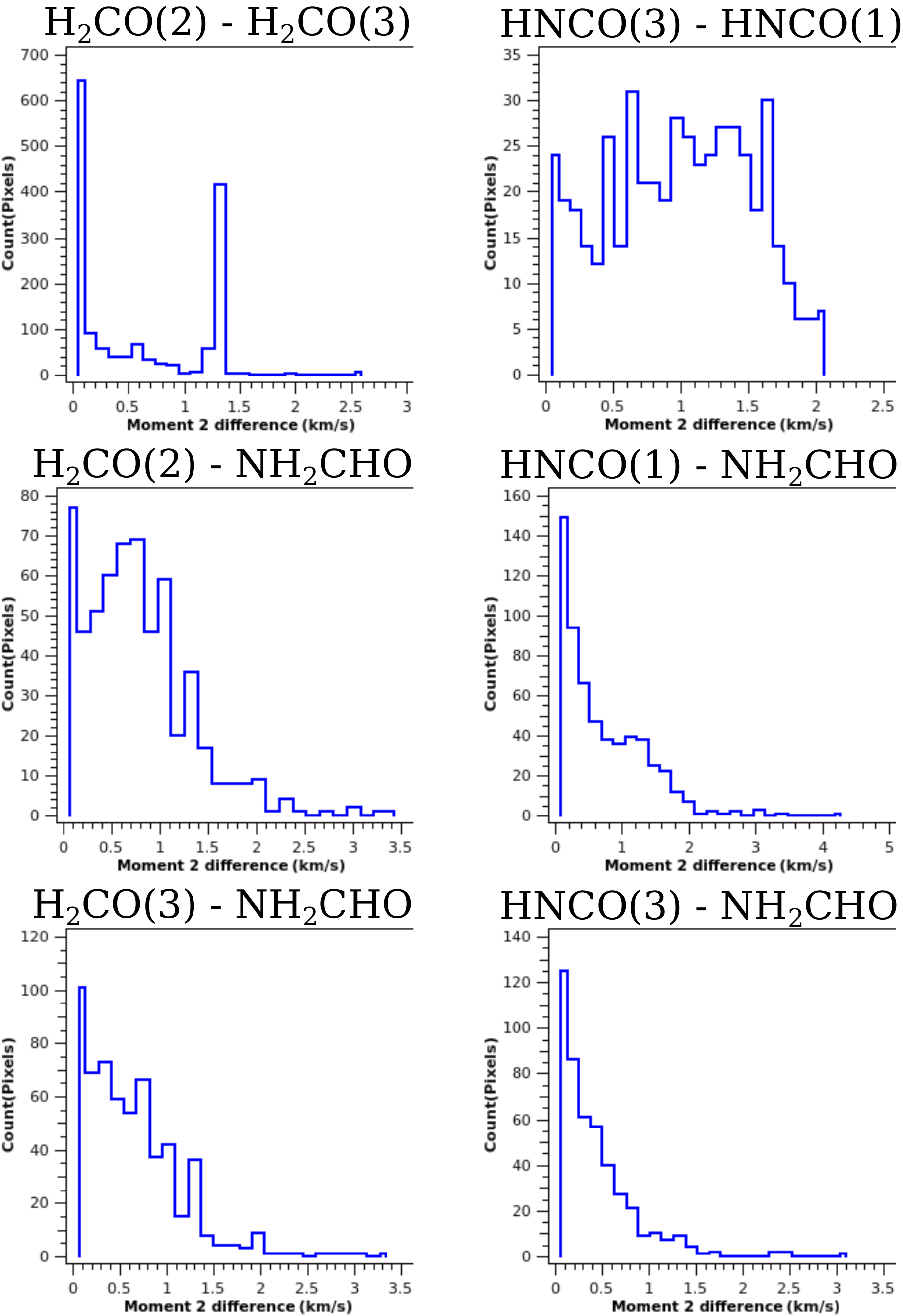}
      \caption{G24A1 second moment difference histogram}
         \label{G24A1hist2}
   \end{figure*}

   \begin{figure*}[!thb]
   \centering
      \includegraphics[width=0.85\hsize]{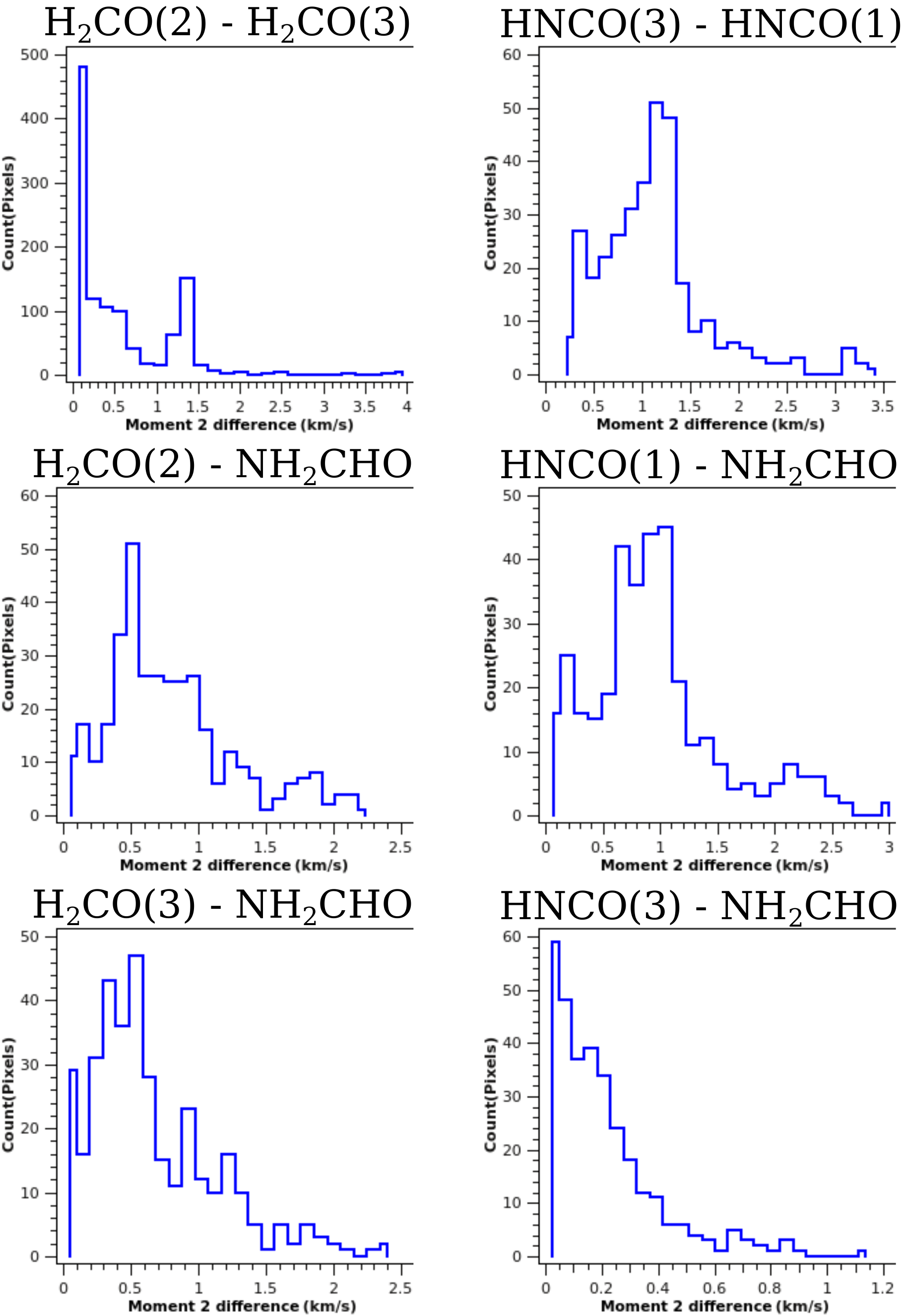}
      \caption{G24A2(N) second moment difference histogram.}
         \label{G24A2Nhist2}
   \end{figure*}

   \begin{figure*}[!thb]
   \centering
      \includegraphics[width=0.85\hsize]{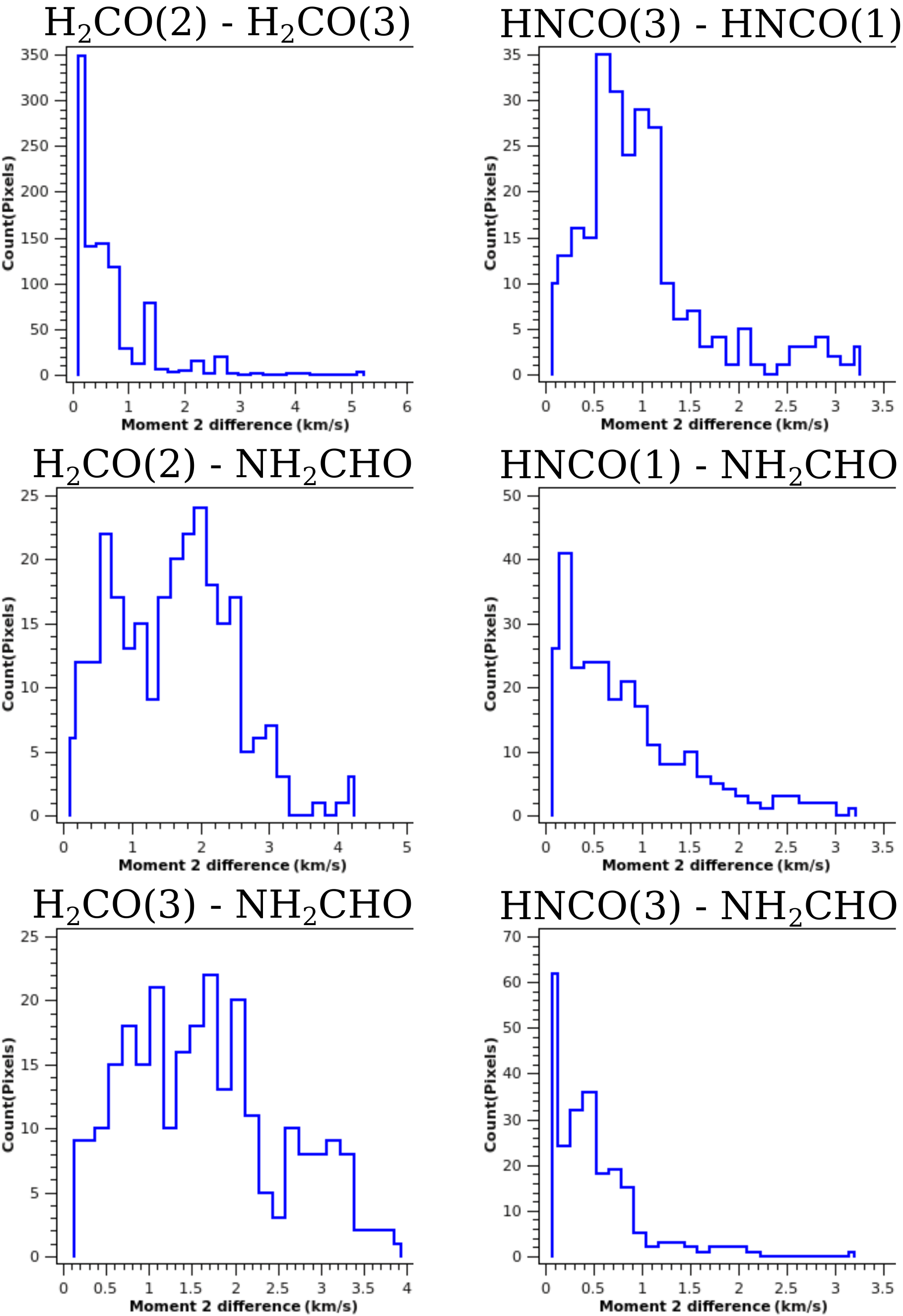}
      \caption{G24A2(S) second moment difference histogram.}
         \label{G24A2Shist2}
   \end{figure*}

   \begin{figure*}[!thb]
   \centering
      \includegraphics[width=0.85\hsize]{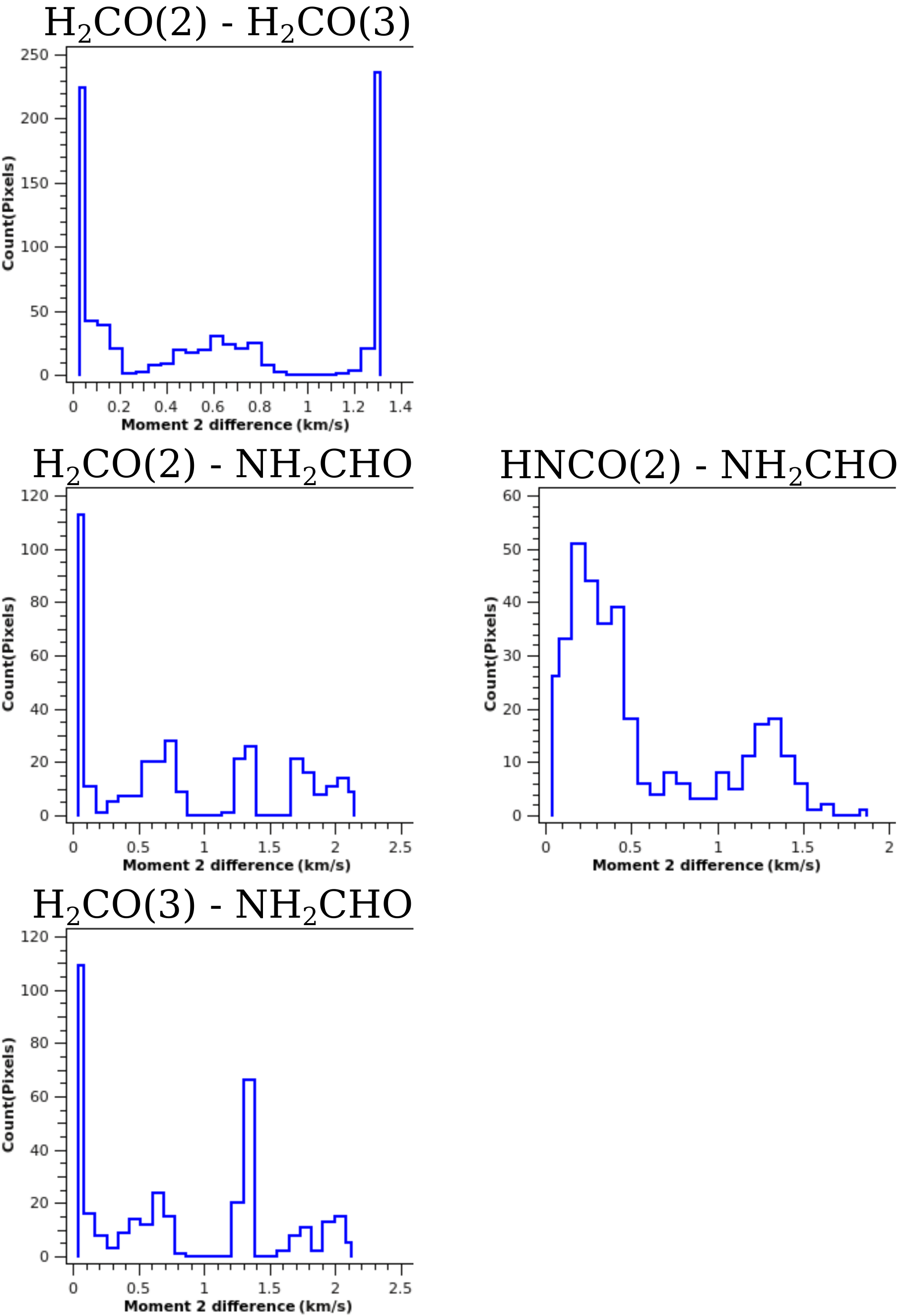}
      \caption{G345 Main second moment difference histogram.}
         \label{G34549mainhist2}
   \end{figure*}

   \begin{figure*}[!thb]
   \centering
      \includegraphics[width=0.85\hsize]{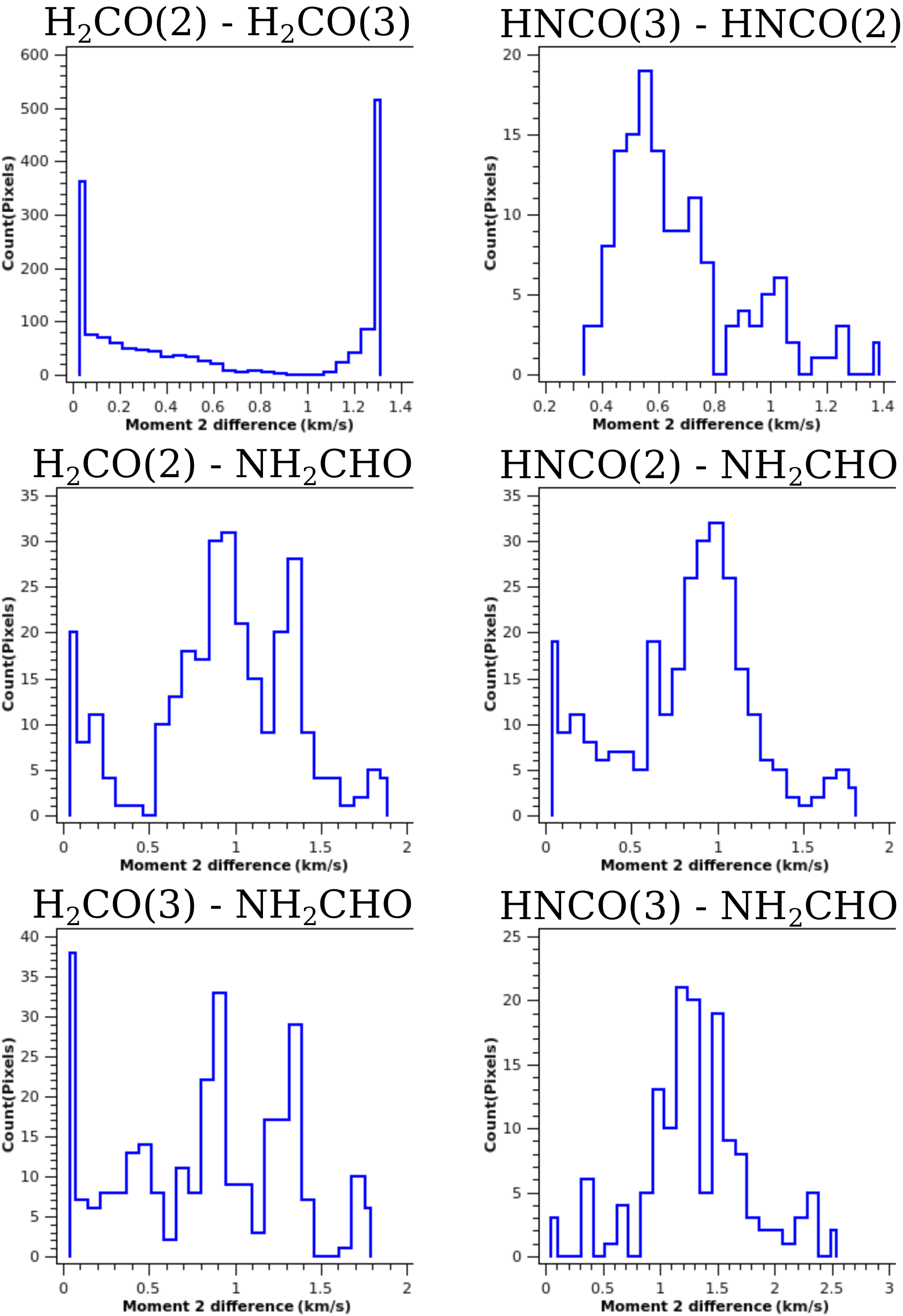}
      \caption{G345 NW spur second moment difference histogram.}
         \label{G34549nwspurhist2}
   \end{figure*}
\end{onecolumn}

\end{appendix}

\end{document}